\documentclass[fleqn,usenatbib]{mnras}

\usepackage{newtxtext,newtxmath}

\usepackage[T1]{fontenc}
\usepackage{ae,aecompl}

\usepackage{graphicx}	
\usepackage{amsmath}	
\usepackage{multicol} 
\usepackage{multirow}
\usepackage{comment}
\usepackage{array}
\usepackage{soul}
\hypersetup{colorlinks=true,citecolor=blue,linkcolor=purple,filecolor=black,runcolor=black,breaklinks=true}
\usepackage{etoolbox}
\usepackage{ae,aecompl}
\usepackage{color}
\usepackage{hyperref}
\usepackage{threeparttable}
\usepackage{mathrsfs}
\usepackage{scalerel}
\usepackage{longtable,booktabs,threeparttablex}
\usepackage{float}

\definecolor{purple}{RGB}{160,32,240}
\definecolor{red}{RGB}{225,50,50}

\definecolor{addchange}{RGB}{215,25,25}

\definecolor{removechange}{RGB}{25,25,215}

\newcommand{\HST}{\emph{HST}}
\newcommand{\JWST}{\emph{JWST}}
\newcommand{\Spitzer}{\emph{Spitzer}}
\newcommand{\Muv}{\ensuremath{\mathrm{M}_{\mathrm{UV}}^{ }}}

\newcommand{\Lya}{\ensuremath{\mathrm{Ly}\alpha}}
\newcommand{\zphot}{\ensuremath{z_{\mathrm{phot}}}}
\newcommand{\zspec}{\ensuremath{z_{\mathrm{spec}}}}

\newcommand{\zLya}{\ensuremath{z_{_{\mathrm{Ly\alpha}}}}}

\newcommand{\xiionObs}{\ensuremath{\xi_{\mathrm{ion}}^{}}}

\newcommand{\logxiionObs}{\ensuremath{\log[\xi_{\mathrm{ion}}^{} / (\mathrm{erg}^{-1}\ \mathrm{Hz})]}}

\newcommand{\fesc}{\ensuremath{f_{\mathrm{esc}}}}

\newcommand{\OIIIHb}{[OIII]+H\ensuremath{\beta}}
\newcommand{\BBrat}{\ensuremath{F_\nu}(4200\,\AA{})\,/\,\ensuremath{F_\nu}(3500\,\AA{})}

\newcommand{\Msol}{\ensuremath{M_{\odot}}}
\newcommand{\Mstar}{\ensuremath{M_{\ast}}}
\newcommand{\logMstar}{\ensuremath{\log\left(M_{\ast}/M_{\odot}\right)}}

\newcommand{\ageCSFH}{\ensuremath{\mathrm{age_{CSFH}}}}

\newcommand{\muEW}{\ensuremath{\mu_{\scaleto{\mathrm{EW}}{4pt}}}}
\newcommand{\sigmaEW}{\ensuremath{\sigma_{\scaleto{\mathrm{EW}}{4pt}}}}

\newcolumntype{P}[1]{>{\centering\arraybackslash}p{#1}}
\newcommand\Tstrut{\rule{0pt}{2.6ex}}         

\title[Properties of EoR Galaxies]{The Star-forming and Ionizing Properties of Dwarf \texorpdfstring{$\mathbf{z\sim6-9}$}{z ~ 6 - 9} Galaxies in JADES: Insights on Bursty Star Formation and Ionized Bubble Growth}

\author[R. Endsley et al.]{Ryan Endsley$^{1}$\thanks{E-mail: ryan.endsley@austin.utexas.edu},
Daniel P. Stark$^{2}$, 
Lily Whitler$^{2}$,
Michael W. Topping$^{2}$,
Benjamin D. Johnson$^{3}$,
\newauthor
Brant Robertson$^{4}$,
Sandro Tacchella$^{5,6}$,
Stacey Alberts$^{2}$,
William M. Baker$^{5,6}$,
Rachana Bhatawdekar$^{7}$, 
\newauthor
Kristan Boyett$^{8,9}$,
Andrew J. Bunker$^{10}$,
Alex J. Cameron$^{10}$,
Stefano Carniani$^{11}$, 
Stephane Charlot$^{12}$,
\newauthor
Zuyi Chen$^{2}$,
Jacopo Chevallard$^{10}$,
Emma Curtis-Lake$^{13}$,
A. Lola Danhaive$^{5,6}$,
Eiichi Egami$^{2}$,
\newauthor
Daniel J. Eisenstein$^{3}$
Kevin Hainline$^{2}$,
Jakob M. Helton$^{2}$,
Zhiyuan Ji$^{2}$,
Tobias J. Looser$^{5,6}$,
\newauthor
Roberto Maiolino$^{5,6,14}$,
Erica Nelson$^{15}$,
D{\'a}vid Pusk{\'a}s$^{5,6}$,
George Rieke$^{2}$,
Marcia Rieke$^{2}$,
Hans-Walter Rix$^{16}$,
\newauthor
Lester Sandles$^{5,6}$
Aayush Saxena$^{10,17}$,
Charlotte Simmonds$^{5,6}$,
Renske Smit$^{18}$,
Fengwu Sun$^{2}$,
\newauthor
Christina C. Williams$^{19}$,
Christopher N. A. Willmer$^{2}$,
Chris Willott$^{20}$, 
and Joris Witstok$^{5,6}$
\\
$^{1}$Department of Astronomy, University of Texas, Austin, TX 78712, USA\\
$^{2}$Steward Observatory, University of Arizona, 933 N Cherry Ave, Tucson, AZ 85721 USA\\
$^{3}$Center for Astrophysics $|$ Harvard \& Smithsonian, 60 Garden St., Cambridge MA 02138 USA\\
$^{4}$Department of Astronomy and Astrophysics University of California, Santa Cruz, 1156 High Street, Santa Cruz CA 96054, USA\\
$^{5}$Kavli Institute for Cosmology, University of Cambridge, Madingley Road, Cambridge, CB3 0HA, UK\\
$^{6}$Cavendish Laboratory, University of Cambridge, 19 JJ Thomson Avenue, Cambridge, CB3 0HE, UK\\
$^{7}$European Space Agency, ESA/ESTEC, Keplerlaan 1, 2201 AZ Noordwijk, NL\\
$^{8}$School of Physics, University of Melbourne, Parkville 3010, VIC, Australia\\
$^{9}$ARC Centre of Excellence for All Sky Astrophysics in 3 Dimensions (ASTRO 3D), Australia\\
$^{10}$Department of Physics, University of Oxford, Denys Wilkinson Building, Keble Road, Oxford OX1 3RH, UK\\
$^{11}$Scuola Normale Superiore, Piazza dei Cavalieri 7, I-56126 Pisa, Italy\\
$^{12}$Sorbonne Universit\'e, CNRS, UMR 7095, Institut d'Astrophysique de Paris, 98 bis bd Arago, 75014 Paris, France\\
$^{13}$Centre for Astrophysics Research, Department of Physics, Astronomy and Mathematics, University of Hertfordshire, Hatfield AL10 9AB, UK\\
$^{14}$Department of Physics and Astronomy, University College London, Gower Street, London WC1E 6BT, UK\\
$^{15}$Department for Astrophysical and Planetary Science, University of Colorado, Boulder, CO 80309, USA\\
$^{16}$Max-Planck-Institut f\"ur Astronomie, K\"onigstuhl 17, D-69117, Heidelberg, Germany\\
$^{17}$Department of Physics and Astronomy, University College London, Gower Street, London WC1E 6BT, UK\\
$^{18}$Astrophysics Research Institute, Liverpool John Moores University, 146 Brownlow Hill, Liverpool L3 5RF, UK\\
$^{19}$NSF’s National Optical-Infrared Astronomy Research Laboratory, 950 North Cherry Avenue, Tucson, AZ 85719, USA\\
$^{20}$NRC Herzberg, 5071 West Saanich Rd, Victoria, BC V9E 2E7, Canada
}

\date{Accepted XXX. Received YYY; in original form ZZZ}

\pubyear{2024}

\begin{document}
\label{firstpage}
\pagerange{\pageref{firstpage}--\pageref{lastpage}}
\maketitle


\begin{abstract}
\noindent Reionization is thought to be driven by faint star-forming galaxies, but characterizing this population has long remained very challenging. Here we utilize deep nine-band JADES/NIRCam imaging to study the star-forming and ionizing properties of 756 $z\sim6-9$ galaxies, including hundreds of very UV-faint objects ($M_\mathrm{UV}>-18$). The faintest ($m\sim30$) galaxies in our sample typically have stellar masses of $M_\ast\sim(1-3)\times10^7\ M_\odot$ and young light-weighted ages ($\sim$50 Myr), though some show strong Balmer breaks implying much older ages ($\sim$500 Myr). We find no evidence for extremely massive galaxies ($>3\times10^{10}\ M_\odot$) in our sample. We infer a strong (factor $>$2) decline in the typical [OIII]$+$H$\beta$ EWs towards very faint $z\sim6-9$ galaxies, yet a weak UV luminosity dependence on the H$\alpha$ EWs at $z\sim6$. We demonstrate that these EW trends can be explained if fainter galaxies have systematically lower metallicities as well as more recently-declining star formation histories relative to the most UV-luminous galaxies. Our data provide evidence that the brightest galaxies are frequently experiencing a recent strong upturn in SFR. We also discuss how the EW trends may be influenced by a strong correlation between $M_\mathrm{UV}$ and Lyman continuum escape fraction. This alternative explanation has dramatically different implications for the contribution of galaxies along the luminosity function to cosmic reionization. Finally, we quantify the photometric overdensities around two $z>7$ strong Ly$\alpha$ emitters. One Ly$\alpha$ emitter lies close to a strong photometric overdensity while the other shows no significant nearby overdensity, perhaps implying that not all strong $z>7$ Ly$\alpha$ emitters reside in large ionized bubbles.
\end{abstract}




\section{Introduction} \label{sec:intro}

The formation and assembly of galaxies within the first billion years of cosmic history directly influenced the large-scale ionization state of the Universe via the process of hydrogen reionization \citep[e.g.][]{Stark2016_ARAA,Dayal2018,Robertson2022}.
Recent Ly$\alpha$ forest measurements from a statistical sample of high-redshift quasars indicate that essentially all hydrogen atoms in the intergalactic medium (IGM) had been reionized by $z=5.3$ \citep{Bosman2022}, approximately 1.1 Gyr after the Big Bang.
Additional quasar and galaxy spectra at $z>6$, along with measurements of the cosmic microwave background, imply that reionization was about halfway complete $\sim$400 Myr earlier at $z\sim7-8$ \citep[e.g.][]{Davies2018,Mason2018_IGMneutralFrac,Planck2020,Yang2020_Poniuaena,Goto2021}.
The primary agents of hydrogen reionization appear most likely to be star-forming galaxies given that constraints on the $z\gtrsim6$ quasar luminosity function imply that active supermassive black holes were very rare at early times \citep[e.g.][]{Shen2020,Jiang2022,Matsuoka2023}. 

Prior to \JWST{}, over 1000 Lyman-break galaxies at $z>6$ had been identified from deep \textit{Hubble Space Telescope} (\HST{}) imaging \citep[e.g.][]{McLure2013,Atek2015a,Bouwens2015_LF,Bouwens2022_lensedLF,Finkelstein2015_LF,Finkelstein2022,McLeod2016,Livermore2017,Ishigaki2018,Oesch2018_z10LF,Salmon2020} as well as wide-area ground-based imaging \citep[e.g.][]{Bowler2014,Bowler2020,Ono2018,Stefanon2017_Brightestz89,Stefanon2019,Endsley2021_OIII,Harikane2022_LF,Donnan2023}.
These studies demonstrated that the $z\sim6-10$ galaxy UV continuum luminosity functions had very steep faint-end slopes ($\alpha \sim -2$; e.g. \citealt{Atek2015a,Livermore2017,Ishigaki2018,Bouwens2022_lensedLF}), clearly indicating that very UV-faint sources ($\Muv{} > -18$) greatly dominated the galaxy number counts during reionization.

The very high relative abundance of the faintest ($\Muv{} > -18$) $z\gtrsim6$ galaxies has long motivated efforts to characterize their physical properties.
Deep \HST{} observations revealed that these systems often exhibit very blue rest-UV continuum slopes ($-2.5 \lesssim \beta \lesssim -2$ where $f_\lambda \propto \lambda^\beta$; e.g. \citealt{Dunlop2012,Finkelstein2012,Rogers2013,Bouwens2014_beta,Bhatawdekar2021}) as well as very compact rest-UV morphologies (effective radii $r_e\lesssim$200 pc; \citealt{Shibuya2015,Bouwens2022_faintSizes,Neufeld2022}).
Based on results from local ($z\sim0.3$) samples, such properties hint that very UV-faint reionization-era galaxies might be efficient at leaking ionizing photons into the IGM \citep{Chisholm2022,Flury2022_LyCdiagnostics}.
However, detailed statistical studies of the star-forming and ionizing properties of this faint, abundant $z\gtrsim6$ population have been precluded by the lack of deep data probing the rest-frame optical portion of their spectral energy distributions (SEDs).

For much of the past 20 years, the only instrument capable of delivering any constraints on the rest-optical SEDs of high-redshift galaxies was the Infrared Array Camera (IRAC) on board the \textit{Spitzer Space Telescope}.
Even so, IRAC could only reach a 5$\sigma$ imaging sensitivity of $m\sim26.5$ ($\Muv{} \lesssim -20.5$ at $z\sim7$) in the deepest pointings (120 hours; \citealt{Labbe2013,Labbe2015,Oesch2013_z9LF,Stefanon2021_GREATS}).
While dedicated observations in lensing fields were able to push this sensitivity to intrinsically fainter UV luminosities \citep{Smit2014,Huang2016_SURFSUP,Strait2020}, statistical constraints on the rest-optical SEDs of the faintest reionization-era galaxies remained very limited.
Consequently, in the lead-up to \JWST{}, much attention was concentrated on understanding the physical properties of relatively bright ($\Muv{} \lesssim -20$) Lyman-break $z\gtrsim6$ galaxies \citep[e.g.][]{Laporte2021,Stefanon2021_SMF,Endsley2021_OIII,Castellano2022_BDF,Schouws2022_dust,Sommovigo2022,Tacchella2022_SFHs,Topping2022_REBELS,Witstok2022,Whitler2023_z7}, though even these analyses suffered from IRAC's poor angular resolution (FWHM$\sim$2 arcsec) and SED sampling (only two imaging filters at 3--5$\mu$m, both with broad bandpasses).

Within the past year alone, data from \JWST{} have greatly advanced our understanding of reionization-era galaxies.
Rest-optical spectra from the Near-Infrared Spectrograph (NIRSpec; \citealt{Jakobsen2022_NIRSpec,Boker2023_NIRSpec}) have generally revealed very strong nebular line emission (e.g. [OIII], H$\alpha$, H$\beta$) coupled with signatures of hard radiation fields ([OIII]/[OII] $>$ 10) and low gas-phase metallicities ($\lesssim$0.1 $Z_\odot$) among $z>6$ galaxies \citep[e.g.][]{ArrabalHaro2023_z8to10,Bunker2023_GNz11,Cameron2023_jadesISM,Curti2023,Fujimoto2023_CEERS,Larson2023_EGSY,Sanders2023_excitation,Saxena2023,Tacchella2023_NIRCamNIRSpec,Tang2023_CEERS}.
This is consistent with expectations from findings prior to \JWST{} \citep[e.g.][]{Labbe2013,Smit2014,Smit2015,Stark2015_CIII,Stark2015_CIV,Stark2017,RobertsBorsani2016,Laporte2017,Mainali2017,Mainali2018,Schmidt2017,deBarros2019,Hutchison2019,Tang2019,Tang2021_UVlines,Endsley2021_OIII,Endsley2021_LyA,Stefanon2022_IRACz8}.
Less expected sub-populations of $z>5$ galaxies are also emerging from early NIRSpec observations, including sources with (sometimes tentative) contributions from active galactic nuclei (AGN; \citealt{Harikane2023_AGN,Larson2023_EGSY,Kocevski2023,Maiolino2023_GNz11_AGN,Onoue2023_CEERS,Ubler2023}) as well as objects that are in a very inactive stage of star formation \citep{Looser2023,Strait2023}.

The Near-Infrared Camera (NIRCam; \citealt{Rieke2005_NIRCam,Rieke2023_NIRCam}) on-board \JWST{} is also proving to be an invaluable tool for studying the demographics of reionization-era galaxies.
This is not only because of NIRCam's dramatic improvement in sensitivity, angular resolution, and SED sampling over IRAC, but also because Lyman-break $z\gtrsim6$ selections with imaging are very efficient at yielding highly complete samples of objects with faint continua.
Much of the early NIRCam studies on $z\gtrsim6$ galaxies focused on data from the Early Release Science (ERS) CEERS \citep{Finkelstein2023_zg9} and GLASS \citep{Treu2022_glass} surveys which immediately pushed down to extremely deep \textit{Hubble} imaging depths ($m_\mathrm{AB}\sim29$ mag at 5$\sigma$) across 1--5$\mu$m with just $\sim$3 hours of observations per photometric band.
These data (among that from other early \JWST{}/NIRCam surveys) have delivered a wealth of insight into faint ($\Muv{} \lesssim -19$) $z\gtrsim6$ galaxies including their ages, stellar masses, nebular line strengths, UV slopes, dust attenuation strengths, and morphologies \citep[e.g.][]{Endsley2023_CEERS,Topping2022_blueSlopes,Barrufet2023,Chen2023,Dressler2023_glass,Furtak2023_SMACS,Hsiao2023,Labbe2023,Laporte2023,Leethochawalit2023,Rodighiero2023,Treu2023_sizes,Whitler2023_z10}.
While these studies have clearly advanced our understanding of reionization-era systems far beyond that possible prior to \JWST{}, a detailed statistical analysis of the properties of very faint ($\Muv{} > -18$) $z\gtrsim6$ galaxies has yet to be undertaken. 
Such an endeavor is clearly warranted as these very faint objects are often though to contribute substantially to cosmic reionization, and also are the likely progenitors of more typical galaxies at lower redshifts.

Here we take steps to address this shortcoming by utilizing NIRCam imaging taken as part of the JWST Advanced Deep Extragalactic Survey (JADES; \citealt{Eisenstein2023}).
JADES is a collaborative effort of the NIRCam, NIRSpec, and U.S. Mid-Infrared Instrument (MIRI; \citealt{Wright2023_MIRI}) teams utilizing coordinated parallels to maximize science outcomes across the $\approx$770 hours of allocated observing time.
All data in JADES are being taken over the Great Observatories Origins Deep Survey (GOODS) fields in the northern (GOODS-N) and southern (GOODS-S) hemispheres \citep{Giavalisco2004} which contain some of the deepest \textit{Hubble} imaging ever obtained \citep[e.g.][]{Ellis2013,Illingworth2013}.
By adding exceptionally deep ($m_\mathrm{AB}\sim30-31$ mag 5$\sigma$) NIRCam imaging in several bands from $\sim$1--5$\mu$m (often including two medium-band filters) across $>$200 arcmin$^2$ of the GOODS fields, JADES is opening a completely new window on the high-redshift Universe \citep[e.g.][]{Baker2023,CurtisLake2023_jades,Dressler2023_jades,Hainline2023,Helton2023_z5p4,Robertson2023_JADES,Tacchella2023_GNz11}.

The primary goal of this paper is to utilize the deep NIRCam imaging from JADES to measure the rest-UV+optical SEDs among a large sample of very faint ($\Muv{} > -18$) $z\sim6-9$ galaxies and statistically characterize their physical properties in detail, comparing with more luminous systems.
One of the key conclusions of this paper relates to the star formation histories of reionization-era galaxies, complementing the JADES/NIRCam investigation in \citet{Dressler2023_jades} which utilized broadband SEDs to infer the presence of multiple bursts by considering non-parametric SFHs.
Here, we utilize the two long-wavelength NIRCam medium bands from JADES to focus on insight from nebular line emission about variations in star formation activity on short ($\sim$3 Myr) timescales.

The structure of this paper is as follows.
We begin by describing the imaging data, source extraction, photometric calculations, sample selection, and photo-ionization modelling in \S\ref{sec:sec2}.
Next, we present and discuss the UV luminosities, photometric redshifts, stellar masses, and light-weighted ages among our sample (\S\ref{sec:properties}).
We then proceed to utilize photometric constraints on strong rest-optical nebular lines (\OIIIHb{} and H$\alpha$; \S\ref{sec:distributions}) to statistically analyze the star-forming and ionizing properties of relatively bright and very faint reionization-era galaxies (\S\ref{sec:discussion}).
Finally, we quantify the photometric overdensities surrounding two strong Ly$\alpha$ emitters at $z>7$ within the JADES footprints to improve our understanding of the connection between strong Ly$\alpha$ and ionized bubbles deep in the epoch of reionization (\S\ref{sec:overdensity}).
Our main conclusions are summarized in \S\ref{sec:summary}.

Throughout this paper, we quote all magnitudes in the AB system, report EWs in the rest frame, assume a \citet{Chabrier2003} initial mass function (IMF) with limits of 0.1--300 \Msol{}, and adopt a flat $\Lambda$CDM cosmology with parameters $h=0.7$, $\Omega_\mathrm{M}=0.3$, and $\Omega_\mathrm{\Lambda}=0.7$. 

\section{Imaging Data and Sample Selection} \label{sec:sec2}

In this section, we first describe the JADES/NIRCam imaging over the GOODS-S and GOODS-N fields as well as the complementary \textit{Hubble} Advanced Camera for Surveys (ACS) imaging that assists with our Lyman-break dropout selection (\S\ref{sec:imaging}).
We next detail our procedure for source extraction and photometric measurements in \S\ref{sec:photometry}, then describe our Lyman-break selections of $z\sim6$ and $z\sim7-9$ galaxies (\S\ref{sec:selection}).
Lastly, we describe the photo-ionization SED models used throughout this work to infer the physical properties of the galaxies within our sample (\S\ref{sec:models}).

\subsection{\JWST{} and \HST{} Imaging} \label{sec:imaging}

We consider all JADES/NIRCam imaging data taken prior to February 10, 2023 (see \citealt{Rieke2023_jades}).
These data include imaging in the broad-band F090W, F115W, F150W, F200W, F277W, F356W, and F444W filters\footnote{F070W imaging is also available over a single pointing as of Feb. 10, 2023. Because this represents a small fraction of the total JADES/NIRCam imaging at the time, we here ignore the F070W data.} as well as the F410M medium-band filter across the full footprints in both the GOODS-S and GOODS-N fields.
Additional medium-band F335M imaging is available over the majority of the imaging area taken up to Feb. 10, 2023.
Because the medium-band filters are particularly useful for determining whether galaxies show strong nebular line emission or continuum discontinuities (e.g. Balmer breaks), we here restrict our analysis to the $\approx$90 arcmin$^2$ of JADES/NIRCam imaging that include data in all nine of the above filters. This imaging is split approximately evenly into the GOODS-S ($\approx$44 arcmin$^2$) and GOODS-N fields ($\approx$46 arcmin$^2$).

The NIRCam imaging reduction algorithm applied here largely follows that described in previous works (e.g. \citealt{Endsley2023_CEERS,Merlin2022,Bezanson2022,Bagley2023,Donnan2023,Rieke2023_jades,Robertson2023_JADES,Tacchella2023_GNz11,Williams2023_jems}) and is detailed below.
We begin by processing the uncalibrated ({\tt *\_rate.fits}) NIRCam exposures through the stage 1 step of the \JWST{} Science Calibration Pipeline\footnote{\url{https://jwst-pipeline.readthedocs.io/en/latest/index.html}} (v1.9.4) using the \JWST{} Calibration Reference
Data System (CRDS) context map {\tt jwst\_1045.pmap}.
During this step, we implement a custom snowball \citep{Rigby2023} masking algorithm largely based on the methods described in \citet{Bagley2023}.
From the output {\tt *\_uncal.fits} files, we subtract prominent artifacts (e.g. wisps; \citealt{Rigby2023}) from the short-wavelength NIRCam imaging data using custom-built templates.
These custom templates were created by combining all F090W, F115W, F150W, and F200W imaging data (as of Feb. 1 2023) from JADES as well as the deep, blank-field public programs CEERS \citep{Finkelstein2023_zg9}, PRIMER (PI J. Dunlop), and NGDEEP \citep{Bagley2023_ngdeep}. 
All snowball-masked and artifact-subtracted {\tt *\_uncal.fits} files are visually inspected; for a very small fraction of imaging exposures, we implement additional masking of remaining prominent artifacts or remove the exposures from the reduction entirely if the overall data quality appears poor.
From the resulting files, we build custom sky flats in all JADES imaging bands considered here, folding in additional data from CEERS, PRIMER, NGDEEP, and FRESCO \citep{Oesch2023_fresco}.

The {\tt *\_uncal.fits} files are then processed through the stage 2 step of the \JWST{} Science Calibration Pipeline with our custom sky flats and the photometric zeropoints from \citet{Boyer2022} that remain the most recent values implemented in CRDS.
From the resulting calibrated {\tt *\_cal.fits}, we then subtract off 1/f noise \citep{Schlawin2020} using the sigma-clipped median values along given rows (on an amplifier-by-amplifier basis) and columns.
Next, we subtract off the 2D background using the \textsc{sep} package \citep{Barbary2016_sep} and follow the methods described in \citet{Bagley2023}.
The final {\tt *\_cal.fits} files are then processed through the stage 3 step of the Calibration Pipeline to create mosaics with a pixel scale of 30 mas/pixel.
Using the \textsc{tweakreg} software\footnote{\url{https://drizzlepac.readthedocs.io/en/latest/tweakreg.html}}, these mosaics are astrometrically-matched to the \HST{} imaging over the GOODS fields that have been registered to the \textit{Gaia} frame (see below).
With the primary intention of improving the astrometric alignment, we fold in F444W data from FRESCO (which significantly extends the covered area in each GOODS field) into our reduction and first align the F444W mosaics to the \HST{}/WFC3 F160W images, and then subsequently align all other NIRCam mosaics to that of F444W in the respective field.

For our dropout selection of $z\sim6-9$ galaxies, we also utilize imaging from \HST{} taken over the GOODS fields.
The \HST{} mosaics adopted here come from the \textit{Hubble} Legacy Field archive (HLF; see \citealt{Illingworth2013,Whitaker2019} and references therein) and are registered to the \textit{Gaia} frame using the astrometry from the CHArGE images (G. Brammer private communication) as described in \citet{Williams2023_jems}.
All HLF mosaics used here (v2.0) have the same pixel scale as the final NIRCam mosaics (30 mas/pixel).
Here we focus on using the optical ACS imaging for our dropout selections, which includes data in the F435W, F606W, F775W, F814W, and F850LP bands.

To obtain reliable photometric colors across the 0.4--5$\mu$m SEDs, we must account for the fact that the angular resolution of the ACS and NIRCam bands considered here differ by up to a factor of $\approx$4.
We therefore convolve all mosaics to the PSF of F444W which has the poorest angular-resolution among these bands (FWHM$\approx$0.15 arcsec).
We refer the interested reader to \citet{Endsley2023_CEERS} for details on our procedure for constructing the empirical PSFs (one per band and field) and creating the convolution kernels using \textsc{photutils} \citep{Bradley2022_photutils}.

\begin{table}
\centering
\caption{Summary of the 5$\sigma$ depths achieved in each band for our final Lyman-break sample of F775W and F090W dropouts. Because the depth for a given source depends not only on its position within the inhomogeneous GOODS HST and NIRCam imaging, but also the source size, we report the 10th, 50th, and 90th percentile 5$\sigma$ depths in each band to show the range relevant to our sample. All photometry is computed on mosaics convolved to the F444W PSF (FWHM=0.15\arcsec{}).}
\begin{tabular}{P{1.0cm}P{1.7cm}P{1.7cm}P{1.7cm}}
\hline
\multirow{2}{*}{Band} & \multicolumn{3}{c}{5$\sigma$ depths of $z\sim6-9$ sample [AB mag]} \Tstrut{} \\[1pt]
 & 10$^\mathrm{th}$ percentile & 50$^\mathrm{th}$ percentile & 90$^\mathrm{th}$ percentile \\[2pt]
\hline
F435W & 28.0 & 28.6 & 30.0 \Tstrut{} \\[1pt]
F606W & 28.1 & 28.7 & 30.3 \\[1pt]
F775W & 27.6 & 28.4 & 30.2 \\[1pt]
F814W & 28.0 & 28.5 & 29.0 \\[1pt]
F850LP & 27.2 & 27.8 & 29.3 \\[1pt]
F090W & 28.1 & 29.2 & 29.9 \\[1pt]
F115W & 28.5 & 29.5 & 30.2 \\[1pt]
F150W & 28.4 & 29.4 & 30.1 \\[1pt]
F200W & 28.5 & 29.5 & 30.1 \\[1pt]
F277W & 28.7 & 29.7 & 30.4 \\[1pt]
F335M & 28.1 & 29.1 & 29.6 \\[1pt]
F356W & 28.8 & 29.7 & 30.2 \\[1pt]
F410M & 28.0 & 29.0 & 29.7 \\[1pt]
F444W & 28.2 & 28.9 & 29.5 \\[1pt]
\hline
\end{tabular}
\label{tab:imaging}
\end{table}

\subsection{Source Extraction and Photometry} \label{sec:photometry}

To identify objects across the JADES footprints, we run \textsc{Source Extractor} \citep{Bertin1996} on an inverse-variance weighted stack of the PSF-convolved F200W, F277W, F335M, F356W, F410M, and F444W mosaics\footnote{We do not expect that adopting such a red-sided detection image is strongly biasing our selection against very blue high-redshift galaxies. \citet{Topping2023_JADES} adopted the same detection image and nonetheless found a large sample of high-redshift galaxies with robust very blue UV slopes ($\beta < -2.8$).}.
The photometry is computed in Kron \citep{Kron1980} apertures following commonly-used procedures for high-redshift galaxies \citep[e.g.][]{Oesch2013_z9LF,Bouwens2015_LF,Finkelstein2015_LF,Endsley2021_OIII,Endsley2023_CEERS}. 
First, we measure the photometry on the PSF-matched mosaics of each band within elliptical apertures with a \citet{Kron1980} factor of $k=1.2$, as these apertures have been shown to yield optimal S/N \citep{Finkelstein2022}.
The photometric errors are determined separately for each object and each band to take into account both the size and shape of the aperture, as well as the local background level from the varying exposure times across the mosaics.
Specifically, the photometric errors are measured as the standard deviation of flux values computed within randomly-placed elliptical apertures of the size/shape of interest) in nearby empty regions of the mosaic (as determined using \textsc{sep}) with similar background pixel flux variance as the source of interest.

The photometry in the $k=1.2$ apertures (and the associated error values) are corrected to total flux in two stages.
The fluxes and their associated errors are first multiplied by the ratio of the flux measured in $k=2.5$ apertures divided by that of the flux in the $k=1.2$ apertures, with these measurements performed on the inverse-variance weighted stack used for the \textsc{source extractor} detection.
The second stage is correcting for flux outside the $k=2.5$ apertures by assuming a point-source light profile outside these apertures and adopting our empirical F444W PSFs.
We have verified that our measured photometry for $z\sim6-9$ galaxies located in the current public release region of JADES (see \citealt{Rieke2023_jades}) is broadly consistent with the photometry released in that associated catalog.

Because of NIRCam's sensitivity, a significant number of our identified Lyman-break dropout $z\sim6-9$ galaxies have $k=2.5$ apertures that contain one or more different objects, artificially boosting their recovered total fluxes.
We therefore utilize the neighbor subtraction algorithm described in \citet{Endsley2023_CEERS} to subtract off any neighboring object within the $k=2.5$ aperture (as determined by the \textsc{source extractor} segmentation map) and recompute the photometry on those images.
Moreover, because the morphologies of $z\gtrsim6$ galaxies are sometimes clumpy \citep[e.g.][]{Chen2023,Hainline2023,Treu2023_sizes}, the galaxies we intend to study can be identified as multiple separate objects by \textsc{source extractor}.
If two or more nearby objects satisfy one of our Lyman-break dropout selections ($z\sim6$ or $z\sim7-9$; see below) and have overlapping $k=2.5$ apertures, we determine the smallest elliptical aperture that contains all pixels from their combined segmentation maps and recompute the photometry in that aperture, subtracting off other nearby objects when appropriate.

In Table \ref{tab:imaging}, we summarize the range of 5$\sigma$ depths in each HST and NIRCam band among our final Lyman-break galaxy sample of F775W and F090W dropouts (see \S\ref{sec:selection}). 
Given the inhomogeneous exposure maps of the GOODS imaging as well as our Kron aperture photometry, the achieved depth for a given galaxy depends on both its on-sky location as well as its angular size. 
We therefore report the 10th, 50th, and 90th percentile 5$\sigma$ depths in each band to show the range relevant to our sample, noting that this is the depth achieved using the PSF-matched mosaics.
The median 5$\sigma$ depth of our final $z\sim6-9$ galaxy sample is $m_\mathrm{AB} \approx 28.5$ in F435W, F606W, F775W, and F814W, though $\approx$0.7 mag shallower in F850LP ($m_\mathrm{AB} = 27.8$).
In NIRCam, the typical 5$\sigma$ depths range from $m_\mathrm{AB} \approx 29.0$ in F410M and F444W to $m_\mathrm{AB} \approx 29.7$ in F277W and F356W.
The deepest imaging in GOODS-S reaches 5$\sigma$ depths of $m_\mathrm{AB} \approx 30.2$ in several ACS and NIRCam bands for compact objects.

\subsection{Selection of Lyman-break \texorpdfstring{$\mathbf{z\sim6-9}$}{z ~ 6 - 9} Galaxies} \label{sec:selection}

Our primary goal is to statistically characterize the physical properties of very UV-faint reionization-era galaxies for the first time, and moreover compare their properties to that of the more UV-luminous population.
We therefore restrict our analysis to galaxies at $z\sim6-9$ where the lower bound reflects when reionization remains significantly incomplete \citep[e.g.][]{Nasir2020,Bosman2022,Zhu2022} while the upper bound of $z\sim9$ is chosen such that the NIRCam SEDs remain sensitive to the \OIIIHb{} emission line set, a key photometric probe of high-redshift galaxy properties \citep[e.g.][]{RobertsBorsani2016,Stark2017,Tang2019,Endsley2021_OIII}.

Following the approach of many previous studies \citep[e.g.][]{Bunker2004,Bouwens2015_LF,Ono2018,Stefanon2019,Endsley2023_CEERS,Leethochawalit2023}, we identify $z\sim6-9$ galaxies via a Lyman-break color selection.
These Lyman-break selections rely on the appearance of a sharp spectral discontinuity at $\lambda_\mathrm{rest} = 1216$ \AA{} that is imposed by very strong absorption blueward of the Ly$\alpha$ line from HI in the intervening IGM (e.g. \citealt{Inoue2014}). 
Other studies have also selected high-redshift galaxies via photometric redshifts, including the study of \citet{Hainline2023} which aimed at establishing a census of $z>8$ galaxies within the JADES/NIRCam data set and providing a first discussion of their colors and morphologies.
Such photometric redshift selections have the advantage of folding in all photometric data points in determining the probability that a given object lies in the redshift interval of interest.
However, here we explicitly choose not to select on photometric redshifts as doing so may bias our sample preferentially towards objects with strong rest-optical lines (which imprint unique long-wavelength NIRCam color patterns) and the goal in this work is to characterize reionization-era galaxy physical properties.
In Appendix \ref{app:photoz}, we compute the photometric redshifts for all objects in our sample excluding the rest-optical photometry during the fits.
The large majority of sources have a high probability ($>$90\%) of lying at high redshift ($z>4$; see Fig. \ref{fig:photozCdf}), and we have verified that accounting for photometric redshifts has no significant impact on our main results regarding the statistical ionizing and star-forming properties of our sample (see Appendix \ref{app:photoz}).

Because the Ly$\alpha$ break spans a considerable range in observed wavelength across our targeted $z\sim6-9$ interval, we divide our selection according to two separate sets of criteria, one for $z\sim6$ and one for $z\sim7-9$, which we discuss in sequence below.
At $z\sim6$, the Ly$\alpha$ break is located at $\approx$0.85$\mu$m and thus galaxies at this redshift should appear as strong ACS/F775W dropouts \citep[e.g.][]{Stanway2003,Bunker2004,Bouwens2004}.
We therefore begin our $z\sim6$ galaxy selection with the following color cuts:
\begin{enumerate}
    \item F775W $-$ F090W $>$ 1.2
    \item F090W $-$ F150W $<$ 1.0
    \item F775W $-$ F090W $>$ F090W $-$ F150W $+$ 1.2.
\end{enumerate}
These criteria enforce the presence of a strong break between the F775W and F090W bands (at least 1.2 mags) as well as a much flatter color between F090W and F150W. 
The second color cut above sets an upper redshift limit of $z\approx6.5$ where the Ly$\alpha$ break redshifts well into F090W. 
In these cuts, the F775W flux is set to its 1$\sigma$ upper limit in cases where the S/N$<$1 in that band following past Lyman-break color cut selections in the literature (e.g. \citealt{Bouwens2015_LF,Endsley2021_OIII}). 

Additional analytic cuts are imposed to verify the presence of a strong \Lya{} break.
First, we enforce S/N$<$2 in ACS/F435W since this band lies fully blueward of the Lyman-continuum break at $z>5.5$ and all Lyman-continuum photons at these redshifts face extremely strong IGM attenuation.
Second, we ensure that a strong dropout is also seen in ACS/F606W with either F606W $-$ F090W $>$ 2.7, or F606W $-$ F090W $>$ 1.8 if the S/N(F606W)$<$2, again setting the F606W flux to its 1$\sigma$ upper limit when S/N$<$1.
However, we ignore the cut in F435W S/N as well as the F606W dropout criteria for objects with extremely strong Ly$\alpha$ breaks (F775W $-$ F090W $>$ 2.5) which can confidently be identified as $z\sim6$ galaxies even if photometric scatter impacts the F435W or F606W fluxes.
Finally, to ensure each source is real, we enforce the criteria that every $z\sim6$ galaxy in our sample be detected at S/N$>$5 in at least one NIRCam band, as well as detected at S/N$>$3 in at least three NIRCam bands in addition to either ACS/F814W or ACS/F850LP.

Our $z\sim7-9$ Lyman-break selection largely follows the same logic of the $z\sim6$ criteria described above. 
Specifically, we begin with the color cuts
\begin{enumerate}
    \item F090W $-$ F115W $>$ 1.5
    \item F115W $-$ F200W $<$ 1.2
    \item F090W $-$ F115W $>$ F115W $-$ F200W $+$ 1.5.
\end{enumerate}
These cuts are satisfied by galaxies at $z\approx7.0-9.0$ with blue rest-UV colors ($-2.5 \lesssim \beta \lesssim -2.0$), as well as extremely red ($-1.0 \lesssim \beta \lesssim 0.0$) objects at $z\approx7-8$ where the Lyman-alpha break has not yet shifted into F115W.
As with the $z\sim6$ criteria, we also impose the condition that S/N$<$2 in F435W.
For the $z\sim7-9$ selection, we additionally utilize the $\chi^2_\mathrm{opt}$ parameter defined in \citet{Bouwens2015_LF} as $\chi^2_\mathrm{opt} \equiv \sum \mathrm{abs}(f)/f\, \left(f/\sigma\right)^2$ where $f$ and $\sigma$ represent, respectively, the measured flux density and its error in a given band, while $\mathrm{abs}(f)$ is the absolute value of the flux density.
After summing over the ACS F435W, F606W, and F775W bands, we enforce $\chi^2_\mathrm{opt} < 5$.
However, we ignore the S/N cut in F435W as well as the $\chi^2_\mathrm{opt}$ cut for objects with extremely strong Ly$\alpha$ breaks (F090W $-$ F115W $>$ 2.5).
Every selected $z\sim7-9$ galaxy must be detected at S/N$>$5 in at least one NIRCam band redward of F090W, as well as detected at S/N$>$3 in at least three such bands.

We enforce a final set of cuts to both the $z\sim6$ and $z\sim7-9$ samples to ensure that the NIRCam data are sufficiently sensitive for rest-optical color measurements (tracing e.g. Balmer breaks or nebular line EWs).
Following the IRAC-based approach of \citet{Endsley2021_OIII}, we require that $f(\mathrm{FUV}) / e(\mathrm{X}) > 3$ where $f(\mathrm{FUV})$ is the far-UV flux density corresponding to the inferred \Muv{} value from the constant star formation history SED fits (see \S\ref{sec:models}) while $e(\mathrm{X})$ is the 1$\sigma$ uncertainty in the flux density of band $X$.
Here, this cut is enforced with the four reddest bands in the JADES filter set (i.e. $X \in $ \{F335M, F356W, F410M, F444W\}).

Every selected $z\sim6-9$ object was visually inspected in all HST and NIRCam mosaics (both at original and PSF-matched resolution) to remove spurious sources due to artifacts (mostly diffraction spikes) or diffuse emission from large, low-redshift objects. 
We also remove bright point-source objects that show colors consistent with brown dwarfs, which can mimic a strong F090W dropout.
The measured colors are compared to the empirical SPEX 0.8--2.5$\mu$m brown dwarf spectral library \citep{Burgasser2014_SPEX}, as well as the Sonora model brown dwarf spectral templates \citep{Marley2021_sonora}.
After performing the final steps of neighbor subtraction and combined aperture photometry for the appropriate set of objects (see \S\ref{sec:photometry}), we end up with a final sample of 280 F775W dropouts ($z\sim6$) and 480 F090W dropouts ($z\sim7-9$) across the $\approx$90 arcmin$^2$ area with coverage in ACS F435W, F606W, F775W, F814W, and F850LP as well as NIRCam F090W, F115W, F150W, F200W, F277W, F335M, F356W, F410M, and F444W. 

\subsection{Photoionization SED Modelling} \label{sec:models}

To infer the physical properties of each Lyman-break $z\sim6-9$ galaxy in our JADES sample, we fit their 14-band ACS+NIRCam photometry with star-forming photoionization models.
One of the primary goals of this work is to quantify the stellar masses implied by the rest-UV+optical SEDs.
As has been discussed previously \citep[e.g.][]{Carnall2019_SFH,Leja2019,Lower2020,Johnson2021,Endsley2023_CEERS,Tacchella2022_SFHs,Topping2022_REBELS,Whitler2023_z7,Whitler2023_z10}, stellar mass estimates can change significantly depending on the assumed star formation history (SFH).
We therefore fit every galaxy in our sample with four different sets of models to quantify the systemic stellar mass uncertainties.
Each of these four models is described in turn below, though we first note the model assumptions applied in every case.

For all four sets of SED model fits described below, we adopt a \citet{Chabrier2003} stellar initial mass function (IMF) with bounds 0.1--300 \Msol{} as well as an SMC dust attenuation curve \citep{Pei1992}.
Moreover, in each case, we fit with log-uniform priors on stellar mass in the range $5 \leq \logMstar{} \leq 12$, on V-band dust optical depth in the range $-3 \leq$ log $\tau_{_\mathrm{V}}$ $\leq 0.7$, on ionization parameter in the range $-4 \leq$ log $U$ $\leq -1$, and on metallicity in the range $-2.2 \leq$ log($Z/Z_\odot$) $\leq -0.3$.
The upper limit of $\approx$50\% $Z_\odot$ is set to avoid unphysically high metallicities for the faint ($\Muv{} \gtrsim -20$) reionization-era galaxies that vastly dominate our sample.
Moreover, the stellar and ISM metallicities are equivalent in all models considered here.
We leave implementations of e.g. a stellar mass to metallicity relation prior or $\alpha$-enhanced metallicity models for future work.

All F775W and F090W dropouts are fit in the redshift range $z=4-8$ and $z=6-10$, respectively, with uniform priors adopting the IGM attenuation model of \citet{Inoue2014}.
We intentionally choose not to make sample cuts based on photometric redshifts as this would likely bias our sample towards $z\sim6-9$ objects with strong nebular lines (and hence young light-weighted ages) given the unique NIRCam color patterns caused by \OIIIHb{} and H$\alpha$ lines at high redshifts.

\textbf{\textsc{beagle} Constant SFH models:} We first consider models adopting a constant SFH (hereafter CSFH) using the BayEsian Analysis of GaLaxy sEds (\textsc{beagle}) SED-fitting code \citep{Chevallard2016}.
\textsc{beagle} fits photometry against the suite of star-forming photoionization SED models described in \citet{Gutkin2016} which utilize isochrones computed by the PAdova and
TRieste Stellar Evolution Code (\textsc{parsec}; \citealt{Bressan2012_parsec,Chen2015_parsec}).
The Bayesian \textsc{multinest} algorithm \citep{Feroz2008,Feroz2009} is implemented in \textsc{beagle} to determine the posterior probability distribution for each free physical parameter in the fit.
In the \textsc{beagle} CSFH fits, the galaxy age is fit between 1 Myr and the age of the Universe at the sampled redshift with a log-uniform prior, where no star formation is assumed to have occurred prior to the fitted age in these models.
Throughout this work, we refer to these \textsc{beagle} CSFH ages as the light-weighted ages of the galaxies.

\textbf{\textsc{beagle} Two-component SFH models:} We allow for more flexible SFH models within context of \textsc{beagle} following the approach of previous works \citep[e.g.][]{Stark2017,Endsley2021_OIII,Castellano2022_BDF}.
Specifically, we adopt a two-component SFH (hereafter TcSFH) composed of 1) a delayed $\tau$ model component (SFR $\propto t\ e^{-t/\tau_{_\mathrm{SF}}}$ where $t$ is the time since the onset of star formation) and 2) a constant SFH component.
The onset of the delayed $\tau$ model component is assumed to have been between 10$^{1.35}$ ($\approx$22) Myr age up to the age of the Universe at the fitted redshift, while the constant SFH component defines the SFH over the most recent 1--20 Myr (log-uniform prior for both components).
The SFR from the delayed component is not added to the constant SFH component over its fitted recent time interval, allowing for both strong recent increases in the SFR, as well as strong recent drops in the SFR (and anything between).
For the delayed $\tau$ component, the $\tau_{_\mathrm{SF}}$ parameter is fit with a log-uniform prior in the range 1 Myr to 30 Gyr.
The SFR of the constant SFH component is determined via the specific star formation rate (sSFR) over the associated time interval, where we adopt a log-uniform prior of $-14 \leq$ log(sSFR/yr$^{-1}$) $\leq -6$. 

\textbf{\textsc{prospector} continuity SFH prior models:} We also consider non-parametric SFH models using the SED-fitting code \textsc{prospector} \citep{Leja2019,Johnson2021}. 
\textsc{prospector} adopts the star-forming photoionization models from \textsc{fsps} \citep{Conroy2009_FSPS,Conroy2010_FSPS} which by default utilizes products from the MESA Isochrones and Stellar Tracks project (\textsc{mist}; \citealt{Choi2016_mist}).
Posterior probability distributions on fitted parameters are determined using the \textsc{dynesty} sampling package \citep{Speagle2020}.
We first consider the `continuity' prior implemented in the non-parametric SFH models in \textsc{prospector}.
This continuity SFH prior weights against strong changes in the SFR over adjacent time bins (see \citealt{Leja2019} for details), thereby preferentially yielding more extended SFHs for galaxies with young light-weighted ages.

The implementation of the non-parametric SFH models in this work largely follows that described in \citet{Whitler2023_z7}.
To summarize, the SFHs are composed of 8 time bins where the SFR in each is a constant value, and the ratios of the SFR in adjacent time bins are fit by \textsc{prospector}.
The earliest time bin extends to a fitted formation redshift in the range $z_\mathrm{form} = 10-30$ (uniform prior), and the two most recent time bins are fixed to 0--3 Myr and 3--10 Myr while the remaining 6 time bins are divided evenly in logarithmic space to the fitted formation redshift.

\textbf{\textsc{prospector} bursty SFH prior models:} Finally, we consider the `bursty' SFH prior \citep{Tacchella2022_SFHs}, a slight modification of the continuity prior described above.
These bursty priors more easily allow for strong deviations in the SFR in adjacent time bins while still permitting very extended star formation histories like the continuity prior.

For each of the four models described above, fiducial values on inferred physical properties for a given galaxy are taken as the median of the posterior probability distribution.
Similarly, the associated $\pm$1$\sigma$ error as the inner 68\% credible interval from the posterior.
Reported absolute UV magnitudes, \Muv{}, are computed here from the continuum flux density at 1500 \AA{} rest-frame using the output redshift and SED posteriors.

Prior to running the SED fits, we add a 5\% systematic error to all photometric measurements, largely with the intention of being conservative about the precision of current photometric zero points.
In general, we find that these star-forming photoionization models yield acceptable fits to the $z\sim6-9$ Lyman-break galaxies in our sample.
For each of the four models sets described above, the median best-fitting $\chi^2$ value over the full sample ranges between 10--14 (14 photometric data points are fit in every case).
Moreover, 90-98\% of the SED fits yield best-fitting $\chi^2 < 28$.

However, there are four objects in our sample with measured photometry that are very poorly fit by these star-forming only models.
These four objects are those with best-fitting $\chi^2 > 100$ with the \textsc{beagle} CSFH models and best-fitting $\chi^2 > 50$ for the other three models.
Two of these objects appear to have steeply rising SEDs between F277W and F444W yet relatively blue colors between F115W and F277W, reminiscent of the `Little Red Dot' (\citealt{Matthee2023_LRD}) population that has garnered much debate within the literature as to whether their light detected in NIRCam is primarily due to emission from stars or AGN \citep[e.g.][]{Akins2023,Endsley2023_CEERS,Greene2023_UNCOVER,Williams2023_HSTdark,PerezGonzalez2024_LRD}.
Both of these systems have been explored in \citet[][IDs 121710 and 132229]{Williams2023_HSTdark} and \citet{PerezGonzalez2024_LRD}, where these works consider both AGN and star-forming SED model fits once adding in deep multi-band MIRI imaging. 
A third object selected in our Lyman-break sample yet very poorly fit with star-forming only models shows very broad H$\alpha$ emission and is a candidate for hosting a black hole merger (ID 10013704 in \citealt{Maiolino2023_mergingBH}).
Because the goal in this paper is to study reionization-era galaxies in context of star-forming models, we remove these four poorly-fit objects from our sample.
Two of these objects were selected as F775W dropouts while the other two were selected as F090W dropouts.
Therefore, our final sample of Lyman-break $z\sim6-9$ galaxies analyzed below consists of 278 F775W dropouts ($z\sim6$) and 478 F090W dropouts ($z\sim7-9$).

\section{The Properties of Galaxies at Redshifts \texorpdfstring{$\mathbf{6-9}$}{6 - 9}} \label{sec:properties}

\begin{figure}
\includegraphics{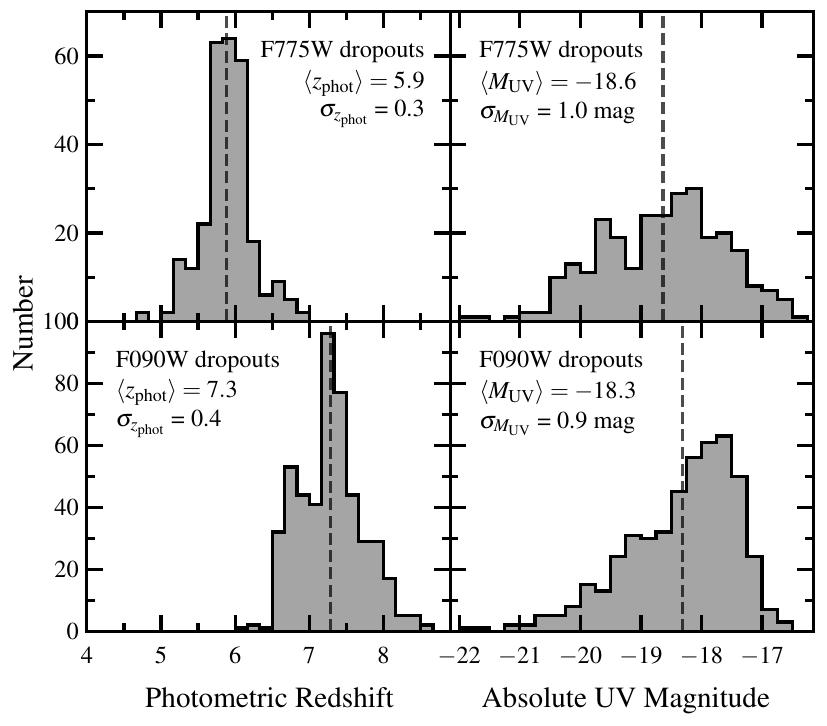}
\caption{Distribution of photometric redshifts (left) and absolute UV magnitudes (right) among the sample of 278 F775W dropouts (top) and 478 F090W dropouts (bottom) considered in this work. Here, the adopted value for each galaxy is taken as the median value from the posterior of the \textsc{beagle} CSFH fit outputs. We quote the mean (also shown with a vertical dashed line) and standard deviation of values for each dropout sample in their respective panel. Even though the photometric redshifts of our F090W dropouts extend to $z\approx8.6$, our selection criteria is sensitive to blue galaxies at redshifts as high as $z\approx9.0$ where the \OIIIHb{} emission lines remain in F444W (see \S\ref{sec:selection}).}
\label{fig:MuvRedshiftDistns}
\end{figure}

In this section we begin by discussing the overall properties of the sample, beginning with two basic parameters -- photometric redshifts and absolute UV magnitudes -- and then shift our focus to the inferred stellar masses (\S\ref{sec:stellar_mass}) and light-weighted ages (\S\ref{sec:ages}).
We include a detailed discussion of the most massive objects in our sample, as well as those with the youngest and oldest light-weighted ages.

The F775W and F090W dropout subsets are largely comprised of galaxies with photometric redshifts in the range $5.5<z_\mathrm{phot}<6.5$ and $6.5<z_\mathrm{phot}<8.5$, respectively (see Fig. \ref{fig:MuvRedshiftDistns}), consistent with expectations given our Lyman-break selection criteria.
The mean photometric redshift is $\langle z_\mathrm{phot} \rangle = 5.9$ and $7.3$ for the F775W and F090W dropout subsets, respectively.
Considering all 756 of these $z\sim6-9$ JADES/NIRCam galaxies, our sample spans a very broad range in absolute UV magnitude from $\Muv{} = -22.0$ at the bright end to $\Muv{} = -16.4$ at the faintest end (see Fig. \ref{fig:MuvRedshiftDistns}).
Using a characteristic UV luminosity of $L^{\ast}_\mathrm{UV} = -20.5$ \citep[e.g.][]{Bowler2017,Harikane2022_LF}, the \Muv{} range of our sample corresponds to $\approx$0.02 -- 4 $L^{\ast}_\mathrm{UV}$ thus covering a factor of $\approx$200 in UV luminosity.
The average absolute UV magnitude of the F775W dropout ($z\sim6$) subset is $\langle \Muv{} \rangle = -18.6$, which is slightly brighter than that of the F090W dropout subset ($\langle \Muv{} \rangle = -18.3$) given that the sensitivity of our F775W selection is limited more by the \HST{}/ACS depth.
Both dropout subsets have very similar standard deviation in \Muv{} (0.9--1.0 mag) and each include a handful of galaxies at very bright UV luminosities ($\Muv{} < -21.25$ or $L^{}_\mathrm{UV} > 2\  L^{\ast}_\mathrm{UV}$) as well as $\approx$100--200 galaxies at the very faint end ($\Muv{} > -18$ or $L^{}_\mathrm{UV} < 0.1\ L^{\ast}_\mathrm{UV}$; see Fig. \ref{fig:MuvRedshiftDistns}).

\begin{figure*}
\includegraphics{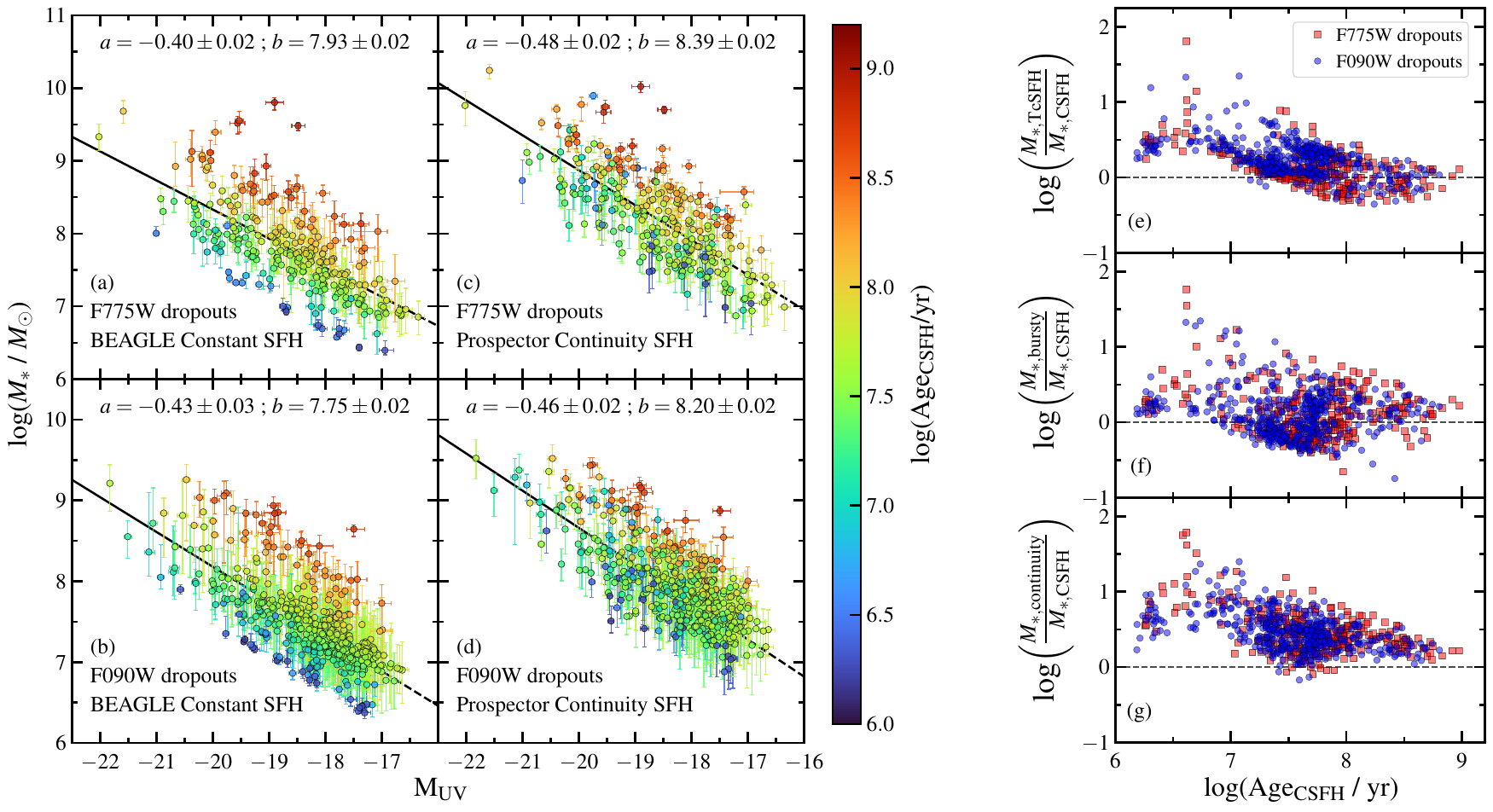}
\caption{\textbf{Left:} The relationship between absolute UV luminosity and stellar mass for our sample of F775W dropouts (top) and F090W dropouts (bottom). We show this relationship adopting the stellar masses from the \textsc{beagle} constant SFH (left panels) and \textsc{prospector} continuity SFH prior fits (see \S\ref{sec:models}). These two mass estimates generally bracket the systematic uncertainties associated with the assumed SFH and we show the fitted relation (black line) for each adopting the form log($M_\ast / M_\odot$) = $a\,$(\Muv{}$+19$) $+$ $b$ and restricting the fit to $\Muv{} \leq -18$ (the dashed line shows the relation extrapolated to fainter UV luminosities). Each point is color-coded by the inferred CSFH age of the associated galaxy. \textbf{Right:} Illustration of systematic differences in stellar mass estimates from adopting different assumptions on the SFH. We compare stellar mass estimates from the two-component SFH (TcSFH) \textsc{beagle} fits (top), the bursty SFH prior \textsc{prospector} fits (middle), and the continuity SFH prior \textsc{prospector} fits (bottom), all relative to the \textsc{beagle} CSFH stellar masses as a function of CSFH age.}
\label{fig:MstarSFH}
\end{figure*}

\subsection{The Stellar Masses of Lyman-break \texorpdfstring{$\mathbf{z\sim6-9}$}{z ~ 6 - 9} Galaxies} \label{sec:stellar_mass}

The 756 Lyman-break $z\sim6-9$ galaxies comprising our JADES sample possess a wide range of inferred stellar masses.
From the \textsc{beagle} CSFH models, we infer stellar masses spanning approximately 3.5 orders of magnitude from $\Mstar{} \approx2\times10^6\ \Msol{}$ up to $\Mstar{} \approx7\times10^{9}\ \Msol{}$ (see Fig. \ref{fig:MstarSFH}a,b).
However, we find that the inferred stellar masses of individual $z\sim6-9$ galaxies can change substantially depending on the assumed SFH, consistent with the results of previous works \citep[e.g.][]{Endsley2023_CEERS,Tacchella2022_SFHs,Topping2022_REBELS,Whitler2023_z7,Whitler2023_z10}. 
In our sample the \textsc{beagle} CSFH stellar masses are, on average, approximately 0.2, 0.1, and 0.5 dex smaller than (i.e. 0.6$\times$, 0.8$\times$, and 0.3$\times$) the stellar masses inferred from the \textsc{beagle} two-component SFH fits, the \textsc{prospector} bursty SFH prior fits, and the \textsc{prospector} continuity SFH prior fits, respectively (see Fig. \ref{fig:MstarSFH}e--g).
Therefore, the \textsc{beagle} CSFH fits tend to yield the lowest stellar masses among our sample while the \textsc{prospector} continuity SFH prior fits typically yield the largest mass estimates.
This is because the bulk of galaxies in our sample have young inferred ages in context of a CSFH ($\sim$50 Myr; see \S\ref{sec:ages}) and in these models no stars are assumed to have formed prior to that inferred age. 
On the other hand, the \textsc{prospector} continuity prior imposes a preference for significant star formation extending back to $z=10-30$ ($\Delta t \sim 300-650$ Myr for a galaxy at $z\sim7$) by weighting against strong time-variability in the SFH.

For galaxies with the youngest light-weighted ages in our sample ($\ageCSFH{} < 10$ Myr), the stellar masses inferred with the \textsc{prospector} continuity SFH prior are typically 0.8 dex (i.e. 6$\times$) higher than that inferred with the \textsc{beagle} CSFH setup, though this difference can rise to factors of $\sim$30--100 in the most extreme cases (see Fig. \ref{fig:MstarSFH}g) as found in \citet{Whitler2023_z7} using a sample of very bright ($\Muv{} \lesssim -21$) $z\sim7$ galaxies with IRAC coverage.
The offset between constant and continuity prior SFH stellar masses is generally more moderate among galaxies with relatively old light-weighted ages ($\ageCSFH{} > 100$ Myr), with a median and maximum offset of approximately 0.3 dex (2$\times$) and 0.7 dex (5$\times$), respectively (see Fig. \ref{fig:MstarSFH}g), again consistent with the IRAC-based findings of \citet{Whitler2023_z7}.
Consequently, we find that the range of stellar masses for our sample shifts upwards to $\Mstar{} \approx 7\times10^6\ \Msol{} - 2\times10^{10}\ \Msol{}$ with the \textsc{prospector} continuity SFH prior fits (see Fig. \ref{fig:MstarSFH}c,d), though we continue to conclude that the faintest ($\Muv{} \sim -17$) and youngest ($\ageCSFH{} \lesssim 30$ Myr) galaxies tend to have the lowest stellar masses of $\sim10^7 \Msol{}$.

We quantify the relationship between UV luminosity and stellar mass among our sample by fitting the equation log$\left(M_\ast / M_\odot\right) = a\ \left(\Muv{} + 19\right) + b$.
This relation is derived at $z\sim6$ and $z\sim7-9$ by fitting the F775W and F090W dropout subsets separately, and we moreover perform the fits using the \textsc{beagle} CSFH and \textsc{prospector} continuity SFH prior models to bracket the systematic uncertainties associated from different SFH assumptions.
In these \Mstar{}--\Muv{} fits, we only include objects with relatively bright UV luminosities ($\Muv{} \leq -18$) as we aim to mitigate potentially significant biases at the faintest end, though acknowledge that future work directly accounting for incompleteness will be required.
The fitted parameters ($a$ and $b$) and their uncertainties are taken as the median and standard deviation of values obtained from 1000 realizations of randomly sampling stellar masses and UV luminosities of each galaxy from the posteriors of their \textsc{beagle} CSFH outputs, again keeping only those with $\Muv{} \leq -18$.
With this simple approach, we derive the \Mstar{}--\Muv{} relations shown in panels Fig. \ref{fig:MstarSFH}a--d.

When adopting the \textsc{beagle} CSFH models, we infer a typical stellar mass of $\Mstar{} \approx 2.1\times10^8\ \Msol{}$ and $\approx3.4\times10^7\ \Msol{}$ for $z\sim6$ galaxies with $\Muv{} = -20$ and $\Muv{} = -18$, respectively.
At $z\sim7-9$, the typical inferred CSFH stellar masses are lower by $\approx$0.15--0.2 dex at these \Muv{} values, resulting in values of $\Mstar{} \approx 1.5\times10^8\ \Msol{}$ and $\approx2.1\times10^7\ \Msol{}$, respectively.
This redshift trend is consistent with previous findings that, at fixed \Muv{}, typical inferred stellar masses decrease slightly with increasing redshift at $z\sim6-9$ implying systematically higher sSFRs at earlier epochs \citep[e.g.][]{Stark2013_NebEmission,Duncan2014,Song2016,Bhatawdekar2019,Stefanon2021_SMF}.

The normalization of our derived \Mstar{}--\Muv{} relation is $\approx$0.5 dex (i.e. $\approx$3$\times$) higher when adopting the stellar masses from the \textsc{prospector} continuity SFH prior fits, as expected from our discussion above.
The slope of the \Mstar{}--\Muv{} relation we derive in each case ($-0.48 \leq a \leq -0.40$) is consistent with results of several previous works \citep[e.g.][]{Stark2013_NebEmission,Duncan2014,Gonzalez2014,Song2016,Bhatawdekar2019,Kikuchihara2020,Stefanon2021_SMF}.
However, we are generally inferring considerably ($\sim$0.3--0.5 dex) lower stellar masses at fixed \Muv{} than found previously with \HST{} and \Spitzer{} data, even when accounting for differences in the assumed SFHs throughout various studies.
We defer a more detailed comparison with literature results to a future work, though quickly note that the $z\sim7-9$ CSFH \Mstar{}--\Muv{} relation derived here is consistent with the early \JWST{} results of \citet{Endsley2023_CEERS}, perhaps reflecting the much richer information gained from \JWST{} on reionization-era galaxy rest-optical SEDs.

\begin{figure*}
\includegraphics{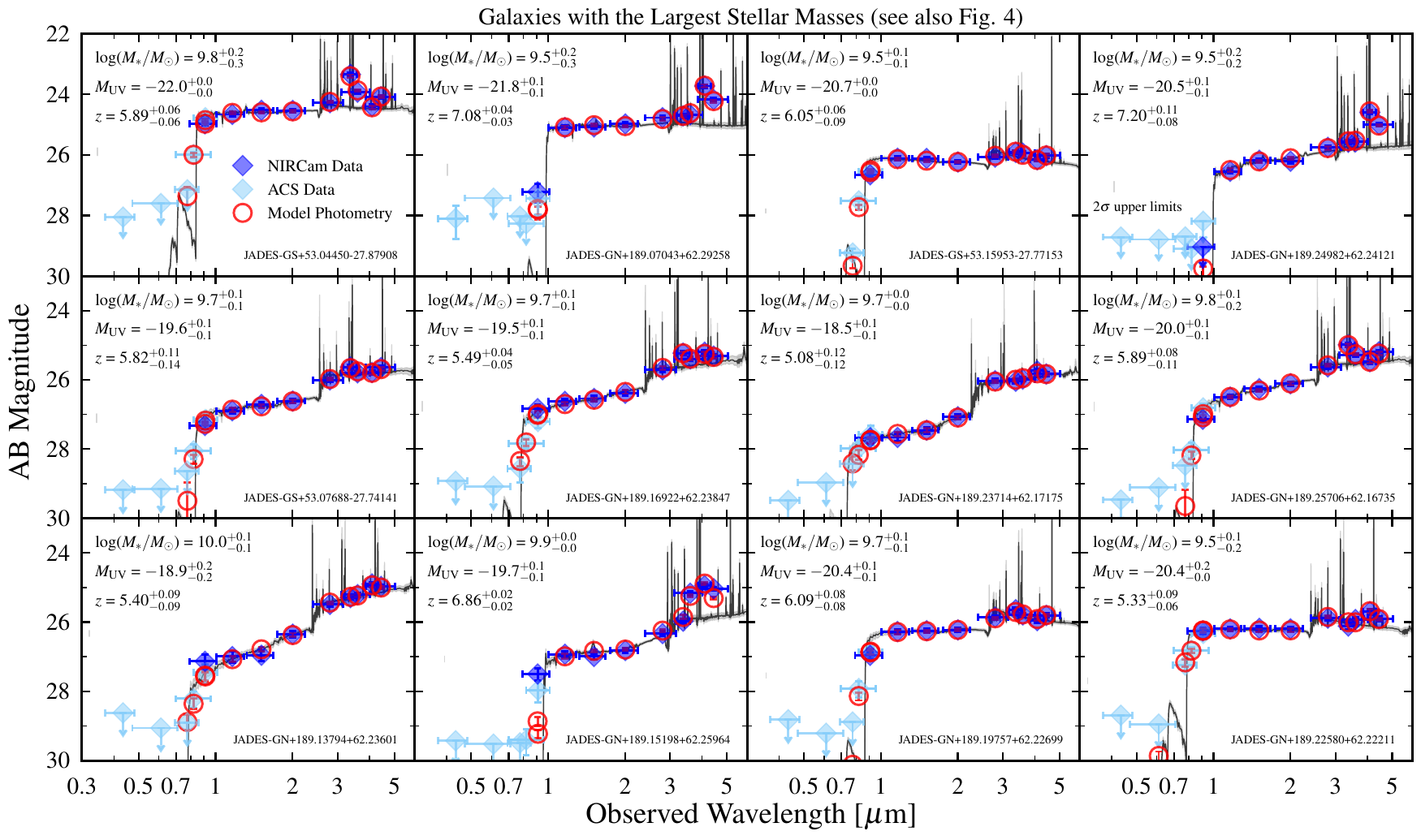}
\caption{The ACS+NIRCam SEDs for 12 of the 13 galaxies with the largest inferred stellar masses in our sample ($\Mstar{}\geq10^{9.5}$ \Msol{} from the \textsc{prospector} continuity SFH prior fits; the last object is shown in Fig. \ref{fig:GS-28012}). The measured photometry is shown with blue diamonds while the model photometry from the \textsc{prospector} continuity SFH prior SEDs is shown with red circles. The black lines and corresponding shaded regions show the median posterior from the SEDs and their 68\% inner credible intervals, respectively, fixed at the median redshift from the posterior of each galaxy. The \textsc{prospector} continuity SFH prior fits generally yield the largest stellar masses (see Fig. \ref{fig:MstarSFH}) and we find a maximum stellar mass of $10^{10.0} \Msol{}$ across our full sample with these fits (see Fig. \ref{fig:GS-28012}).}
\label{fig:mostMassive}
\end{figure*}

\begin{figure}
\includegraphics{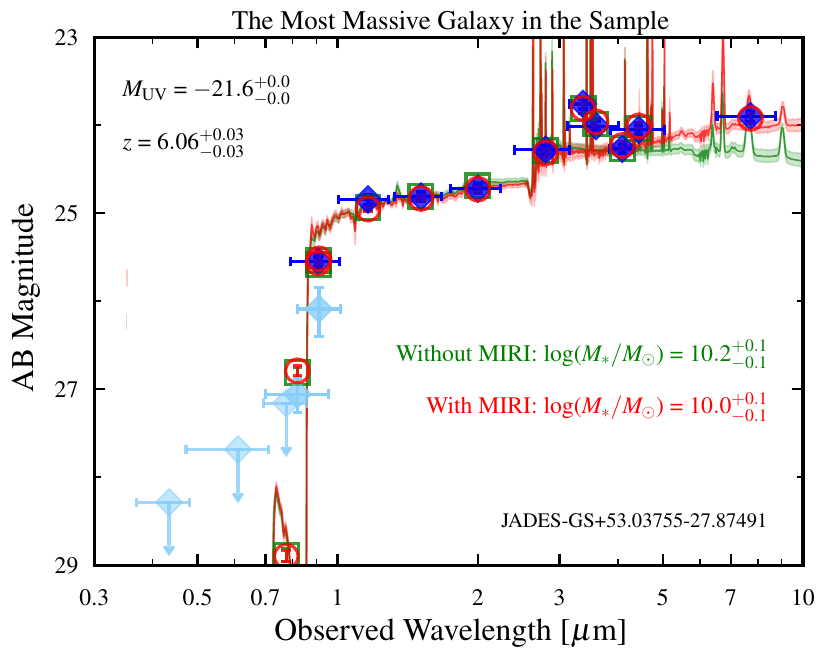}
\caption{The SED fit of the most massive $z\sim6-9$ Lyman-break galaxy in our sample, JADES-GS+53.03755$-$27.87491. We show the fit both excluding (green) and including (red) the MIRI F770W photometry for this object, with both adopting the \textsc{prospector} continuity SFH prior. The inclusion of the MIRI data results in a consistent (yet slightly lower) stellar mass estimate for this galaxy, yielding a maximum stellar mass of $10^{10.0}$ \Msol{} across our full sample (see also JADES-GN+189.13794+62.23601 in Fig. \ref{fig:mostMassive}).}
\label{fig:GS-28012}
\end{figure}

\subsubsection{The Most Massive Lyman-break \texorpdfstring{$z\sim6-9$}{z ~ 6 - 9} Galaxies in JADES} \label{sec:mostMassive}

Shortly after the first \JWST{}/NIRCam images were released, there were reports of very high-redshift ($z\sim7-11$) candidates with extremely large stellar masses of $\Mstar{} \sim10^{11}$ \Msol{} identified over small areas ($\approx$40 arcmin$^2$; \citealt{Labbe2023}) which clearly challenged models of galaxy formation \citep[e.g.][]{BoylanKolchin2023,Lovell2023}.
Since these initial reports\footnote{\url{https://arxiv.org/abs/2207.12446v2}}, stellar mass estimates for nearly all the candidates have been lowered to $<3\times10^{10}$ \Msol{}, though one candidate $\Mstar{} = 10^{10.9}$ \Msol{} object remains in the \citet{Labbe2023} sample.
However, due the limited filter set available in these earliest \JWST{} data, solutions with 10--200$\times$ lower stellar masses can be obtained for the most massive candidate in \citet{Labbe2023} by adopting different model assumptions that yield similar (if not considerably better) $\chi^2$ (see \citealt{Endsley2023_CEERS}).
Here, we utilize the deep 9-band NIRCam imaging of JADES (including two long-wavelength medium bands) to build upon these initial studies by investigating the potential abundance of extremely massive $z\gtrsim6$ galaxies.

\begin{figure*}
\includegraphics{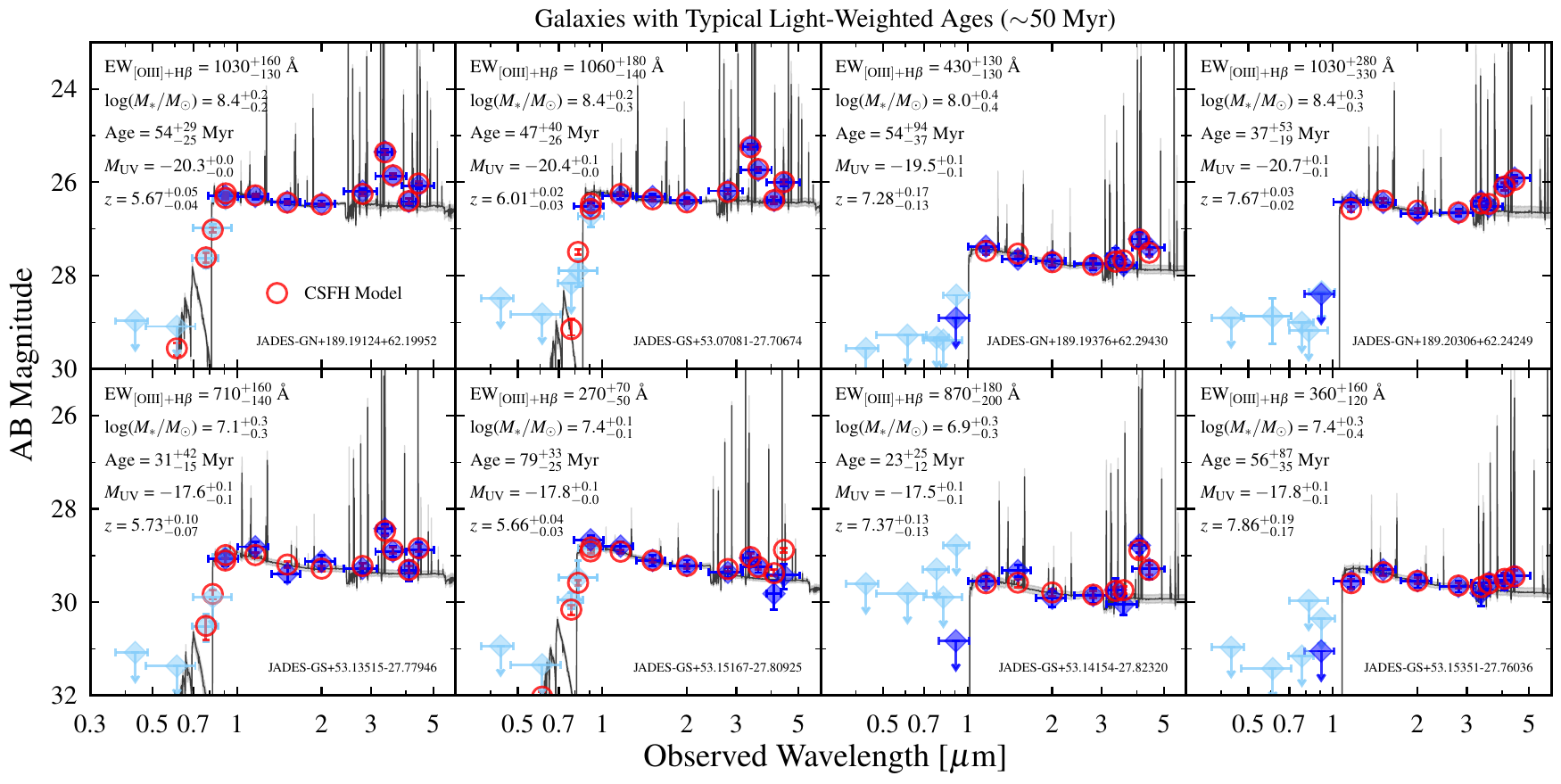}
\caption{ACS+NIRCam SEDs for a subset of galaxies with fairly typical light-weighted ages among our sample ($\ageCSFH{} \sim 50$ Myr). We show galaxies at bright ($\Muv{} \leq -19.5$; upper panels) and faint ($\Muv{} > -18$; lower panels) UV luminosities as well as those selected as F775W dropouts ($z\sim6$; left panels) and F090W dropouts ($z\sim7-9$; right panels). The model photometry from the \textsc{beagle} constant star formation history (CSFH) fits are shown with red circles; all quoted properties in this figure are inferred from the \textsc{beagle} CSFH models.}
\label{fig:typicalAge}
\end{figure*}

There are no galaxies in our sample where the photometric data clearly imply stellar masses of $>$3$\times$10$^{10}$ \Msol{}.
Only a small number of galaxies have inferred stellar masses around $\sim$10$^{10}$ \Msol{} when applying the continuity SFH prior with \textsc{prospector} which, as discussed in \S\ref{sec:stellar_mass}, generally yields a maximum stellar mass estimate for a given galaxy (see Fig. \ref{fig:MstarSFH}). 
Here, we discuss these most massive $z\sim6-9$ candidates in our JADES sample.

There are 13 galaxies with inferred stellar masses of $\geq$3$\times$10$^9$ \Msol{} from the \textsc{prospector} continuity SFH prior fits among our JADES sample.
Eleven and two of these thirteen galaxies are in the F775W and F090W dropout subsets, respectively, and all of their SEDs are shown in Figs. \ref{fig:mostMassive} and \ref{fig:GS-28012}.
Unsurprisingly, many of these objects show NIRCam colors consistent with strong Balmer breaks implying that a relatively old ($\gtrsim$300 Myr) stellar population is contributing significantly to the emergent light.
However, a few of these galaxies show colors consistent with relatively weak Balmer breaks and are simply so luminous ($-22 \lesssim \Muv{} \lesssim -21.5$) that a substantial population of relatively old stars can be hidden within the observed SED if outshined by more recently-formed stars.

The galaxy with the highest inferred stellar mass in the sample (JADES-GS+53.03755$-$27.87491; $\Mstar{} = 10^{10.2\pm0.1} \Msol{}$) lies at $z_\mathrm{phot} \approx 6.1$, is very luminous in the rest UV ($\Muv{} = -21.6$), and shows a significant Balmer break as well as moderate nebular line emission from \OIIIHb{} and H$\alpha$ (Fig. \ref{fig:GS-28012}).
Fortuitously, this galaxy falls within the deep MIRI F770W parallel imaging of JADES (see \citealt{Eisenstein2023}) and is clearly detected at high S/N, granting improved constraints on the total amount of stellar mass allowed by the SED by extending measurements into the rest-frame near infrared \citep[e.g.][]{Tang2022,Papovich2023}.
With the MIRI measurement included \citep{Alberts2024}, we infer a consistent though slightly ($\approx$0.2 dex) lower stellar mass of $\Mstar{} = 10^{10.0\pm0.1} \Msol{}$ (see Fig. \ref{fig:GS-28012}).
Because the F770W flux density (largely probing the rest-NIR continuum) is significantly higher than that in F410M (probing the rest-optical continuum), the model posteriors push to higher metallicity solutions ($\approx$0.3 $Z_\odot$) over that without the MIRI data ($\approx$0.05--0.1 $Z_\odot$).
Such higher metallicity solutions alter the light-to-mass ratio across the rest-frame $\approx$0.4-1$\mu$m SED, resulting in a slightly lower stellar mass.
A more systematic assessment of how including MIRI data impacts stellar mass inferences among the high-redshift JADES sample will be presented in upcoming works (Florian et al. in prep.; Helton et al. in prep; Ji et al. in prep).

Aside from JADES-GS+53.03755$-$27.87491, there is only one other galaxy in the $z\sim6-9$ sample considered here with a comparable stellar mass: JADES-GN+189.13794+62.23601 with an estimated $\Mstar{} = 10^{10\pm0.1} \Msol{}$ from the ACS+NIRCam data.
JADES-GN+189.13794+62.23601 does not possess any MIRI constraints in the rest-frame near-infrared so we adopt this estimated mass as fiducial (in context of an extended SFH), though we note that previous work has shown that the addition of MIRI data generally lowers estimated stellar masses of high-redshift galaxies \citep{Papovich2023}.
Overall, we find that the maximum inferred stellar mass (from the continuity \textsc{prospector} fits) among all galaxies in our sample is 1.0$\times$10$^{10}$ \Msol{}.

We remind the reader that, in this work, we are only considering Lyman-break selected $z\sim6-9$ galaxies with photometry that can be reasonably explained by star-forming photoionization models.
There are four Lyman-break $z\sim6-9$ objects in JADES with photometry that is very poorly matched by star-forming only models and hence have been excluded from the sample considered here (see \S\ref{sec:models}).
Notably, these include compact objects with NIRCam colors similar to the most massive candidate reported in \citet{Labbe2023} and others subsequently identified in more recent data sets \citep{Furtak2022_uncover,Akins2023,Barro2023}, including within JADES \citep{Williams2023_HSTdark,PerezGonzalez2024_LRD}.
The contribution of AGN emission to the observed light in these systems remains heavily debated, and additional data is crucial to better estimate their stellar masses.

\subsection{The Constant SFH Ages of Lyman-break \texorpdfstring{$\mathbf{z\sim6-9}$}{z ~ 6 - 9} Galaxies} \label{sec:ages}

In this sub-section, we discuss the light-weighted ages (defined here as CSFH ages) inferred among our sample of 756 Lyman-break galaxies.
A wide variety of rest UV+optical SED shapes are clearly found across our sample, indicating a large diversity in light-weighted ages among reionization-era galaxies.
The typical galaxy in our sample shows a NIRCam SED consistent with a roughly fixed power-law continuum extending from the rest-UV to rest-optical, implying a relatively weak Balmer break and hence a young light-weighted age ($\ageCSFH{} \sim 50$ Myr) consistent with previous findings from early \JWST{} data \citep[e.g.][]{Endsley2023_CEERS,Furtak2023_SMACS,Laporte2023,Leethochawalit2023,Morishita2023,Whitler2023_z10}.
The SEDs for a subset of these typical, young galaxies are shown in Fig. \ref{fig:typicalAge} including a variety of redshifts and UV luminosities.
Many of the young galaxies show a significant photometric excess in at least one long-wavelength band consistent with strong \OIIIHb{} and/or H$\alpha$ emission (EW$\sim$700--1000 \AA{}; see e.g. IDs JADES-GN+189.19124+62.19952 and JADES-GS+53.14154$-$27.82320 in Fig. \ref{fig:typicalAge}) implying a substantial contribution of hot, massive stars to the SED.
However, a subset of the young ($\ageCSFH{} \sim 50$ Myr) galaxies have measured photometry consistent with surprisingly weak nebular line emission (e.g. IDs JADES-GN+189.19376+62.29430 and JADES-GS+53.15167$-$27.80925 in Fig. \ref{fig:typicalAge}); this sub-population is discussed further in \S\ref{sec:bursty}.

In addition to the typical young ($\ageCSFH{} \sim 50$ Myr) galaxies, several sources in our sample show long-wavelength photometric excess patterns indicating extremely strong nebular line emission (\OIIIHb{} EW $\sim 2000-5000$ \AA{} or H$\alpha$ EW $\sim 1000-2500$ \AA{}), consistent with exceptionally young light-weighted ages (\ageCSFH{} $\sim$ 3 Myr). 
Other galaxies show strong Balmer breaks with signatures of weak-to-no nebular line emission (\OIIIHb{} EW$\lesssim$300 \AA{}), consistent with relatively old light-weighted ages ($\ageCSFH{} \sim 300-1000$ Myr).
These sub-populations representing the extreme ends of the light-weighted age distribution are analyzed in greater detail below.

\begin{figure*}
\includegraphics{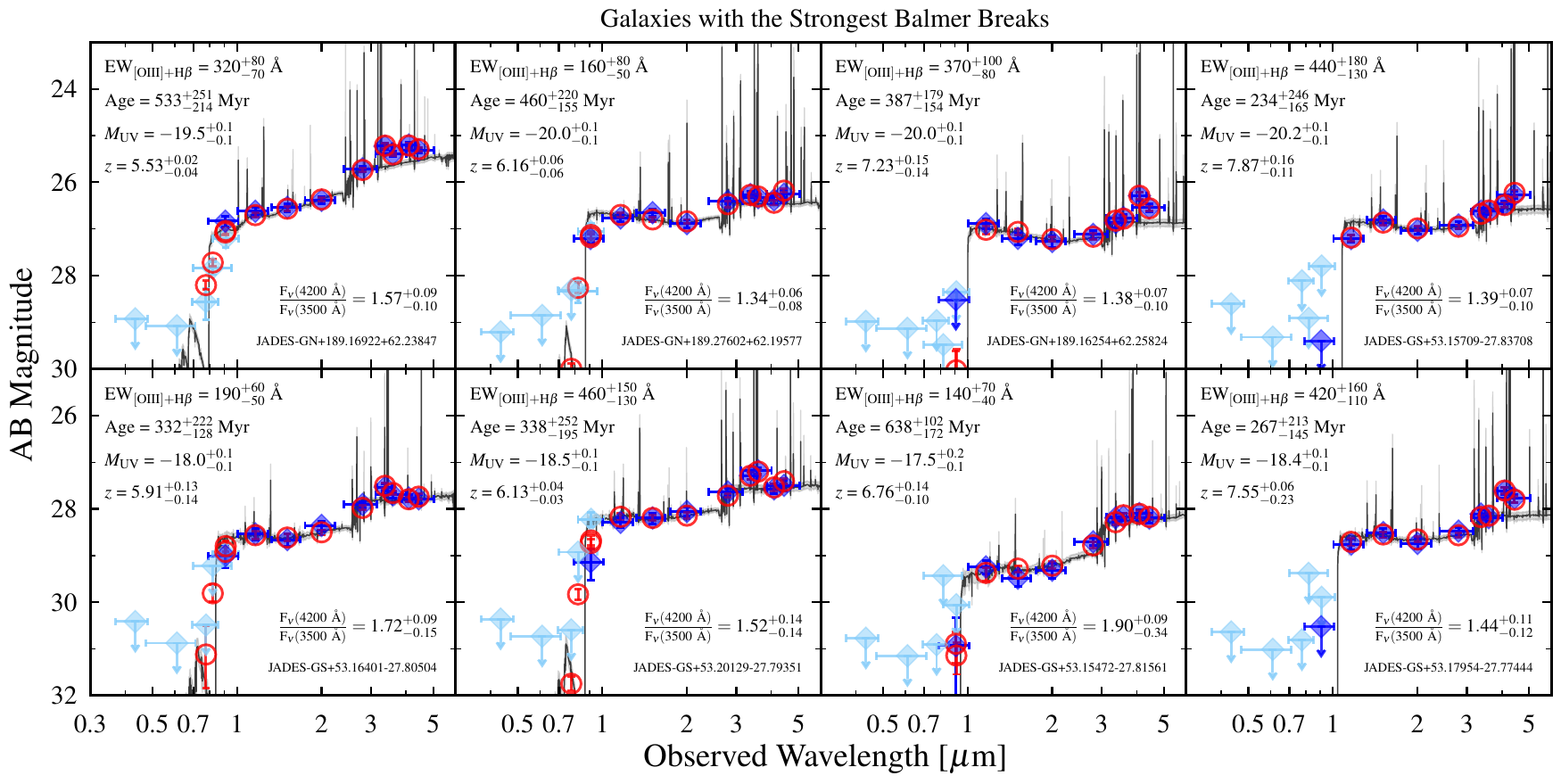}
\caption{SEDs for a subset of galaxies that confidently show strong Balmer breaks (\BBrat{} $>$ 1.25). All quoted properties are inferred from the \textsc{beagle} CSFH fits, though the inferred Balmer breaks remain strong at the $>$84\% credible interval with the two-component SFH \textsc{beagle} fits. The figure format is very similar to Fig. \ref{fig:typicalAge}.}
\label{fig:balmerBreaks}
\end{figure*}

\subsubsection{Galaxies with the Strongest Balmer Breaks} \label{sec:balmerBreaks}

\noindent
The task of identifying reionization-era galaxies with prominent Balmer breaks has long been hindered by the sparse rest-optical SED sampling afforded by \Spitzer{}/IRAC photometry, leading to strong degeneracy with SED solutions of extremely strong nebular line emission (e.g. \citealt{Schaerer2010,Stark2013_NebEmission,RobertsBorsani2020}).
This degeneracy could be alleviated for galaxies situated in specific redshift intervals, leading to the early identification of a small number of $z\sim7-9$ galaxies showing colors consistent with strong Balmer breaks \citep[e.g.][]{Hashimoto2018_z9,Strait2020,Laporte2021,Endsley2021_OIII,Tacchella2022_SFHs}.
The list of $z\sim6-9$ galaxies reported to exhibit prominent Balmer breaks has now steadily grown since the delivery of the first \JWST{} data \citep[e.g.][]{Endsley2023_CEERS,Labbe2023,Laporte2023}, though in some cases the available NIRCam photometry could not confidently rule out solutions with extremely strong nebular line emission.
Now equipped with deep imaging in two non-overlapping medium bands as well as three broad bands at $\approx$3--5$\mu$m, we identify and characterize the $z\sim6-9$ Lyman-break galaxies in our JADES data that confidently show significant Balmer breaks. 

For ease of comparison with existing literature \citep{Binggeli2019}, we quantify the strength of the Balmer break as \BBrat{} where $F_\nu$ is the (continuum) flux density at the specified rest-frame wavelength.
These flux densities are taken from the posterior SEDs from the \textsc{beagle} fits and are corrected for dust using the inferred $A_V$ value associated with each step of the nested sampling chain.
Here we focus on the subset of galaxies with inferred \BBrat{}$>$1.25 at $>$84\% probability from the posteriors of both the CSFH and two-component SFH \textsc{beagle} fits.
These objects are classified as those with confident strong Balmer breaks throughout this work.

We identify 13 and 9 galaxies with confident strong Balmer breaks (\BBrat{}$>$1.25) in the F775W and F090W dropout samples, respectively, a subset of which is shown in Fig. \ref{fig:balmerBreaks}.
Notably, four of these 22 galaxies fall in the very UV-faint regime ($-18 < \Muv{} < -17$; see e.g. JADES-GS+53.15472$-$27.81561 in Fig. \ref{fig:balmerBreaks}) indicating that Balmer breaks do exist among reionization-era galaxies that were previously only identifiable in very deep \HST{} imaging.
The full sample of 22 galaxies with confident Balmer breaks spans absolute UV magnitudes of $-20.2 \leq \Muv{} \leq -17.1$ and redshifts of $z_\mathrm{phot} = 5.1-7.9$.
While we do not identify any objects with confident strong Balmer breaks at the brightest UV luminosities ($-22 \lesssim \Muv{} \lesssim -21$), we emphasize that our sample size is very limited at this end of the luminosity function given the search area considered in this work ($\approx$90 arcmin$^2$).
IRAC studies covering $>$deg$^2$ fields have revealed $z\sim7-8$ galaxy candidates in this luminosity regime with potential strong Balmer breaks \citep[e.g.][]{Stefanon2017_Brightestz89,Stefanon2019,Topping2022_REBELS,Whitler2023_z7}.

The JADES galaxies with confident strong Balmer breaks are among those with the oldest light-weighted ages in the sample with CSFH ages spanning $\ageCSFH{} \approx 250 - 1000$ Myr. 
In context of these CSFH models, we may expect to see significant rest-optical emission line signatures in the photometry from O stars that have been produced over the past $\sim$10 Myr.
Indeed, a subset of galaxies with confident strong Balmer breaks show long-wavelength photometric excesses implying \OIIIHb{} EWs $\approx$ 300--400 \AA{} (e.g. IDs JADES-GN+189.16254+62.25824 and JADES-GS+53.17954$-$27.77444 in Fig. \ref{fig:balmerBreaks}).
These emission line signatures imply moderate sSFRs over the most recent 10 Myr of sSFR$_{10\,\mathrm{Myr}} \approx 1-10$ Gyr$^{-1}$ with all four SED fitting procedures described in \S\ref{sec:models}.
However, we also identify a number of galaxies with strong Balmer breaks showing colors consistent with very weak to no rest-optical line emission (\OIIIHb{} and H$\alpha$ EWs $\lesssim$ 100 \AA{}; e.g. JADES-GN+189.27602+62.19577 and JADES-GS+53.15472$-$27.81561 in Fig. \ref{fig:balmerBreaks}).
For these galaxies, the SED fits allowing for more flexible SFHs tend to yield very low recent specific star formation rates (sSFR$_{10\,\mathrm{Myr}} \lesssim 0.3$ Gyr$^{-1}$) due to a declining SFH.
The CSFH fits are forced to sSFR$_{10\,\mathrm{Myr}} > 1$ Gyr$^{-1}$ solutions given the age of the Universe at $z>6$, and thus always have some modest line emission in the models (minimum \OIIIHb{} EW$\sim$200 \AA{}).
Due to the very minor effect such weak emission lines can have on the observed broadband photometry, we cannot reliably conclude whether any of these objects are in fact experiencing a declining SFH.
However, NIRSpec spectra has proven to be highly effective at confirming the existence of such systems in the early Universe, with two post-starburst/micro-quenched galaxies now known at $z=5.2-7.3$ \citep{Looser2023,Strait2023}. 
Dedicated follow-up of the candidates identified in this work would help characterize and clarify the abundance of these relatively inactive galaxies during the reionization era.

\begin{figure*}
\includegraphics{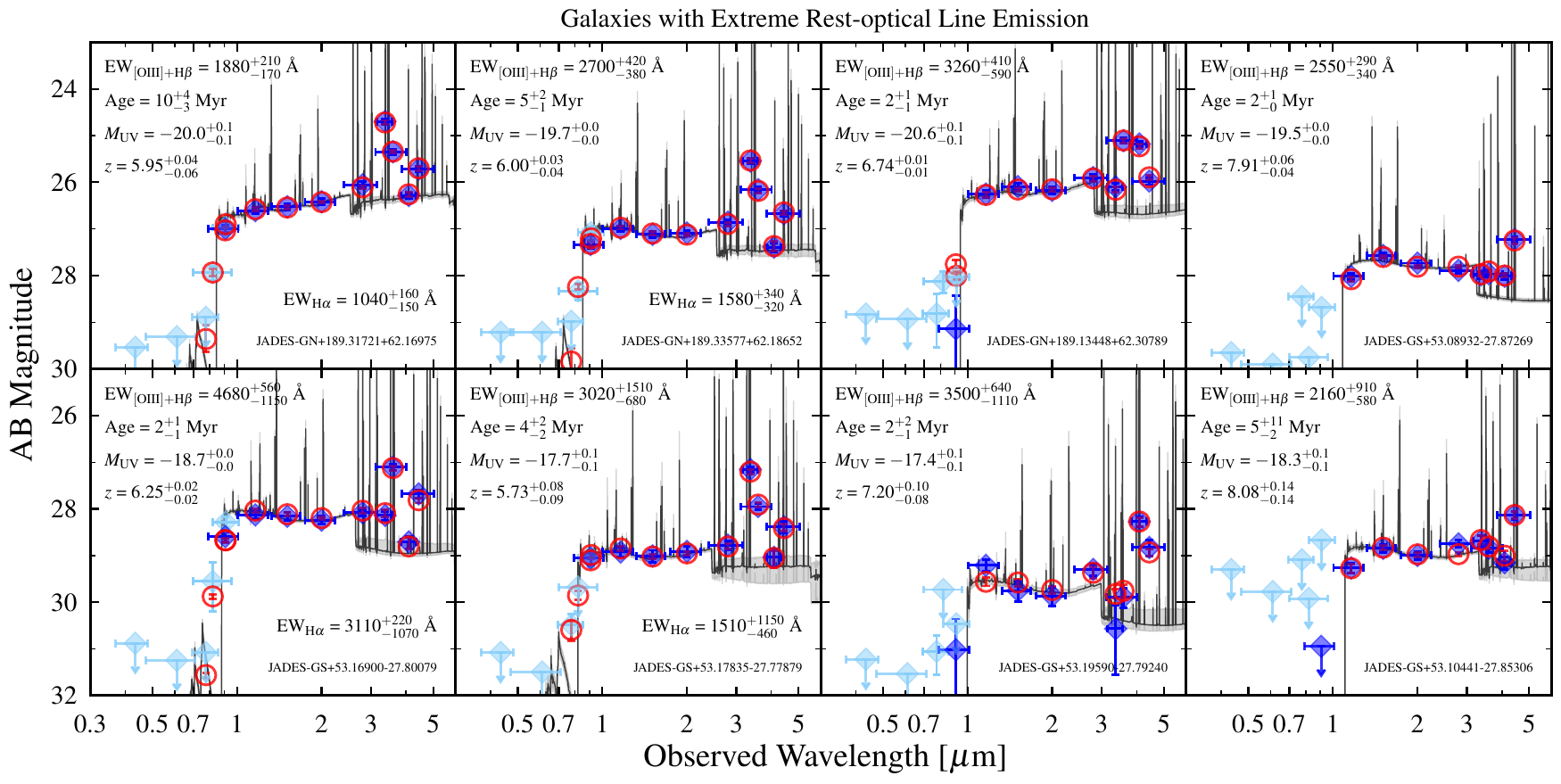}
\caption{SEDs for a subset of galaxies that confidently exhibit extreme rest-optical line emission (\OIIIHb{} EW$>$1500 \AA{} and/or H$\alpha$ EW$>$800 \AA{}). All quoted properties are inferred from the \textsc{beagle} CSFH fits, though the optical line EWs remain extreme at the $>$84\% credible interval with the two-component SFH \textsc{beagle} fits. The figure format is very similar to Fig. \ref{fig:typicalAge} and inferred H$\alpha$ EWs are only quoted for objects at $z_\mathrm{phot} < 6.5$ where NIRCam is sensitive to this line.}
\label{fig:EELGs}
\end{figure*}

\subsubsection{The Most Extreme Line Emitters} \label{sec:EELGs}

\noindent
For about a decade, it has been known that $z>4$ galaxies typically exhibit high EW ($\gtrsim$500 \AA{}) rest-optical nebular emission lines (i.e. \OIIIHb{} or H$\alpha$) from their impact on broadband \textit{Spitzer}/IRAC colors (e.g. \citealt{Schaerer2009,Ono2012,Labbe2013,Stark2013_NebEmission,Rasappu2016}).
As more observations were dedicated to deep IRAC imaging, it became clear that a surprisingly large number of bright ($\Muv{} \lesssim -20$) $z\sim7-8$ Lyman-break galaxies exhibited very high EW \OIIIHb{} emission ($\gtrsim$1500 \AA{}; \citealt{Smit2014,Smit2015,RobertsBorsani2016,deBarros2019,Endsley2021_OIII,Stefanon2022_sSFR}), implying efficient ionizing photon production and very young light-weighted ages.
In the past year, \JWST{} imaging has proven to be highly efficient at identifying many more of these $z\gtrsim6$ extreme line emitters at fainter continuum luminosities, particularly when the nebular lines are situated in a medium band (e.g. \citealt{Endsley2023_CEERS,Bouwens2023_JEMS,Rinaldi2023,Withers2023}).
In this sub-section, we build upon these previous works by utilizing the deep 9-band NIRCam imaging from JADES (including F335M and F410M) to identify and characterize a statistical sample of $z\sim6-9$ galaxies which confidently exhibit extreme rest-optical line emission.

Across our full sample of 756 $z\sim6-9$ Lyman-break GOODS+JADES galaxies, we identify several systems with long-wavelength NIRCam color patterns indicating extremely high EW rest-optical emission lines.
For the purpose of explicitly investigating this population, we restrict our attention to galaxies where the \textsc{beagle} posteriors yield $>$84\% probability of extremely high EWs from both the constant SFH and two-component SFH fits.
By imposing this probability cut on the results from two SFH models we better ensure that the inferred presence of extreme line emission is likely not being confused with a very strong Balmer break.

There are 20 and 29 galaxies that confidently exhibit extreme \OIIIHb{} emission (EW$>$1500 \AA{}) within our F775W and F090W dropout samples, respectively. 
We show the SEDs of a subset of these systems in Fig. \ref{fig:EELGs}, where objects were chosen to reflect a range of UV magnitudes and redshifts.
The confident extreme \OIIIHb{} emitters in GOODS+JADES have absolute UV magnitudes spanning $-21.0 \leq \Muv{} \leq -16.9$ and thus encompass nearly the entire \Muv{} range of our full $z\sim6-9$ sample.
Remarkably, this includes galaxies in the very UV-faint regime ($\Muv{} > -18$; see e.g. JADES-GS+53.17835$-$27.77879 and JADES-GS+53.19590$-$27.79240 in Fig. \ref{fig:EELGs}).
The vast majority of these extreme \OIIIHb{} emitters show color patterns indicating they lie at redshifts consistent with the targeted windows of our dropout selections ($z_\mathrm{phot} \approx 5.5 - 8.4$).
Only two outlier sources show strong excesses in F277W suggesting $z_\mathrm{phot} \sim 5$ which presumably entered our $z\sim6$ Lyman-break selection due to photometric noise in F775W and/or F090W.

The confident extreme \OIIIHb{} emitters all show SEDs consistent with exceptionally young light-weighted ages (\ageCSFH{} $\approx 3$ Myr) where the emergent light is heavily dominated by both stellar and nebular emission powered by recently-formed O stars.
Several of these galaxies exhibit a significant drop in flux density between the short-wavelength NIRCam bands and at least one of the long-wavelength medium bands implying a significant Balmer jump between the rest-UV and optical continua (see e.g. IDs JADES-GS+53.16900$-$27.80079 and JADES-GN+189.33577+62.18652 in Fig. \ref{fig:EELGs}).
Such Balmer jumps are caused by strong nebular continuum emission which is most prominent at very young ages and low metallicities (e.g. \citealt{Byler2017,Topping2022_blueSlopes}).
It is only for these objects that the inferred \OIIIHb{} EWs confidently reach values of $>$3000 \AA{} (up to $\approx$5000 \AA{}) as the data give direct evidence for weak rest-optical continuum emission yet extremely strong line emission from excesses in other bands. 

The very young light-weighted ages indicate that these extreme \OIIIHb{} emitters have recently experienced a dramatic rise in star formation rate \citep[e.g.][]{Tang2019,Tang2022,Endsley2021_OIII,Carnall2023_SMACS,Tacchella2023_NIRCamNIRSpec}.
In context of a constant star formation history, we infer an extremely large typical sSFR$_{10\,\mathrm{Myr}} \sim 300$ Gyr$^{-1}$ for these extreme \OIIIHb{} emitters, which is simply the inverse of their typical light-weighted age ($\ageCSFH{} \sim 3$ Myr).
But even with the \textsc{prospector} continuity SFH prior fits (which generally yield the lowest sSFR estimates; see \S\ref{sec:properties}), we continue to infer a very high median sSFR averaged over the past 10 Myr for these systems (26 Gyr$^{-1}$) with values up to 72 Gyr$^{-1}$.

At $z\sim6$, the JADES/NIRCam photometry provides constraints on not only the EW of \OIIIHb{} emission, but also H$\alpha$.
We identify 31 galaxies within our F775W dropout sub-sample that have a $>$84\% posterior probability of H$\alpha$ EW$>$800 \AA{} and $z<6.5$ from both the CSFH and two-component \textsc{beagle} SFH fits.
Here, we have added the $z<6.5$ criteria as the H$\alpha$ line begins to redshift out of F444W at higher redshifts.
This sample of confident extreme H$\alpha$ emitters overlaps significantly with the confident extreme \OIIIHb{} emitters (20 in both subsets) and thus has similar characteristics.
The confident H$\alpha$ emitters span absolute UV magnitudes of $-21.0 \leq \Muv{} \leq -17.4$ and light-weighted ages of \ageCSFH{} $\approx$ 1--10 Myr.
A subset show colors implying prominent Balmer jumps in their continuum and have inferred H$\alpha$ EWs confidently reaching values up to $\approx$2000-3000 \AA{} (see e.g. JADES-GS+53.16900$-$27.80079 in Fig. \ref{fig:EELGs}). 
Their typical specific star formation rates range from sSFR$_{10\,\mathrm{Myr}} = 150$ Gyr$^{-1}$ with the \textsc{beagle} CSFH fits down to sSFR$_{10\,\mathrm{Myr}} = 14$ Gyr$^{-1}$ with the \textsc{prospector} continuity prior SFH models.
We place both the extreme \OIIIHb{} and extreme H$\alpha$ emitter populations in context of the broader sample in the following section.

\section{The Dependence of [OIII]+H\texorpdfstring{$\mathbf{\beta}$}{beta} and H\texorpdfstring{$\mathbf{\alpha}$}{alpha} Equivalent Width on UV Luminosity} \label{sec:distributions}

The EWs of rest-optical nebular lines have long provided key insight into the physical properties of reionization-era galaxies \citep[e.g.][]{Schaerer2009,Stark2013_NebEmission,Smit2014}.
Previous studies have found that relatively bright ($\Muv{} \lesssim -20$) $z\sim7$ systems commonly show far higher \OIIIHb{} EWs than galaxies even at $z\sim2$ \citep[e.g.][]{deBarros2019,Endsley2021_OIII,Boyett2022_OIII}, implying generally less chemically enriched gas and more vigorous star formation at earlier times.
However, the nature of very UV-faint ($\Muv{} \gtrsim -18$; $L_\mathrm{UV}^{} \lesssim 0.1\ L_\mathrm{UV}^\ast$) $z\gtrsim6$ galaxies has remained far less clear given the difficulties in obtaining sensitive rest-UV+optical SED measurements among a statistical sample of such systems.
Photometric constraints on the nebular emission line EWs of this relatively numerous galaxy population would provide a powerful first glimpse into their physical properties, as these lines are sensitive to several physical parameters including metallicity, star formation history, and ionizing photon production and escape.

With our JADES/NIRCam sample of 756 $z\sim6-9$ Lyman-break galaxies, we now investigate how the reionization-era \OIIIHb{} and H$\alpha$ EW distributions correlate with UV luminosity over a large dynamic range ($-22 \lesssim \Muv{} \lesssim -16.5$).
We focus on investigating how the EWs trend with UV luminosity given that the galaxies with the lowest stellar masses in our sample are biased towards those with the youngest light-weighted ages (see Fig. \ref{fig:MstarSFH}), and thus will be weighted towards higher EWs.
We first describe the methodology for the EW distribution inference as a function of \Muv{} in \S\ref{sec:EW_methods}, following with the results on the \OIIIHb{} (\S\ref{sec:OIIIHb_EW_results}) and H$\alpha$ (\S\ref{sec:Halpha_EW_results}) EW distributions. 

\begin{figure}
\includegraphics{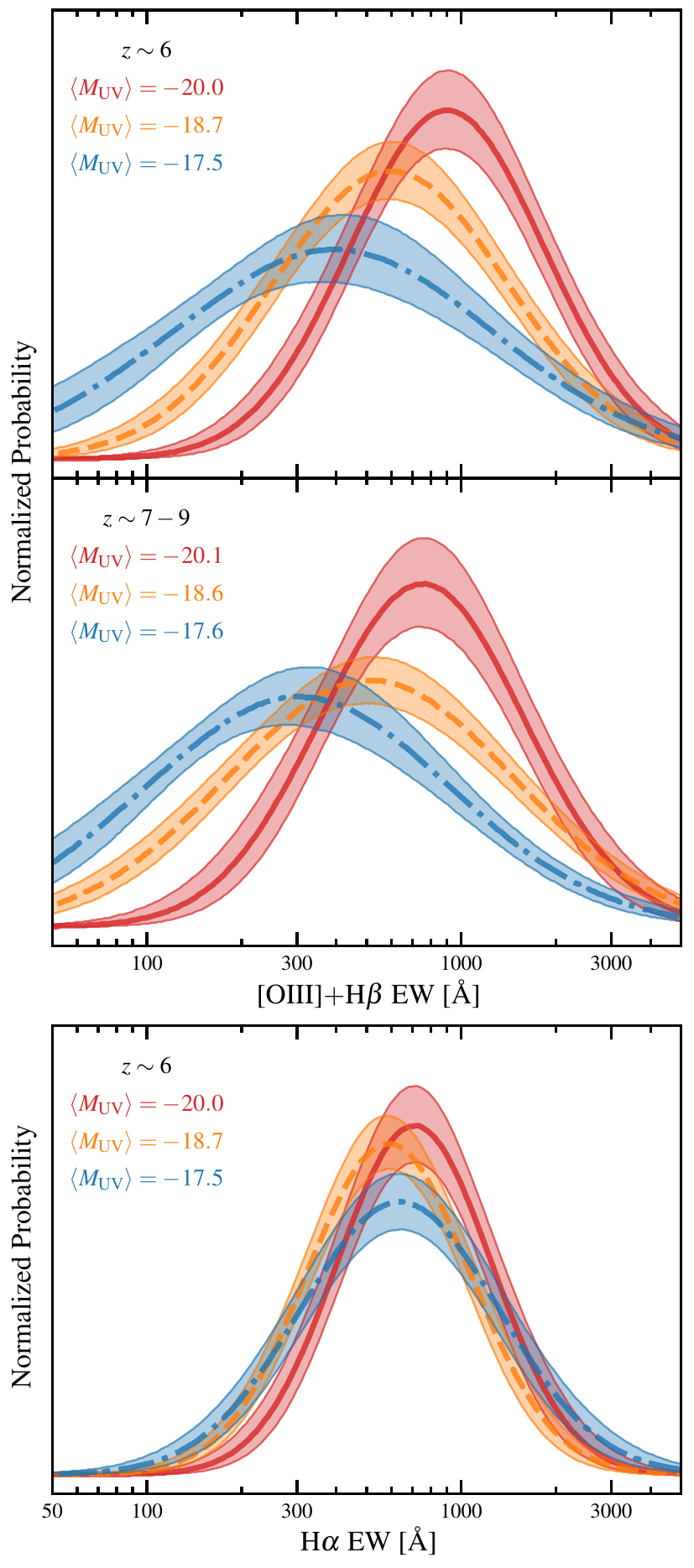}
\caption{The inferred distributions of \OIIIHb{} EW (top) and H$\alpha$ EW (bottom) among our sample using the two-component SFH \textsc{beagle} SED fit outputs. The \OIIIHb{} EW distributions are inferred for both the $z\sim6$ and $z\sim7-9$ subsets while the H$\alpha$ EW distribution is inferred only at $z\sim6$ given this line redshifts out of NIRCam at $z>6.5$. The different colors show the EW distribution inferred among three different UV luminosity bins (see Table \ref{tab:inferredEWparameters}). We find a strong, highly-significant decline in the typical \OIIIHb{} EW with UV luminosity, yet a very weak UV luminosity dependence on H$\alpha$ EW (at least at $z\sim6$).}
\label{fig:EWdistns}
\end{figure}

\begin{table*}
\centering
\caption{Table summarizing the inferred \OIIIHb{} and H$\alpha$ EW distributions among our sample of 756 Lyman-break $z\sim6-9$ galaxies. The EW distributions are assumed to be log-normal and are parametrized by the median EW, \muEW{}, and standard deviation, \sigmaEW{}. We divide our sample into six sub-populations divided by F775W dropouts ($z\sim6$) and F090W dropouts ($z\sim7-9$), as well as into bright ($\Muv{} < -19.5$), faint ($-19.5 < \Muv{} < -18$), and very faint ($\Muv{} > -18$) subsets. For each of the six sub-populations, we infer the EW distribution from the posteriors of the \textsc{beagle} constant star formation history (CSFH) and two-component star formation history (TcSFH) models. We limit the H$\alpha$ EW distribution results to the $z\sim6$ sample given that H$\alpha$ redshifts out of NIRCam at $z>6.5$. We recommend using the EW distributions derived from the TcSFH model outputs given our findings in \S\ref{sec:bursty} (see Fig. \ref{fig:youngWeakLines}).}
\begin{tabular}{P{1.0cm}P{1.5cm}P{1.8cm}P{1.0cm}P{1.0cm}P{2.0cm}P{2.0cm}P{1.5cm}P{1.5cm}}
\hline
Redshift & \Muv{} Bin & SED Model & $\langle \Muv{} \rangle$ & $N_\mathrm{gal}$ & \muEW{}(\OIIIHb{}) & \sigmaEW{}(\OIIIHb{}) & \muEW{}(H$\alpha$) & \sigmaEW{}(H$\alpha$) \Tstrut{} \\[2pt]
 & & & & & [\AA{}] & [dex] & [\AA{}] & [dex] \\[4pt]
\hline
\multirow{6}{*}{$z\sim6$} & \multirow{2}{*}{bright} & \textsc{beagle} CSFH & $-20.0$ & 64 & $930^{+70}_{-80}$ & $0.26^{+0.03}_{-0.03}$ & $760^{+50}_{-30}$ & $0.17^{+0.02}_{-0.02}$ \Tstrut{} \\[3pt]
 & & \textsc{beagle} TcSFH & $-20.0$ & 65 & $890^{+90}_{-80}$ & $0.31^{+0.03}_{-0.03}$ & $710^{+50}_{-60}$ & $0.24^{+0.03}_{-0.02}$ \\[6pt]

& \multirow{2}{*}{faint} & \textsc{beagle} CSFH & $-18.7$ & 138 & $680^{+50}_{-50}$ & $0.33^{+0.03}_{-0.03}$ & $740^{+30}_{-30}$ & $0.18^{+0.01}_{-0.02}$ \\[3pt]
 & & \textsc{beagle} TcSFH & $-18.7$ & 138 & $590^{+40}_{-50}$ & $0.37^{+0.04}_{-0.03}$ & $580^{+40}_{-30}$ & $0.26^{+0.01}_{-0.02}$ \\[6pt]

& \multirow{2}{*}{very faint} & \textsc{beagle} CSFH & $-17.4$ & 76 & $460^{+80}_{-80}$ & $0.49^{+0.09}_{-0.07}$ & $850^{+40}_{-40}$ & $0.14^{+0.03}_{-0.04}$ \\[3pt]
 & & \textsc{beagle} TcSFH & $-17.5$ & 75 & $380^{+80}_{-60}$ & $0.51^{+0.08}_{-0.07}$ & $630^{+60}_{-60}$ & $0.31^{+0.03}_{-0.03}$ \\[6pt]

\hline

\multirow{6}{*}{$z\sim7-9$} & \multirow{2}{*}{bright} & \textsc{beagle} CSFH & $-20.1$ & 52 & $780^{+70}_{-70}$ & $0.26^{+0.04}_{-0.03}$ & - & - \Tstrut{} \\[3pt]
 & & \textsc{beagle} TcSFH & $-20.1$ & 55 & $740^{+90}_{-80}$ & $0.32^{+0.04}_{-0.04}$ & - & - \\[6pt]

& \multirow{2}{*}{faint} & \textsc{beagle} CSFH & $-18.6$ & 218 & $580^{+60}_{-40}$ & $0.42^{+0.04}_{-0.03}$ & - & - \\[3pt]
 & & \textsc{beagle} TcSFH & $-18.6$ & 223 & $500^{+50}_{-40}$ & $0.45^{+0.04}_{-0.03}$ & - & - \\[6pt]

& \multirow{2}{*}{very faint} & \textsc{beagle} CSFH & $-17.5$ & 208 & $330^{+50}_{-40}$ & $0.43^{+0.07}_{-0.05}$ & - & - \\[3pt]
 & & \textsc{beagle} TcSFH & $-17.6$ & 200 & $300^{+40}_{-40}$ & $0.48^{+0.06}_{-0.05}$ & - & - \\[6pt]
\hline
\end{tabular}
\label{tab:inferredEWparameters}
\end{table*}

\subsection{Methodology} \label{sec:EW_methods}

Here, we describe how we quantify the UV luminosity dependence on the \OIIIHb{} and H$\alpha$ EW distributions in the reionization era.
For this analysis, we divide our JADES/NIRCam sample into three different subsets separated by absolute UV magnitude: a bright ($\Muv{} \leq -19.5$) subset, a faint ($-19.5 < \Muv{} \leq -18$) subset, and a very faint ($\Muv{} > -18$) subset.
We also aim to test whether there is any significant redshift evolution in the \OIIIHb{} EW distribution during the reionization era and thus infer this distribution for the F775W dropouts ($z\sim6$) and F090W dropouts ($z\sim7-9$), separately.
Because the H$\alpha$ line redshifts out of the reddest JADES/NIRCam band (F444W) at $z>6.6$, we restrict our analysis of the H$\alpha$ EW distribution to the F775W dropout sample.

To infer the EW distribution among a given sub-population of our sample, we follow the  Bayesian formalism described in \citet{Schenker2014} (see also \citealt{Endsley2021_OIII,Boyett2022_OIII}):
\begin{equation} \label{eq:bayes_int}
    P \left(\theta\right) \propto \prod_i \int P_i\left(\mathrm{EW}\right) P\left(\mathrm{EW}|\theta\right)\  d\mathrm{EW}.
\end{equation}
In this equation, $\theta$ encapsulates the parameters describing the assumed functional form of the distribution (e.g. the median and variance for a log-normal distribution) while $P_i\left(\mathrm{EW}\right)$ is the probability distribution function (PDF) on EW for the $i^\mathrm{th}$ galaxy in the sub-population of interest.
This formalism has two highly-desired features.
First, the expression in the integral of Eq. \ref{eq:bayes_int} is small when the assumed EW distribution ($P\left(\mathrm{EW}|\theta\right)$) is inconsistent with the PDF of galaxy $i$ while the expression is large when the assumed distribution and PDF strongly overlap.
Second, galaxies with narrow (i.e. precise) PDFs will have much stronger constraining power on the inferred $P \left(\theta\right)$ than galaxies with very broad (i.e. largely unconstrained) PDFs.

The PDFs for each galaxy are taken from the posterior of the SED fits, but are explicitly corrected for the prior on EW imposed by the SED fitting approach of interest (i.e. $P_i\left(\mathrm{EW}\right) \equiv \mathrm{posterior}_i\left(\mathrm{EW}\right) /\ \mathrm{prior}\left(\mathrm{EW}\right)$).
In doing so, we ensure that our resulting inferred EW distributions are being driven by evidence from the photometry rather than the choice of priors in the SED modelling.
These model priors have a stronger influence on the output posterior probability distribution functions for galaxies with lower signal-to-noise data, and hence would systematically impact the fainter populations more if we did not divide out the priors when computing $P_i\left(\mathrm{EW}\right)$.
To obtain the EW priors for each SED modelling setup, we fit a single photometric data point in the rest-UV thereby allowing the sampling algorithms to completely explore the allowed parameter space on \OIIIHb{} and H$\alpha$ EW.

In practice, we treat the integral in Eq. \ref{eq:bayes_int} as a Riemann sum to avoid having to approximate $P_i \left(\mathrm{EW}\right)$ as some functional form via, e.g., spline interpolation. 
Therefore, we rewrite Eq. \ref{eq:bayes_int} as
\begin{equation} \label{eq:bayes}
    P \left(\theta\right) \propto \prod_i \left[ \sum_j P_{i,j}\left(\mathrm{EW}\right) P_j\left(\mathrm{EW}|\theta\right) \right]
\end{equation}
where the index $j$ represents a bin in EW. $P_{i,j}\left(\mathrm{EW}\right)$ is the probability that object $i$ has an EW in bin $j$, while $P_j\left(\mathrm{EW}|\theta\right)$ is the integrated probability of the assumed functional EW distribution within bin $j$. Therefore, both $P_{i,j}\left(\mathrm{EW}\right)$ and $P_j\left(\mathrm{EW}|\theta\right)$ implicitly account for the width of bin $j$.

For the \OIIIHb{} EW inference, we divide the EW bins by values of log(EW/\AA{}) = 2.5--3.7 with spacing of 0.05 dex, yielding 26 total bins (i.e. log(EW/\AA{}) = [$\leq$2.50, 2.50--2.55, 2.55-2.60, ..., 3.65--3.70, $\geq$3.70]).
We chose a lower bound of log(EW/\AA{}) = 2.5 to reflect the fact that the photometric data can only place an effective upper limit on the inferred EWs of $\approx$300 \AA{} for galaxies with the lowest S/N of $\sim$3-5 in the long-wavelength NIRCam bands.
The upper bound of log(EW/\AA{}) = 3.7 was chosen given that the maximum inferred \OIIIHb{} EW in our sample is $\approx$5000 \AA{} (see \S\ref{sec:EELGs}).
Similarly, the H$\alpha$ EW bins applied in Eq. \ref{eq:bayes} are divided by values of log(EW/\AA{}) = 2.5--3.5 with spacing of 0.05 dex, where the upper bound of log(EW/\AA{}) = 3.5 reflects the maximum inferred H$\alpha$ EW of $\approx$3000 \AA{} in the sample.
We note that, with this approach, the inferred EW distribution values at $\lesssim$300 \AA{} are effectively constrained by the high-EW tail of the distribution given that we assume a log-normal functional form.

For each galaxy sub-population, we infer the EW distribution assuming a log-normal functional form defined by two parameters: the median of the EW distribution, \muEW{}, and its standard deviation, \sigmaEW{}. 
This assumption of a log-normal EW distribution is motivated by spectroscopic results at lower redshifts \citep{Lee2007,Lee2012,Ly2011}.
Finally, we infer the EW distributions using both the \textsc{beagle} constant star formation history (CSFH) and two-component star formation history (TcSFH) models.
Due to our findings in Appendix \ref{app:chisq} and \S\ref{sec:bursty} (see also Figs. \ref{fig:youngWeakLines} and \ref{fig:chisq}), we adopt the results from the TcSFH models as fiducial.

\subsection{Results on [OIII]+H\texorpdfstring{$\mathbf{\beta}$}{beta} EW Distributions} \label{sec:OIIIHb_EW_results}

We begin by quantifying the inferred \OIIIHb{} EW distribution among the brightest galaxies in our sample, comparing to existing results from the literature.
We then proceed to describe our findings for the very faint population that is now possible with the depth and area of JADES/NIRCam imaging.
From our fiducial TcSFH \textsc{beagle} SED fits, we infer a median \OIIIHb{} EW of \muEW{} = 740$^{+90}_{-80}$ \AA{} among the bright ($\langle \Muv{} \rangle = -20.1$) subset at $z\sim7-9$.
This is consistent with median \OIIIHb{} EWs reported in the literature for similarly UV-luminous \HST{} and NIRCam selected galaxy samples at $z\sim7-8$ \citep{Labbe2013,deBarros2019,Endsley2021_OIII,Endsley2023_CEERS,Stefanon2022_sSFR}, as well as samples of very UV-bright ($-22.5 \lesssim \Muv{} \lesssim -21.5$) $z\sim7$ galaxies selected from ground-based imaging \citep{Endsley2021_OIII,Endsley2023_CEERS}.
The inferred standard deviation of the EW distribution among our bright $z\sim7-9$ JADES/NIRCam subset (0.32$^{+0.04}_{-0.04}$ dex) is also consistent with values reported in the literature at similar luminosities \citep{Endsley2021_OIII,Endsley2023_CEERS}.

Due to the strong degeneracy between Balmer breaks and nebular line emission on IRAC colors at $z\sim6$, the \OIIIHb{} EW distribution has never been quantified in the literature at this epoch.
With the rich (five-band) 3--5$\mu$m SED sampling now afforded by JADES/NIRCam, we here infer that the median \OIIIHb{} EW among bright ($\langle \Muv{} \rangle = -20.0$) $z\sim6$ galaxies is \muEW{} = 890$^{+90}_{-80}$ \AA{}, $\approx$0.08 dex ($\approx$1.2$\times$) higher than that inferred from our bright $z\sim7-9$ sub-sample.
The standard deviation of \OIIIHb{} EWs among the bright $z\sim6$ subset is inferred to be very similar to that of the bright $z\sim7-9$ subset (see Table \ref{tab:inferredEWparameters}).

Now equipped with a sample of over 600 Lyman-break $z\sim6-9$ galaxies with $\approx$2--30$\times$ fainter UV luminosities ($-19.5 \leq \Muv{} \lesssim -16.5$), we investigate whether the \OIIIHb{} EW distribution correlates significantly with UV luminosity in the reionization era.
Using the fiducial TcSFH \textsc{beagle} SED fits, we infer a median \OIIIHb{} EW of \muEW{} = 590$^{+40}_{-50}$ \AA{} and 380$^{+80}_{-60}$ \AA{} among the faint ($\langle \Muv{} \rangle = -18.7$) and very faint ($\langle \Muv{} \rangle = -17.5$) $z\sim6$ subsets, respectively.
This indicates a smooth and strong (factor of $\approx$2.3) decline in the typical \OIIIHb{} EW of $z\sim6$ galaxies between the bright and very faint subsets, which are separated by an order of magnitude in average UV luminosity (see Fig. \ref{fig:EWdistns}).
Similarly, at $z\sim7-9$ we infer a median \OIIIHb{} EW of \muEW{} = 500$^{+50}_{-40}$ \AA{} and 300$^{+40}_{-40}$ \AA{} among the faint ($\langle \Muv{} \rangle = -18.6$) and very faint ($\langle \Muv{} \rangle = -17.6$) subsets, respectively, indicating a $\approx$2.5$\times$ decline in typical \OIIIHb{} EW over $\approx$1 dex in $L_\mathrm{UV}^{}$ (Fig. \ref{fig:EWdistns}).
Therefore, in both the $z\sim6$ and $z\sim7-9$ samples, we infer a strong UV luminosity dependence on the median \OIIIHb{} EW at $z\sim6-9$ with high confidence ($\gtrsim$5$\sigma$).

\begin{figure*}
\includegraphics{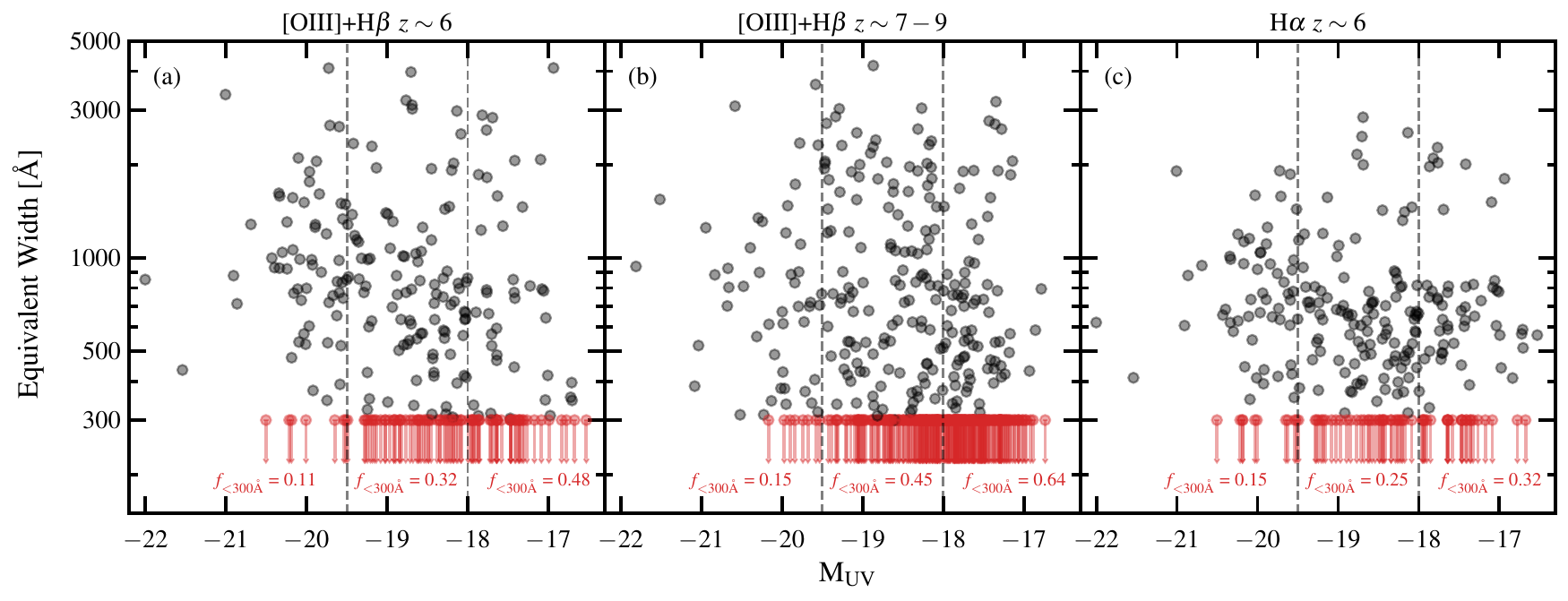}
\caption{The inferred fiducial EWs of each individual galaxy in our $z\sim6-9$ LBG sample plotted as a function of \Muv{}. These fiducial EWs correspond to the median of the posterior probability distribution from the \textsc{beagle} TcSFH fits. Those with EW$<$300 \AA{} are plotted as upper limits at that value (see \S\ref{sec:EW_methods} with red markers, and we quote the fraction of such objects in each \Muv{} bin. The trends obtained from these fiducial EWs alone (i.e., without correcting for the SED-fitting priors or directly accounting for object uncertainties) are broadly consistent with the results of the more robust EW distribution inference method (Fig. \ref{fig:EWdistns}).}
\label{fig:EWMuv}
\end{figure*}

To help visualize these trends, we plot the fiducial \OIIIHb{} EWs (the median of the posterior probability distribution) for each individual galaxy as a function of \Muv{} in Fig. \ref{fig:EWMuv}a,b.
In the UV-bright ($\Muv{} < -19.5$) bins, only a small fraction ($\approx$10--15\%) of $z\sim6-9$ LBGs have inferred \OIIIHb{} EWs of $<$300 \AA{} while this fraction rises to $\approx$50--65\% in the very UV-faint ($-18 \leq \Muv{} \lesssim -16.5$) bins.
Therefore, even when just considering the fiducial EWs of each galaxy (i.e., not accounting for the priors or uncertainties on EW), we find that the median \OIIIHb{} EW is $\lesssim$300 \AA{} at very low UV luminosities. 
With this very simple approach, we obtain much higher median \OIIIHb{} EWs of 880 \AA{} and 700 \AA{} in the UV-bright bins at $z\sim6$ and $z\sim7-9$, respectively.
Nonetheless, we emphasize that the EW distribution inference approach (\S\ref{sec:EW_methods}) should yield more robust constraints on the median EWs given that this method accounts for the SED-fitting priors and EW uncertainties for each object.

To understand why the NIRCam SEDs are driving the models to yield a strong decline in \OIIIHb{} EW with decreasing UV luminosity, we analyze the long-wavelength NIRCam colors sensitive to these lines.
We begin by considering the F775W dropout ($z\sim6$) sample where the large majority of galaxies lie at $5.5<z<6.5$ (see Fig. \ref{fig:MuvRedshiftDistns}).
For this sample, the F410M band provides a relatively clean probe of the rest-optical continuum given that it is free of contamination from the strongest rest-optical lines ([OIII], H$\beta$, and H$\alpha$) at $z\approx5.6-6.7$.
Meanwhile. the F356W and F335M bands are contaminated by \OIIIHb{} at redshifts associated with the majority of the $z\sim6$ subset ($z\approx5.4-6.9$ and $z\approx5.5-6.1$, respectively).
Therefore, we investigate whether the measured photometry of the $z\sim6$ sample supports significant flattening in the F410M$-$F335M and F410M$-$F356W colors as we move from the bright to very faint subsets.
In the bright $z\sim6$ subset, the median F410M$-$F335M and F410M$-$F356W colors are $0.98^{+0.08}_{-0.10}$ and $0.63^{+0.02}_{-0.07}$ mag, respectively, consistent with typical contamination from strong \OIIIHb{} emission.
These colors are typically flatter among the very faint sub-population ($0.67^{+0.10}_{-0.05}$ and $0.54^{+0.06}_{-0.13}$ mag, respectively), consistent with generally weaker \OIIIHb{} EWs at lower UV luminosities.
The uncertainties on median colors are derived using bootstrap resampling.

We conduct a similar color analysis with the F090W dropout sample.
Across nearly the entire redshift range of our F090W dropout selection, the rest-optical continuum is probed free of strong lines by either F335M ($z\approx6.3-8.2$) or F356W ($z\approx7.2-8.8$) while the \OIIIHb{} emission is captured by either F410M ($z\approx7.0-7.6$) or F444W ($z\approx7.0-9.0$).
We therefore investigate whether the measured photometry supports significant flattening in the F335M$-$F410M, F335M$-$F444W, F356W$-$F410M, and F356W$-$F444W colors as we move from the bright to very faint subsets of our $z\sim7-9$ sample. 
In the bright subset, the median values for all four above colors are noticeably red ($0.49^{+0.06}_{-0.12}$, $0.34^{+0.03}_{-0.07}$, $0.46^{+0.03}_{-0.09}$, and $0.27^{+0.08}_{-0.02}$, respectively), consistent with typical contamination of strong \OIIIHb{} emission.
On the other hand, in the very faint subset we measure much flatter median colors ($0.20^{+0.04}_{-0.06}$, $0.00^{+0.05}_{-0.02}$, $0.23^{+0.04}_{-0.06}$, and $-0.01^{+0.08}_{-0.03}$, respectively) suggestive of typically much weaker \OIIIHb{} EWs.

\begin{figure*}
\includegraphics{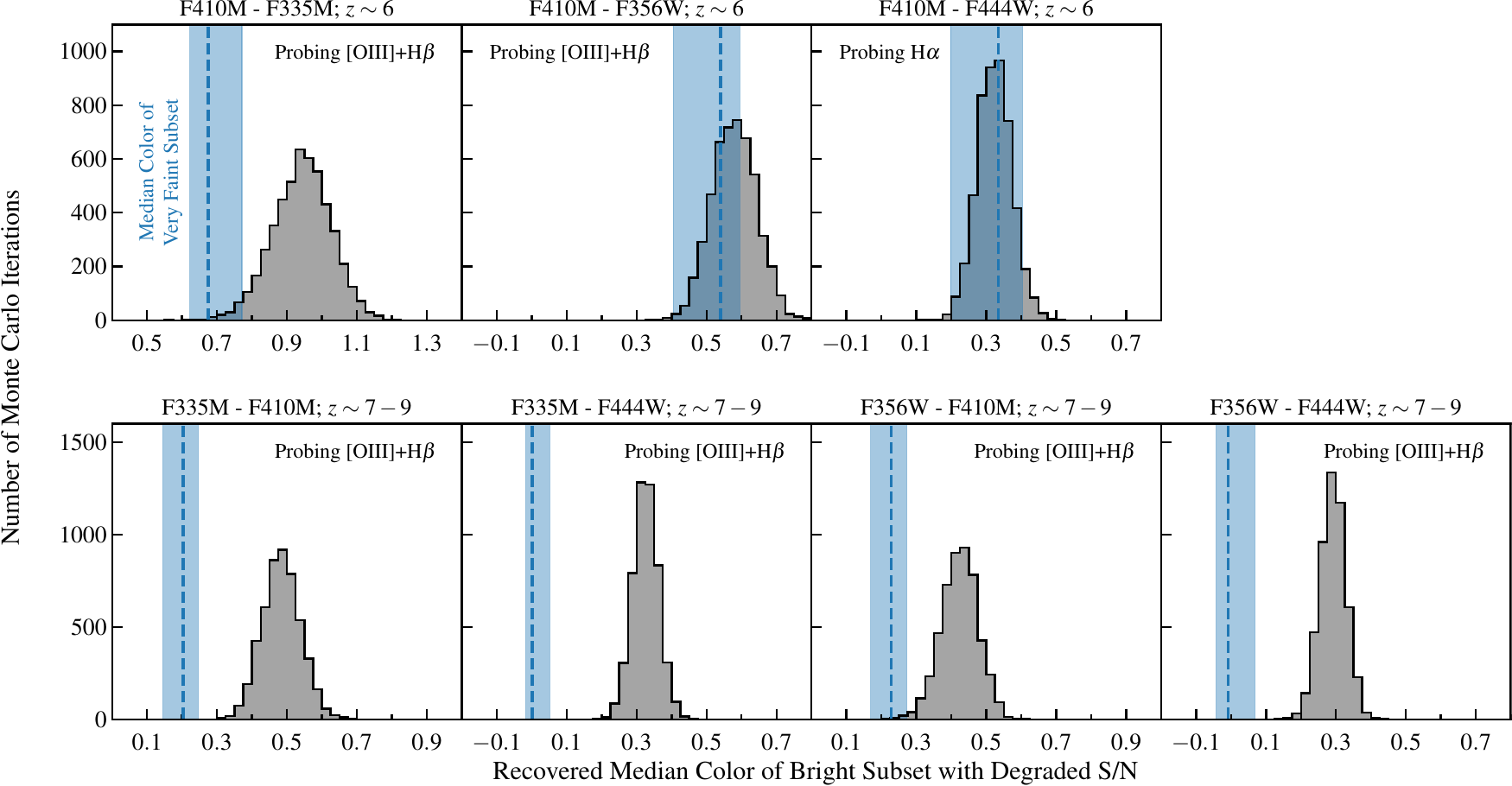}
\caption{Illustration of results from Monte Carlo simulation test validating that the substantial decline in median \OIIIHb{} EW at fainter UV luminosities is very unlikely to be caused by photometric noise. The shaded histograms show the distribution of recovered median colors (one color combination per panel) from 5000 iterations of degrading the photometry of objects in our brightest ($\Muv{} < -19.5$) bin to the S/N of galaxies in the faintest bin ($\Muv{} > -18$). For nearly all colors sensitive to \OIIIHb{} contamination, very few ($\leq$0.1\%) of the 5,000 Monte Carlo iterations yield median recovered colors as flat (or flatter) as that measured among the very faint subset (vertical dashed lines). The median F410M$-$F444W color of the very faint $z\sim6$ subset are however consistent with that recovered from the bright subset (after S/N degradation), implying little to no UV luminosity dependence on the median H$\alpha$ EW.}
\label{fig:monteCarloColors}
\end{figure*}

Because the sample size of each \Muv{} bin is fairly large ($N_\mathrm{gal}$ = 55--200; see Table \ref{tab:inferredEWparameters}), we do not expect these median colors to be strongly impacted by noise.
Nonetheless, we employ a Monte Carlo simulation to quantify the likelihood of measuring the median colors of the very faint subsets (accounting for noise) under the assumption that they indeed have true colors represented by their respective bright subset of galaxies.
For this test, we take the measured photometry of all galaxies in the bright subset for a given dropout sample (randomly sampled with replacement), uniformly re-normalize the photometry of each bright galaxy such that it has an \Muv{} value equivalent to a galaxy in the very faint subset in the same dropout sample (again randomly sampled with replacement), and add Gaussian noise to their photometry in each band using the uncertainties from the associated very faint galaxy.
This Monte Carlo simulation is performed 5000 times and, in each iteration, we compute the median colors of interest.

The distributions of median colors from the Monte Carlo simulations are shown in Fig. \ref{fig:monteCarloColors}, along with the actual measured median colors of the very faint subsets plotted as vertical dashed lines in each panel. 
For nearly every color probing \OIIIHb{} emission, we find that very few ($\leq$0.1\%) of the 5,000 Monte Carlo iterations yield median recovered colors as flat (or flatter) than that measured among the very faint subset.
While the median F410M$-$F356W Monte Carlo colors in the F775W dropout subset are consistent (at the $\approx$1$\sigma$ level) with that measured among the very faint subset, F356W is less sensitive to strong \OIIIHb{} emission relative to F335M given its broader bandwidth and thus the significance in color difference is expected to be considerably weaker.
Moreover, the smaller size of the very faint F775W dropout sample makes its median color uncertainties $\approx$2$\times$ higher relative to those of the F090W dropout sample (see Fig. \ref{fig:monteCarloColors}).
This higher noise likely contributes to the lack of a strong difference in the median F410M$-$F356W Monte Carlo colors relative to that measured for the very faint F775W dropout sample.

Having now validated that the inferred trend in median \OIIIHb{} EW with UV luminosity is consistent with expectations given the measured long-wavelength colors, we consider whether the data imply any significant changes in the width of the EW distribution towards fainter UV luminosities.
Even though the median EW of the very faint population declines substantially relative to the brightest \Muv{} bin, we continue to identify a number of very faint galaxies with prominent long-wavelength colors implying extremely high \OIIIHb{} EWs of $>$1500 \AA{} (see, e.g., ID JADES-GS+53.17835$-$27.77879 and JADES-GS+53.19590$-$27.79240 in Fig. \ref{fig:EELGs}).
Accordingly, our assumed log-normal EW distributions are inferred to broaden considerably towards fainter UV luminosities at both $z\sim6$ and $z\sim7$, with \sigmaEW{} increasing from $\approx$0.30$\pm$0.03 dex in the bright subsets to $\approx$0.50$\pm$0.06 dex in the very faint subsets (see Table \ref{tab:inferredEWparameters}).
From these distributions, we infer that the fraction of galaxies with extremely high \OIIIHb{} EWs ($>$1500 \AA{}) declines significantly towards fainter UV luminosities, going from $24^{+5}_{-4}$\% ($18^{+5}_{-4}$\%) in the bright $z\sim6$ ($z\sim7-9$) subset to $13^{+4}_{-3}$\% ($8^{+2}_{-2}$\%) in the respective very faint subset.
Conversely, the fraction of galaxies with relatively very low \OIIIHb{} EWs ($<$300 \AA{}) rises dramatically towards the faint-end of the luminosity function, increasing from $6^{+3}_{-2}$\% ($11^{+5}_{-4}$\%) in the bright $z\sim6$ ($z\sim7-9$) subset to $42^{+7}_{-7}$\% ($50^{+5}_{-5}$\%) in the respective very faint subset.

\begin{figure*}
\includegraphics{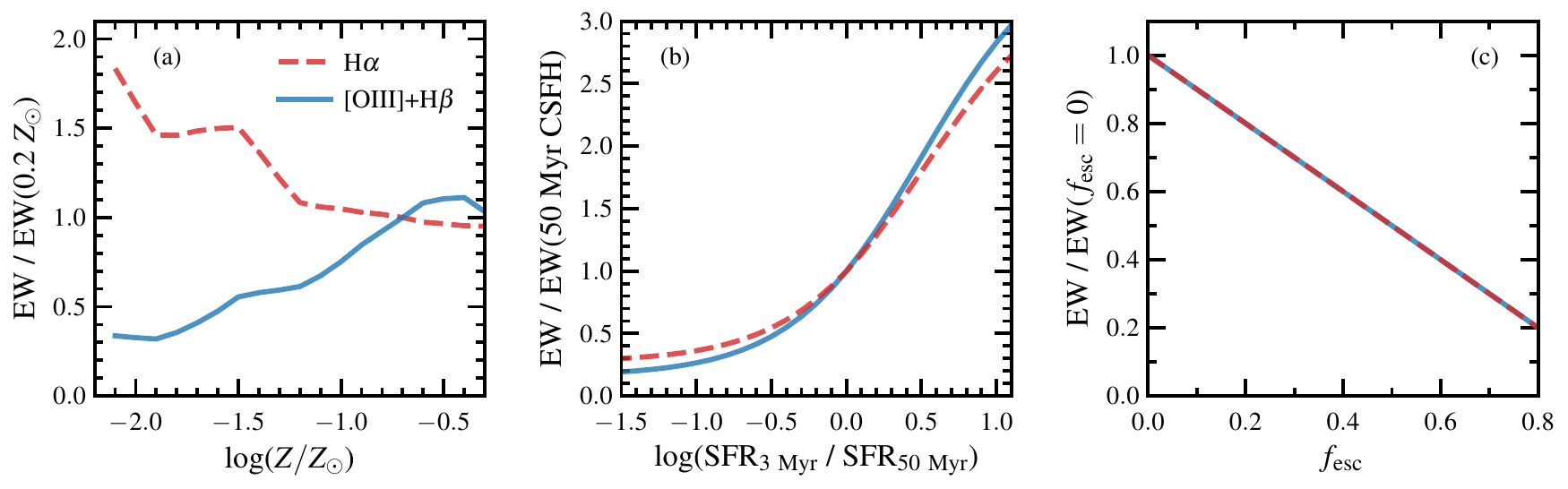}
\caption{Illustration of how the EWs of \OIIIHb{} (blue solid lines) and H$\alpha$ emission correlate with metallicity (panel a), recent star formation history, quantified by the ratio of SFR averaged over the past 3 Myr divided by the SFR averaged over the past 50 Myr (panel b), and LyC photon escape fraction assuming ionization-bounded conditions (i.e. `picket-fence' models; panel c). In both panels, the EWs are normalized by their values adopting a 50 Myr CSFH model with 0.2 $Z_\odot$, log $U = -2.5$, \fesc{} = 0, and $\tau_{_V}$ = 0 with the \citet{Gutkin2016} models implemented in \textsc{beagle}.}
\label{fig:EWratioParameterDependence}
\end{figure*}

\subsection{Results on H\texorpdfstring{$\mathbf{\alpha}$}{alpha} EW Distributions} \label{sec:Halpha_EW_results}

Due to the poor rest-optical SED sampling of IRAC photometry, previous efforts to quantify H$\alpha$ EWs at high redshifts have primarily been limited to the redshift range $z\approx4.0-5.0$ where several works have inferred typical H$\alpha$ EWs of $\sim$400 \AA{} among \HST{}-selected galaxies \citep{Stark2013_NebEmission,Smit2016,Rasappu2016,Faisst2019,Lam2019}.
Most recently, \citet{Stefanon2022_IRACz8} stacked the IRAC 5.8$\mu$m data of $\approx$100 $z\sim8$ galaxies and recovered a considerably larger typical H$\alpha$ EW ($\gtrsim$1000 \AA{}), perhaps implying a significant redshift evolution in the median H$\alpha$ EW at high redshifts.
With the JADES NIRCam data, we are now able to infer the H$\alpha$ EW distribution at $z\sim6$ and moreover investigate whether it correlates strongly with UV luminosity as may be expected given our results on the \OIIIHb{} EW distribution above.

From the \textsc{beagle} TcSFH SED fits, we infer very similar median H$\alpha$ EWs among all three UV luminosity bins at $z\sim6$ ($\approx$600--700 \AA{} in each; see Table \ref{tab:inferredEWparameters}).
Combined with the literature results discussed above, this implies that the typical H$\alpha$ EWs increase by $\approx$0.2 dex ($\approx$1.6$\times$) between $z\sim4.5$ and $z\sim6$.  
But perhaps the most striking result here is that we find no evidence for a strong change in the typical $z\sim6$ H$\alpha$ EW across 1 dex in UV luminosity, in marked contrast with our findings on the \OIIIHb{} EW distribution in \S\ref{sec:OIIIHb_EW_results}.
Even when just using the fiducial H$\alpha$ EWs for each object (Fig. \ref{fig:EWMuv}c), we find that the median H$\alpha$ EW lies in a narrow (0.1 dex) range of 530--670 \AA{} for every \Muv{} bin.

We again verify that the measured long-wavelength NIRCam colors are consistent with little-to-no change in the median H$\alpha$ EW between the bright and very faint $z\sim6$ subsets following the same approach as described in \S\ref{sec:OIIIHb_EW_results}.
At $z\sim6$, the F410M band provides a relatively clean probe of the rest-optical continuum (see \S\ref{sec:OIIIHb_EW_results}) while the F444W band contains H$\alpha$.
The median F410M$-$F444W color of the bright $z\sim6$ subset ($0.35^{+0.05}_{-0.04}$ mag) is nearly identical to that of the very faint subset ($0.33^{+0.07}_{-0.14}$ mag), consistent with a weak trend in typical H$\alpha$ EW with UV luminosity.
The Monte Carlo simulation test described in \S\ref{sec:OIIIHb_EW_results} also reveals that even accounting for signal-to-noise differences in the bright versus very faint $z\sim6$ subsets often yields nearly identical median F410M$-$F444W colors between the two populations (see Fig. \ref{fig:monteCarloColors}).
We thus conclude that the photometry is consistent with little change in typical $z\sim6$ H$\alpha$ EWs over the UV luminosity range spanned by our sample ($-22 \leq \Muv{} \leq -16.4$).

There is moderate evidence ($\approx$2$\sigma$) for a slight ($\approx$0.05 dex) increase in the width of the $z\sim6$ H$\alpha$ EW distribution towards very faint UV luminosities (see Table \ref{tab:inferredEWparameters}). This is consistent with the slightly larger fraction of very faint $z\sim6$ galaxies with fiducial H$\alpha$ EWs $<$300 \AA{} (see Fig. \ref{fig:EWMuv}c) even though the median EW remains effectively unchanged. Nonetheless, overall the inferred H$\alpha$ EW distributions for the three UV luminosity bins are nearly indistinguishable (see Fig. \ref{fig:EWdistns}).
If we assume that the H$\alpha$ EW distribution can be described as a single log-normal form across the entire UV luminosity range of our $z\sim6$ sample ($-22 \leq \Muv{} \leq -16.4$), we obtain \muEW{} = $600^{+40}_{-10}$ \AA{} and \sigmaEW{} = 0.27$\pm$0.01 dex.
Given the strong UV luminosity dependence on \OIIIHb{} EW inferred in \S\ref{sec:OIIIHb_EW_results}, it is no surprise that we then infer a strong UV luminosity dependence on the [OIII]$\lambda$5007/H$\alpha$ ratio (hereafter [OIII]/H$\alpha$) as well.
In the bright $z\sim6$ subset, we infer a median [OIII]/H$\alpha$ $\approx$ 1.6 while in the faint and very faint subsets we infer median [OIII]/H$\alpha$ ratios of $\approx$1.3 and $\approx$0.9, respectively.

\section{Implications for Star Formation and Ioinizing Efficiency in the Reionization Era} \label{sec:discussion}

In the previous section, we demonstrated that the JADES/NIRCam data indicate a strong decline in \OIIIHb{} EW with UV luminosity at $z\sim6-9$, yet an approximately constant H$\alpha$ EW distribution across an order of magnitude in L$_\mathrm{UV}^{}$ at $z\sim6$ (Fig. \ref{fig:EWdistns}).
Here, we begin by discussing what physically may be driving the different L$_\mathrm{UV}^{}$ trends for \OIIIHb{} and H$\alpha$ EWs then explore the results implied from the SED fits, with the ultimate goal of considering potential implications for the nature of very UV-faint reionization-era galaxies.

\subsection{The Role of Metallicity in Explaining the EW Trends} \label{sec:metallicity}

One factor likely influencing the UV luminosity trends in nebular EWs is metallicity.
Due to the deficit of oxygen atoms at very low metallicity, the strength of nebular \OIIIHb{} emission declines substantially at $\lesssim$0.2 $Z_\odot$ (see Fig. \ref{fig:EWratioParameterDependence}a).
Because fainter, lower-mass galaxies are generally expected to be less chemically enriched, this likely contributes to much weaker \OIIIHb{} EWs in our faintest subset.
We consider how much lower the typical metallicity of very faint ($\langle \Muv{} \rangle \approx -17.5$) $z\sim6-9$ galaxies would have to be relative to the bright subset ($\langle \Muv{} \rangle \approx -20$) to completely explain the $\approx$2.4$\times$ decline in median \OIIIHb{} EW inferred in \S\ref{sec:OIIIHb_EW_results}.
For this test, we utilize the \citet{Gutkin2016} models employed in \textsc{beagle} which we note assume equal stellar and gas-phase metallicities (i.e. no $\alpha$-enhancement).
From the \citet{Gutkin2016} models, we find that the metallicity would need to drop by $\approx$1 dex to solely explain the substantial decline in \OIIIHb{} EW (see Fig. \ref{fig:EWratioParameterDependence}a).

Such a dramatic ($\approx$1 dex) decline in metallicity would have a significant impact on the H$\alpha$ EWs.
Due to changes in stellar opacity, models with lower metallicity lead to more efficient ionizing photon production and hence larger H$\alpha$ EWs.
From the \citet{Gutkin2016} star-forming models employed in our \textsc{beagle} fits, we expect that lowering the metallicity by 1 dex from 0.2 $Z_\odot$ to 0.02 $Z_\odot$ would result in a $\approx$50\% increase in H$\alpha$ EWs.
Therefore, while lower metallicities at fainter UV luminosities may partially explain the decreasing \OIIIHb{} EWs, something else must also be changing between the bright and very faint populations to explain why the H$\alpha$ EW distribution remains nearly constant at $z\sim6$.

\subsection{Evidence for Bursty Star Formation Histories at \texorpdfstring{$\mathbf{z\gtrsim6}$}{z > 6}} \label{sec:bursty}

One potential way to offset the expected increase in H$\alpha$ EW at fainter $L_\mathrm{UV}$ due to lower metallicity is by invoking differences in the recent star formation histories of bright versus very faint reionization-era galaxies (we discuss an alternative solution with differing ionizing photon escape fractions below\footnote{Other mechanisms may also be responsible for decreasing the [OIII] and Balmer line EWs at fainter UV luminosities, including changes in IMF, contributions from low-luminosity AGN, or $\alpha$-enhanced metallicities.}).
Because the Balmer and [OIII] nebular lines are primarily powered by hot, massive O stars that have formed in the most recent $\approx$3 Myr, a strong drop in the SFR over this timescale results in weaker emission lines, decreasing the EWs of H$\alpha$ and \OIIIHb{} (see Fig. \ref{fig:EWratioParameterDependence}b).
If UV-faint galaxies are more often caught in a recent downturn of star formation (relative to the UV-bright population), this may explain why their H$\alpha$ EWs remain comparable to that of the more UV luminous population.
Such a change in recent SFHs would also contribute to the reduced \OIIIHb{} EWs in the very UV-faint population (Fig. \ref{fig:EWdistns}), working in tandem with the shift toward lower metallicities.

To better assess whether the photometric data support systematic changes in metallicity and recent star formation histories across the wide UV luminosity range spanned by our sample ($-22 \lesssim \Muv{} \lesssim -16.5$), we utilize the results of our two-component star formation history (TcSFH) \textsc{beagle} fits.
For each galaxy, we compute the inferred ratio of the star formation rate averaged over the past 3 Myr (SFR$_{3\ \mathrm{Myr}}$) to that over the past 50 Myr (SFR$_{50\ \mathrm{Myr}}$) as a proxy for whether their SFH has recently declined or risen on timescales most relevant for nebular line emission.
The 50 Myr timescale adopted in the denominator of this ratio was chosen to reflect the typical light-weighted age of the sample, though we note that our findings below are qualitatively unchanged if we instead adopt a 30 Myr or 100 Myr timescale.
We quantify the distributions of SFR$_{3\ \mathrm{Myr}}$ / SFR$_{50\ \mathrm{Myr}}$ and metallicity in the same three UV luminosity bins as considered for the EW distribution analysis (see \S\ref{sec:EW_methods}), again adopting a log-normal distribution for each \Muv{} bin and applying Eq. \ref{eq:bayes}.
Here, we focus our analysis solely on the $z\sim6$ sample given the lack of H$\alpha$ information for the large majority of the $z\sim7-9$ sample.
Because the SFR$_{3\ \mathrm{Myr}}$ / SFR$_{50\ \mathrm{Myr}}$ ratio is by definition restricted to values $\leq\frac{50}{3}$ (this maximum value occurs when no stars are assumed to form between 3--50 Myr ago), we truncate the log-normal distributions at this upper limit (i.e. $P$(SFR$_{3\ \mathrm{Myr}}$ / SFR$_{50\ \mathrm{Myr}} > \frac{50}{3}\ |\ \theta$) = 0).
Similarly, the log-normal distributions for metallicity are truncated outside the fitted parameter space in our SED fits ($-2.2 \leq$ log($Z/Z_\odot$) $\leq -0.3$).
When quoting median inferred values for metallicity and SFR$_{3\ \mathrm{Myr}}$ / SFR$_{50\ \mathrm{Myr}}$ below we account for this truncation, though the standard deviations are quoted directly from the assumed log-normal functional form.

\begin{figure*}
\includegraphics{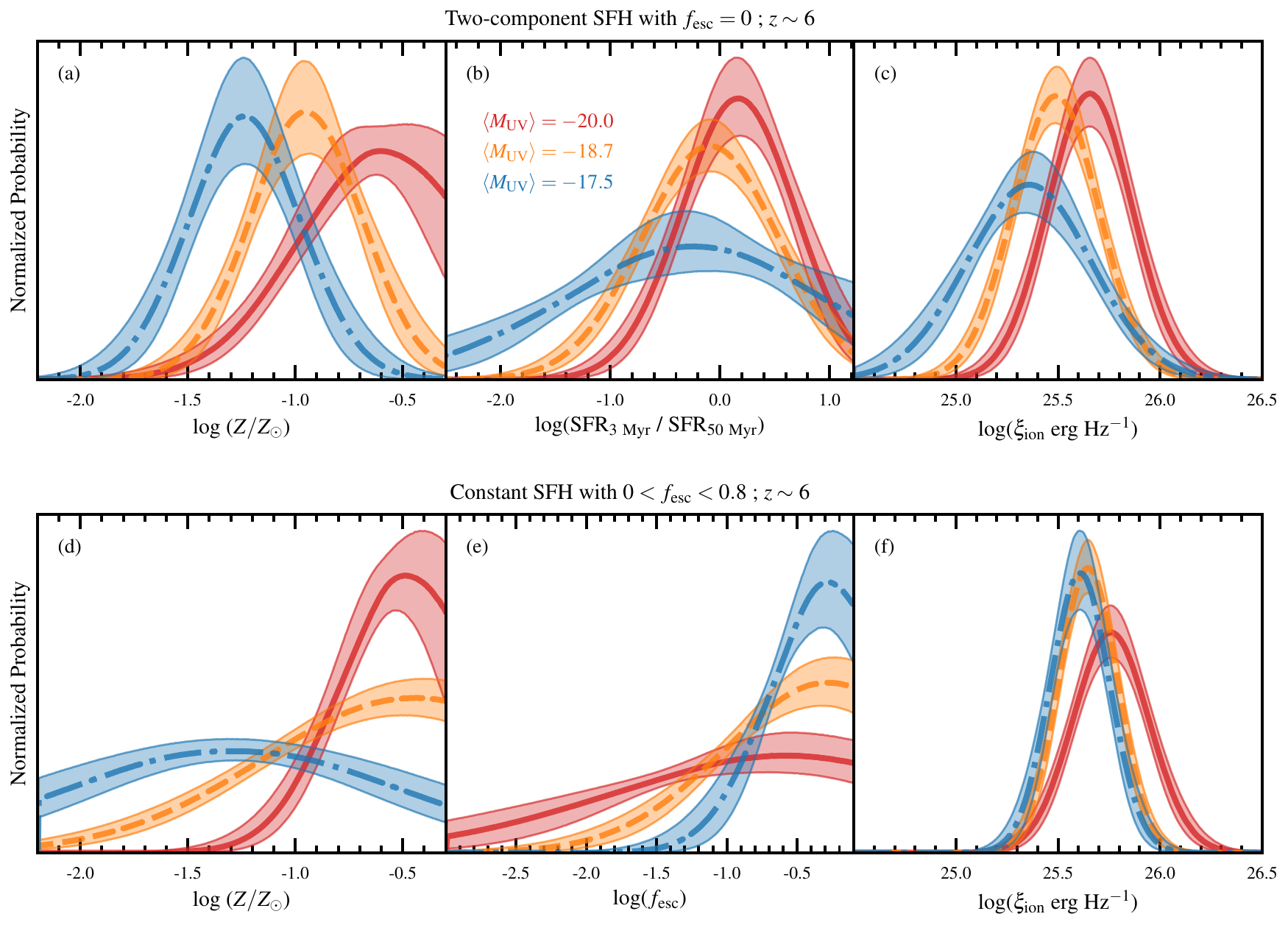}
\caption{Illustration of how the strong decline in \OIIIHb{} EW with UV luminosity yet approximately constant H$\alpha$ EW distributions at $z\sim6$ (see Fig. \ref{fig:EWdistns}) can be explained by either the combination of lower metallicity and more recently declining star formation histories in the very UV-faint population (top) or lower metallicities and higher LyC escape fractions in the very UV-faint population (bottom). These two scenarios have different implications for contributions of galaxies along the UV luminosity function to cosmic reionization, as also illustrated in the inferred ionizing photon production efficiency distributions in the right panels. Deep spectra will be required to better distinguish the mechanisms dominating the UV-luminosity dependent EW trends found here.}
\label{fig:metSFRratFescDistns}
\end{figure*}

The TcSFH \textsc{beagle} outputs suggest a significant anti-correlation between UV luminosity and metallicity at $z\sim6$ (Fig. \ref{fig:metSFRratFescDistns}a), with a median inferred metallicity of $0.27^{+0.13}_{-0.08}$ $Z_\odot$, $0.11^{+0.01}_{-0.01}$ $Z_\odot$, and $0.06^{+0.01}_{-0.01}$ $Z_\odot$ in the bright ($\langle \Muv{} \rangle = -20.0$), faint ($\langle \Muv{} \rangle = -18.7$), and very faint ($\langle \Muv{} \rangle = -17.5$) subsets, respectively.
Therefore, in context of these models, changes in metallicity contribute significantly to the different UV luminosity trends with \OIIIHb{} and H$\alpha$ EWs.
We check whether these inferred metallicities are consistent with predictions from simulations under very simple assumptions on the connection between \Muv{} and stellar mass and assuming that our inferred metallicities largely reflect the gas-phase metallicity.
In our $z\sim6$ sample, the typical stellar mass at $\Muv{} \approx -20$ is \Mstar{}$\sim$4$\times$10$^{8}$ \Msol{} while it is $\sim$2$\times$10$^{7}$ \Msol{} at $\Muv{} \approx -17.5$ (see Fig. \ref{fig:MstarSFH}).
The inferred metallicities above from the TcSFH models are reasonably consistent (factor of $\lesssim$2 difference) with the predicted $z\sim6$ gas-phase metallicities at the respective stellar masses from the \textsc{illustris tng} \citep{Torrey2019}, \textsc{FirstLight} \citep{Langan2020}, and \textsc{astreaus} \citep{Ucci2023} simulations, though systematically $\approx$5$\times$ higher than that predicted from the \textsc{fire} simulations \citep{Ma2016_MZR}.

We now investigate whether the TcSFH \textsc{beagle} fit outputs also imply differences in the recent SFHs of bright vs. very faint $z\sim6$ galaxies.
Using Eq. \ref{eq:bayes}, we infer a median SFR$_{3\ \mathrm{Myr}}$ / SFR$_{50\ \mathrm{Myr}}$ ratio of $1.5^{+0.3}_{-0.2}$, $0.8^{+0.1}_{-0.1}$, and $0.6^{+0.9}_{-0.2}$ in the bright, faint , and very faint subsets, respectively.
Therefore, under context of the adopted SED models, we do find evidence that brighter $z\sim6$ galaxies tend to have more very-recently rising SFHs than the faintest sources (Fig. \ref{fig:metSFRratFescDistns}b).
Notably, not all bright (very faint) galaxies are inferred to have very-recently rising (declining) SFHs as we recover substantial scatter in the SFR$_{3\ \mathrm{Myr}}$ / SFR$_{50\ \mathrm{Myr}}$ ratio at fixed UV luminosity.
The standard deviation of the assumed log-normal distribution is inferred to be $0.52^{+0.09}_{-0.07}$ dex, $0.62^{+0.08}_{-0.06}$ dex, $1.21^{+0.66}_{-0.33}$ dex in the bright, faint, and very faint subsets, respectively, implying generally more variation in recent SFHs at fainter luminosities (and hence lower stellar masses).

\begin{figure}
\includegraphics{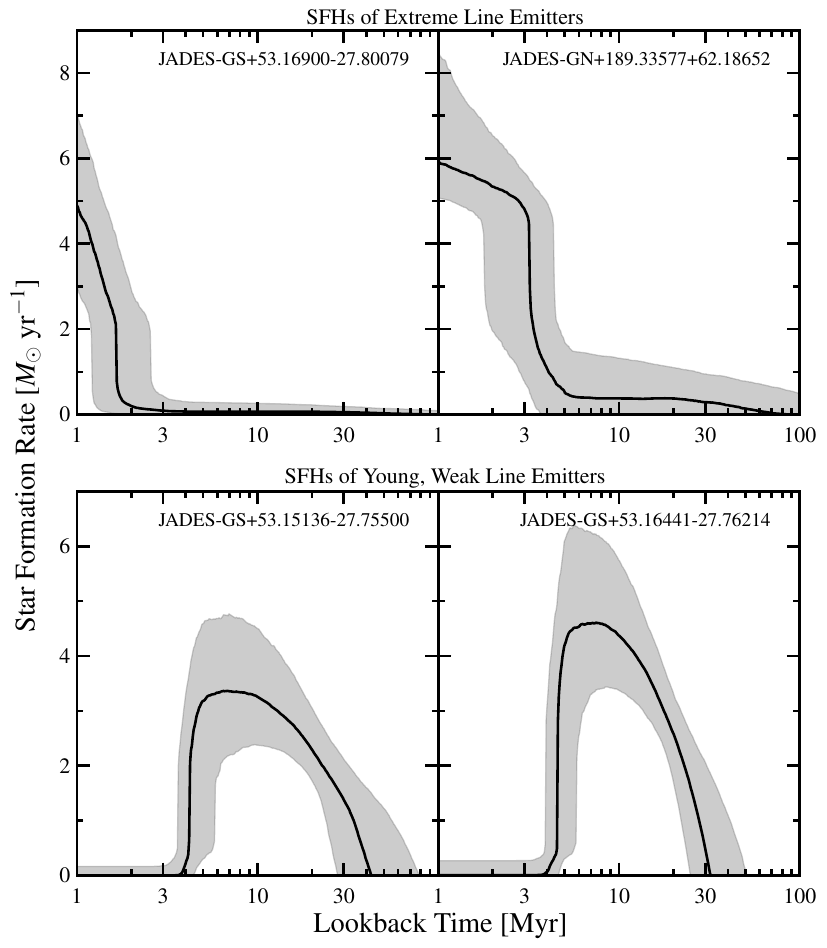}
\caption{Example inferred star formation histories of two $z\sim6$ extreme emission line galaxies (top; SEDs shown in Fig. \ref{fig:EELGs}) and two $z\sim6$ galaxies showing SEDs consistent with young light-weighted ages yet weak line emission (bottom; SEDs shown in Fig. \ref{fig:youngWeakLines}). The median SFHs (solid lines) along with their inner 68\% credible intervals (gray regions) are taken from the \textsc{beagle} TcSFH models. The extreme emission line galaxies are experiencing a rapid upturn in SFR while the young, weak line emitters are inferred to have recently experienced a strong downturn in SFR, consistent with bursty SFHs among reionization-era galaxies.}
\label{fig:exampleSFHs}
\end{figure}

There are two particular subsets of our sample that provide evidence in favor of bursty SFHs among reionization-era galaxies.
The first subset consists of galaxies showing extremely strong nebular line emission.
The very high \OIIIHb{} EWs ($\sim$2000-5000 \AA{}) confidently inferred among $\approx$50 galaxies in our sample (see \S\ref{sec:EELGs}) imply that a considerable fraction of Lyman-break $z\sim6-9$ galaxies have recently experienced a rapid strong upturn in SFR, yielding SEDs that are completely dominated by hot, massive O stars.
We plot the inferred star formation histories of two of the confident $z\sim6$ extreme emission line galaxies in the top panels of Fig. \ref{fig:exampleSFHs}, where both are inferred to have substantially ($\gtrsim$5$\times$) higher SFRs over the most recent 3 Myr relative to that over the past 10--50 Myr (see also e.g. \citealt{Tacchella2023_NIRCamNIRSpec}).
Such a dramatic recent increase in SFR is consistent with a scenario in which these extreme emission line galaxies are undergoing a burst of star formation.
We note that a variety of recent papers are building empirical evidence for bursty SFHs among $z\gtrsim6$ galaxies using different methods \citep{Dome2023,Dressler2023_glass,Dressler2023_jades,Looser2023,Looser2023b,Strait2023}.
Bursty star formation in the early Universe is also predicted in several simulations and models of galactic evolution \citep[e.g.][]{Kimm2015,Ceverino2018,FaucherGiguere2018,Ma2018_burstySFH,Furlanetto2022}.

\begin{figure*}
\includegraphics{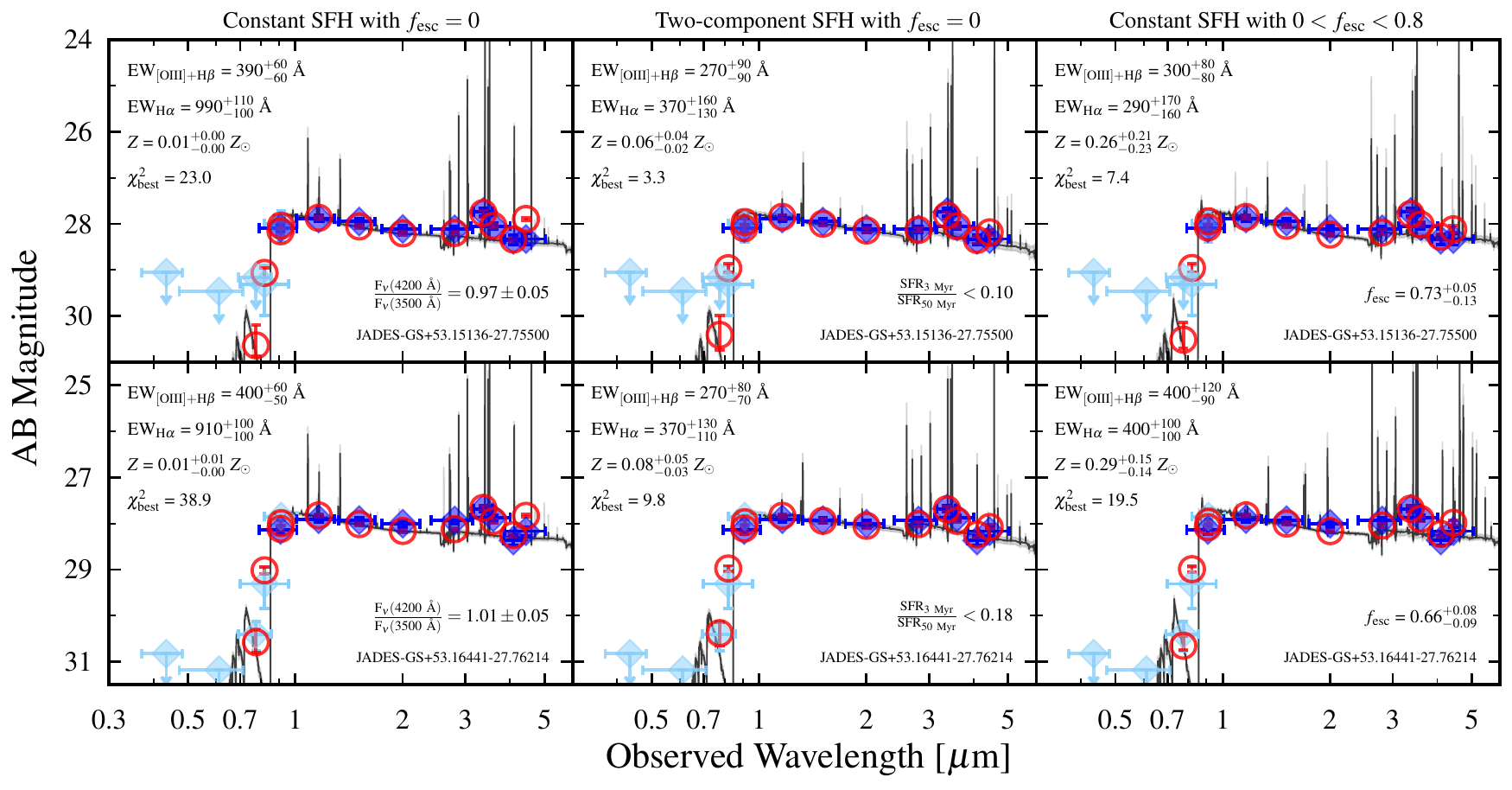}
\caption{Illustration of two $\zphot{} \approx 6.0$ galaxies (one per row) showing SEDs consistent with young light-weighted ages as well as weak \OIIIHb{} and H$\alpha$ emission. We find that our fiducial CSFH models (with \fesc{} = 0; left panels) struggle to match the data while models allowing for strong recent drops in SFR (middle panels; see Fig. \ref{fig:exampleSFHs} for inferred SFHs) or very high LyC escape fractions (right panels) provide considerably better fits to the data.}
\label{fig:youngWeakLines}
\end{figure*}

Another subset of galaxies in our JADES sample may be caught during a relatively inactive phase of star formation.
We identify $z\sim6$ galaxies exhibiting NIRCam SEDs consistent with relatively weak Balmer breaks (thus implying young light-weighted ages) yet also weak \OIIIHb{} and H$\alpha$ emission. 
The SEDs of two of these galaxies are shown in Fig. \ref{fig:youngWeakLines} for illustrative purposes (we discuss the abundance of similar objects within our full $z\sim6$ JADES sample below).
In the left panels of Fig. \ref{fig:youngWeakLines}, we show the fitted SEDs from the \textsc{beagle} CSFH models which clearly struggle to reproduce the observed photometry.
For both galaxies, the CSFH models are forced to extremely low-metallicity solutions ($\approx$0.01 $Z_\odot$) to simultaneously explain the weak Balmer breaks\footnote{While the CSFH model fits to these two galaxies yield fairly high light-weighted ages ($\sim$100 Myr), the resulting Balmer breaks in the models remain weak due to strong nebular continuum emission. As has been discussed elsewhere (e.g. \citealt{Byler2017,Topping2022_blueSlopes}), nebular continuum emission is stronger at lower metallicities and the fractional contribution of nebular emission to the total (stellar+nebular) continuum is larger immediately blueward of the Balmer break than redward of this break.} ($\BBrat{} \approx 1$ as implied by the fairly flat F200W$-$F277W colors) as well as the low \OIIIHb{} EWs ($\approx$300--400 \AA{} as implied by the weak photometric excesses in F335M and F356W).
The extremely low-metallicity solutions in turn cause the models to drastically over-predict the H$\alpha$ EWs relative to that implied by the observed F410M$-$F444W colors, suggesting that the SEDs of these galaxies cannot be well explained by our fiducial CSFH models.

When allowing for more flexible star formation histories, we obtain SED models that yield much better fits to the photometry of these two illustrative $z\sim6$ galaxies.
As shown in the middle panels of Fig. \ref{fig:youngWeakLines}, the \textsc{beagle} TcSFH models can reproduce the NIRCam photometry in all nine bands with moderately low-metallicity solutions ($\approx$0.05--0.1 $Z_\odot$) for these two $z\sim6$ galaxies.
In these TcSFH solutions, the two galaxies are currently experiencing a strong decline in SFR that has followed a burst of star formation which occurred $\sim$5--50 Myr ago (see bottom panels of Fig. \ref{fig:exampleSFHs}).
Such star formation histories yield relatively few O stars that are still alive to power the [OIII] and Balmer lines, while the B and A stars that formed $\gtrsim$10 Myr ago continue to dominate the rest UV+optical SEDs.
It is thus possible that these galaxies were in an extreme emission line phase just $\sim$10--30 Myr earlier, and are now caught in a relatively passive phase of star formation yielding weak nebular lines yet also young SEDs from the continua.
We note that a significant population of reionization-era galaxies with young light-weighted ages and weak \OIIIHb{} emission was first tentatively identified in \citet{Endsley2023_CEERS} from early \JWST{} data, though the lack of H$\alpha$ constraints for the $z\sim6.5-8$ sample considered therein left much uncertainty on the physical origin of the weak line emission (see also \citealt{Simmonds2023,PrietoLyon2023}).
With the improved imaging sensitivity and SED sampling of JADES and GOODS, we have now extended this investigation to a $z\sim6$ sample and find that our fiducial CSFH models cannot adequately explain the photometry of at least some of these very early galaxies.

In addition to the two illustrative objects shown in Fig. \ref{fig:youngWeakLines}, we identify a few other $z\sim6$ galaxies that confidently show SEDs consistent with young light-weighted ages and weak line emission, perhaps indicating that they have recently experienced a strong downturn in SFR.
To formally quantify the number of such candidate objects in our JADES sample, we begin by selecting $z\sim6$ galaxies with $>$84\% probability of having \OIIIHb{} and H$\alpha$ EWs of $<$800 \AA{} from the TcSFH posterior outputs.
We then apply the color-color cut F200W$-$F277W $<$ F150W$-$F200W $+$ 0.3 to ensure that the selected galaxies do not show a strong Balmer break.
While we could apply cuts on the inferred Balmer break strength from the TcSFH fits directly, we find that such an approach is not ideal given that the Balmer break strength can increase dramatically on short ($\sim$10 Myr) timescales when allowing for strong drops in SFR.
Finally, because our goal is to identify galaxies that are relatively poorly fit with the CSFH models yet considerably better fit by the TcSFH models, we only include those with a best-fitting $\chi^2_\mathrm{CSFH} > 14$ (we are fitting 14 data photometric data points) and $\chi^2_\mathrm{CSFH} - \chi^2_\mathrm{TcSFH} > 10$.
This rather strict selection results in a total of seven $z\sim6$ galaxies, including the two illustrative examples shown in Fig. \ref{fig:youngWeakLines}.
The majority of these seven galaxies have inferred SFR$_{3\ \mathrm{Myr}}$ / SFR$_{50\ \mathrm{Myr}}$ ratios of $<0.2$ with $>$84\% confidence from the TcSFH posterior outputs.

We emphasize that the seven objects described above are not an exhaustive census of $z\sim6$ galaxies in our sample showing potential evidence of a recent strong downturn in SFR.
They are simply the most confident cases and hence are limited to galaxies with the highest signal-to-noise photometry (S/N(F277W)$\gtrsim$30 before adding the 5\% systematic error). 
The inferred SFR$_{3\ \mathrm{Myr}}$ / SFR$_{50\ \mathrm{Myr}}$ ratio distributions shown in Fig. \ref{fig:metSFRratFescDistns} demonstrate the collective evidence in our sample for a significant population of $z\sim6$ galaxies (particularly at the UV-faint end) with recently-declining SFHs.
This statistical, photometric evidence for strongly declining SFHs among some reionization-era galaxies is further supported by the spectroscopic confirmation of a $z=7.3$ `(mini-)quenched' galaxy from recent \JWST{} observations (\citealt{Looser2023,Dome2023}; for similar examples at $z\sim5$ see also \citealt{Looser2023b,Strait2023}).

\subsubsection{The Different Nature of UV-bright and UV-faint Reionization-era Galaxies}

The TcSFH models potentially provide new insight into the nature of reionization-era galaxies along the UV luminosity function. 
In context of these models, galaxies at the bright end are weighted more towards systems that have very recently experienced an increase in star formation rate (SFR$_{3\ \mathrm{Myr}}$ / SFR$_{50\ \mathrm{Myr}} > 1$).
Objects with such star formation histories will have a greater fraction of emergent light arising from hot and exceptionally luminous O stars, resulting in a temporary boost to their light-to-mass ratio.
Therefore, UV-luminous reionization-era galaxies may more often consist of relatively low-mass objects that have been up-scattered in luminosity due to their SFH, particularly when considering the exponential decline of the mass function at high masses.
However, given the large systematic uncertainties in stellar mass estimates among $z\sim6-9$ galaxies (see \S\ref{sec:properties}), drawing such a conclusion directly from the data is quite challenging.
If early galaxies do indeed have very bursty SFHs, this would help explain the apparent abundance of bright $z>10$ galaxies recently identified by \JWST{} as one would then expect a higher fraction of low-mass galaxies to be up-scattered above the detection threshold \citep[e.g.][]{Mason2023,Mirocha2023,Sun2023_burstyFIRE}.

In contrast, galaxies much further down the UV luminosity function would be less weighted towards systems with recent upturns in their SFR in context of the TcSFH models.
Nonetheless, the substantial scatter in the SFR$_{3\ \mathrm{Myr}}$ / SFR$_{50\ \mathrm{Myr}}$ ratio inferred among our very UV-faint $z\sim6$ subset ($\approx$1 dex) does imply considerable diversity in the recent SFHs of this population.
This is directly demonstrated by the presence of galaxies with extremely strong \OIIIHb{} EWs ($\gtrsim$1500 \AA{}) within the very UV-faint bin (e.g. JADES-GS+53.17835$-$27.77879 in Fig. \ref{fig:EELGs} and the $\Muv{} = -16.7$ galaxy at $\zspec{} = 7.28$ described in \citealt{Saxena2023}) indicating a strong rise in their SFHs on $\sim$3 Myr timescales.
It may be the case that galaxies considerably fainter than the exponential cut-off of the UV luminosity (and mass) function are more equally likely to be a relatively low-mass object experiencing a recent strong rise in SFR (resulting in high $L_\mathrm{UV}^{}/\Mstar{}$) as they are to be a relatively high-mass object experiencing a recent strong decline in SFR (yielding relatively low $L_\mathrm{UV}^{}/\Mstar{}$), thereby yielding the broader SFR$_{3\ \mathrm{Myr}}$ / SFR$_{50\ \mathrm{Myr}}$ distribution at fainter \Muv{}.

\subsection{An Alternative Explanation: Efficient LyC Escape} \label{sec:fesc}

Above, we have demonstrated how bursty SFHs (when coupled with lower metallicities at fainter \Muv{}) can explain the $z\sim6$ \OIIIHb{} and H$\alpha$ EW trends with UV luminosity found in \S\ref{sec:OIIIHb_EW_results}--\ref{sec:Halpha_EW_results}.
In this sub-section, we discuss how these EW trends can alternatively be explained by allowing for a large fraction of ionizing photons to escape the reionization-era galaxies in our sample.
Because the [OIII] and Balmer nebular lines are powered by ionizing photons interacting with dense gas in galaxies, removing a substantial fraction of these ionizing photons will result in fewer recombinations and hence weaker lines.
We estimate the Lyman-continuum (LyC; $\lambda_\mathrm{rest} < 912$ \AA{}) photon escape fractions required to explain the photometric data (and resulting \OIIIHb{} and H$\alpha$ EW trends) under the assumption of a constant SFH by using the `picket-fence' model within \textsc{beagle}. 
In these picket fence models \citep[e.g.][]{Conselice2000,Heckman2011,RiveraThorsen2017}, a given fraction (hereafter \fesc{}) of LyC photons are assumed to escape the HII regions without ionizing the gas while still assuming ionization-bounded conditions. 
Therefore, all nebular line and continuum emission (independent of wavelength) are reduced by a factor of 1-\fesc{} relative to the base CSFH \textsc{beagle} models (see Fig. \ref{fig:EWratioParameterDependence}c).

In the picket-fence \textsc{beagle} SED fits, we adopt a log-uniform prior on \fesc{} between 0.001 and 0.8 with all other priors equivalent to the nominal CSFH fits.
We then infer log-normal distributions for both metallicity and \fesc{} using Eq. \ref{eq:bayes} in the three different UV luminosity bins for our $z\sim6$ sample.
While density-bounded photoionization models may be more appropriate at the highest \fesc{} values ($\gtrsim$0.5), our goal here is to simply gain an initial sense of the escape fractions required to explain the photometric data (and the implied EW distributions) under the assumption of constant star formation histories.
Adopting density-bounded models \citep{Plat2019} would generally result in a slower decline of the [OIII] EWs with increasing \fesc{} relative to the Balmer line EWs given that the high-ionization zones are the last to be significantly impacted by ionizing photon escape in this context.
Nonetheless, the difference in EWs between the ionization-bounded and density-bounded \textsc{beagle} models is $\lesssim$10\% at $\fesc{} < 0.8$.

The picket-fence CSFH \textsc{beagle} fits also yield an anti-correlation between UV luminosity and metallicity at $z\sim6$ (see Fig. \ref{fig:metSFRratFescDistns}d) as physically expected, with median values of $0.35^{+0.09}_{-0.09}$ $Z_\odot$, $0.38^{+0.08}_{-0.10}$ $Z_\odot$, and $0.05^{+0.03}_{-0.02}$ $Z_\odot$ among the bright ($\langle \Muv{} \rangle = -20.0$), faint ($\langle \Muv{} \rangle = -18.7$), and very faint ($\langle \Muv{} \rangle = -17.5$) $z\sim6$ subsets, respectively.
However, the inferred distribution of metallicities in a given UV luminosity bin is considerably different in shape from that inferred from the TcSFH \textsc{beagle} fits adopting \fesc{} = 0 (c.f. panel (a) in Fig. \ref{fig:metSFRratFescDistns}).
With the picket-fence models, both the bright and faint subsets have inferred metallicity distributions that peak at fairly high values ($\gtrsim$0.3 $Z_\odot$) while the metallicity distribution for the very faint bin is exceptionally broad with an inferred standard deviation of $>$0.8 dex resulting in a fairly uniform distribution across the $-2.2 \leq$ log($Z/Z_\odot$) $\leq -0.3$ parameter space allowed in our SED fits. 
It is unclear whether such a distribution in $z\sim6$ galaxy metallicity is physical at such faint UV luminosities and low stellar masses (maximum $\Mstar{} = 10^9$ \Msol{}; see Fig. \ref{fig:MstarSFH}).
Nonetheless, we proceed with investigating what correlation between UV luminosity and \fesc{} is required in context of these models to explain the photometry of the $z\sim6$ sample.

Under the assumption of constant star formation histories, the trends in \OIIIHb{} and H$\alpha$ EW found in \S\ref{sec:OIIIHb_EW_results}--\ref{sec:Halpha_EW_results} imply a strong increase in \fesc{} towards very faint UV luminosities at $z\sim6$ (Fig. \ref{fig:metSFRratFescDistns}d).
With the picket-fence \textsc{beagle} SED models, we infer median \fesc{} values of $0.07^{+0.05}_{-0.03}$, $0.21^{+0.04}_{-0.04}$, and $0.35^{+0.05}_{-0.05}$ in the bright, faint, and very faint subsets, respectively, when adopting log-normal distributions for \fesc{} in each luminosity bin.
Moreover, the fraction of galaxies with substantial ionizing photon leakage (\fesc{} $>$ 20\%) is inferred to be $28^{+8}_{-6}$\%, $52^{+6}_{-6}$\%, $76^{+9}_{-10}$\% in each of the respective \Muv{} bins.
As can be seen in Fig. \ref{fig:metSFRratFescDistns}, the inferred distribution of \fesc{} simultaneously narrows and peaks closer to the allowed upper bound of \fesc{} = 0.8 towards fainter UV luminosity.
The generally bluer UV slopes, smaller rest-UV sizes, and lower inferred dust attenuation among fainter $z\gtrsim6$ galaxies \citep[e.g.][]{Bouwens2014_beta,Shibuya2015,Bhatawdekar2021} do imply that these galaxies would typically be more efficient leakers of LyC photons based on local constraints \citep[e.g.][]{Chisholm2022,Flury2022_LyCdiagnostics}.
However, it is unclear whether a dramatic (factor of $\sim$5) change in median \fesc{} between $\Muv{} \approx -20$ and $\Muv{} \approx -17.5$ at $z\sim6$ would be expected.
No evidence yet exists among local samples that the LyC escape fraction strongly correlates with far-UV luminosity nor stellar mass \citep{Izotov2021,Flury2022_LyCdiagnostics}.

We consider whether the picket-fence CSFH models can explain the photometry of the seven $z\sim6$ galaxies confidently showing SEDs consistent with young light-weighted ages as well as relatively weak nebular lines discussed in \S\ref{sec:bursty}.
Five of these seven galaxies are considerably better fit with the picket-fence models relative to standard CSFH models ($\Delta \chi^2 > 10$) and, in each of these five cases, the inferred LyC escape fraction is \fesc{}$>$50\% with $>$84\% probability from the output posterior.
We show the picket-fence CSFH model SEDs for the two illustrative example galaxies in the rightmost panels of Fig. \ref{fig:youngWeakLines}.
For one of these objects, the best-fitting picket-fence CSFH model has $\chi^2 = 19.5$ across the 14 fitted data points, implying a somewhat poorer fit relative to the best-fitting TcSFH model ($\chi^2 = 9.8$).
However, in four of the seven galaxies under consideration here, there is little-to-no preference in the data for the TcSFH models over the picket-fence CSFH models ($\Delta \chi^2 < 5$).
As discussed in the following sub-section, these two different scenarios (bursty SFH or substantial LyC escape) have very different implications for reionization, and deep spectroscopic observations will ultimately be required to better determine the physical origin of very early galaxies showing young light-weighted ages and relatively weak line emission.

\subsection{Potential Implications for Reionization} \label{sec:reionization}

Above, we have demonstrated that the $z\sim6$ \OIIIHb{} and H$\alpha$ EW trends found in \S\ref{sec:OIIIHb_EW_results}--\ref{sec:Halpha_EW_results} can be explained by a combination of decreasing metallicity with UV luminosity, in addition to either 1) systematically more very-recently rising SFHs among brighter systems with small LyC leakage at all luminosities or 2) systematically much greater LyC leakage among fainter systems with constant SFHs at all luminosities (Fig. \ref{fig:metSFRratFescDistns}). 
To better understand the potential implications for the ionizing efficiency of bright vs. very faint galaxies in these two limiting cases of model assumptions (i.e. bursty SFH with $\fesc{} \approx 0$ or CSFH with $0<\fesc{}<0.8$), we infer the distribution of ionizing photon production efficiencies (\xiionObs{}) as a function of UV magnitude in each case.
Here, we are considering the \xiionObs{} defined as the production rate of LyC photons divided by the observed luminosity at rest-frame 1500 \AA{} (i.e. no correction is made to $L_{1500}$ for dust attenuation or nebular continuum emission).

With our TcSFH \textsc{beagle} models that adopt \fesc{} = 0, we infer median ionizing photon production efficiencies of \logxiionObs{} = $25.65^{+0.03}_{-0.03}$, $25.48^{+0.03}_{-0.02}$, and $25.35^{+0.05}_{-0.05}$ in the bright ($\langle \Muv{} \rangle = -20.0$), faint ($\langle \Muv{} \rangle = -18.7$), and very faint ($\langle \Muv{} \rangle = -17.5$) $z\sim6$ subsets, respectively (Fig. \ref{fig:metSFRratFescDistns}c).
This is consistent with expectations given that the very faint population is inferred to generally have more very-recently declining SFHs with these models (Fig. \ref{fig:metSFRratFescDistns}b), resulting in a lower fraction of O stars producing the bulk of ionizing photons.
We note that the typical \xiionObs{} we infer among our faintest subset is in excellent agreement with the value reported in \citet{Atek2024} for a sample of four $\zspec{}=6.2-6.8$ galaxies with $-17.2 \leq \Muv{} \leq -16.5$ (\logxiionObs{} = 25.4$\pm$0.2).
The standard deviation of the assumed log-normal distribution in \xiionObs{} is inferred to moderately increase from $\approx$0.20 dex in the bright and faint bins to $\approx$0.3 dex in the very faint bin, also consistent with the inferred larger variation in recent SFHs among the very faint population.
This broadening reflects that, while the median \xiionObs{} decreases with UV luminosity, there remains a significant subset of a very UV-faint $z\sim6$ galaxies with very efficient production of hydrogen ionizing photons ($\xiionObs{} \gtrsim 10^{25.7}$ erg Hz$^{-1}$; see Fig. \ref{fig:EELGs}). 
Similar sources at extremely low UV luminosities ($\Muv{} \sim -16$) have now been spectroscopically confirmed \citep{Atek2024}.

The UV luminosity dependence on \xiionObs{} is inferred to be much weaker with the CSFH models that allow for significant LyC leakage (Fig. \ref{fig:metSFRratFescDistns}f).
From these fits, the median inferred \logxiionObs{} is $25.75^{+0.03}_{-0.02}$, $25.64^{+0.02}_{-0.01}$, and $25.60^{+0.03}_{-0.02}$ in the respective UV luminosity bins, with the standard deviation in the range $\approx$0.15--0.18 dex in all bins.
Such a small UV luminosity dependence on \xiionObs{} yet large dependence on \OIIIHb{} is accommodated in the models by pushing \fesc{} to the very large values ($\gtrsim$0.5) in the very faint population. 
We also note that the CSFH models allow for less dynamic range in the inferred \xiionObs{} value of an individual galaxy given that a significant population of O stars is always assumed to be present (given the allowed age of a galaxy at $z\gtrsim6$), in contrast to scenarios with the TcSFH models where the SFR can sharply decline in the most recent $\sim$3 Myr (see Fig. \ref{fig:exampleSFHs}).
These differences in model assumptions contribute to the systematically higher and narrower inferred \xiionObs{} distributions with the CSFH models.

The outputs of the TcSFH and picket-fence CSFH models therefore have very different implications for the contribution of galaxies to reionization along the UV luminosity function.
The picket-fence CSFH models yield a substantial, systematic increase in \fesc{} towards very faint UV luminosities yet a weak UV-luminosity dependence on \xiionObs{}, implying that the faintest galaxies vastly dominate the ionizing photon budget for reionization.
However, the TcSFH models assume that all galaxies regardless of \Muv{} have small \fesc{}, but that the most UV-luminous galaxies have systematically larger \xiionObs{}.
While the assumption that all reionization-era galaxies have small \fesc{} may very well be invalid, so too may be the assumption that all such galaxies have a CSFH given the abundance of extreme line emitters.
Our primary intent here is to demonstrate that existing uncertainties on what is physically driving the EW trends with \Muv{} directly result in large uncertainties on the relative contribution of galaxies along the UV luminosity function to reionization.

In summary, the UV-luminosity dependence on \OIIIHb{} and H$\alpha$ EWs implied by the NIRCam data provide empirical evidence on a statistical basis for distinct differences in the nature of UV-bright and UV-faint $z\sim6-9$ galaxies.
We have demonstrated that these trends can be explained by either lower metallicity and more recently declining star formation histories in the very UV-faint population, or lower metallicities and higher LyC escape fractions in the very UV-faint population.
The degeneracy of bursty SFHs and LyC leakage on the nebular lines cannot be broken with photometry alone and it may very well be the case that both factors play a significant, perhaps connected, role (e.g. \citealt{Fletcher2019,Tang2019,Barrow2020,Ma2020b}).
As noted previously in this sub-section, additional mechanisms such as different IMFs, low-luminosity AGN, or $\alpha$-enhanced metallicities could also play a significant role in driving the EW distribution trends found in \S\ref{sec:OIIIHb_EW_results}--\ref{sec:Halpha_EW_results}.
These various scenarios will have different implications for the contributions of galaxies along the UV luminosity function to cosmic reionization, and thus deep spectroscopic efforts are critical to better determine the physical mechanisms responsible for the \OIIIHb{} and H$\alpha$ EW trends.

\section{Overdensities Around Ly\texorpdfstring{$\mathbf{\alpha}$}{alpha} Emitters at \texorpdfstring{$\mathbf{z>7}$}{z > 7}} \label{sec:overdensity}

It has long been proposed that galaxies exhibiting strong Ly$\alpha$ emission (EW$>$25 \AA{}) at $z>7$ often trace strong galaxy overdensities that are capable of generating large ionized bubbles prior to the completion of reionization \citep[e.g.][]{Wyithe2005,Castellano2016,Hutter2017,Stark2017,Weinberger2018,Jung2020,Tilvi2020,Endsley2022_bubble,Larson2022,Leonova2022,Lu2023}.
In this scenario, strong $z>7$ Ly$\alpha$ emitters (LAEs) would commonly act as signposts of the earliest sites of structure formation.
Here, we use the improved photometric redshifts enabled by the JADES/NIRCam GOODS-N and GOODS-S imaging to better test the connection between overdensities and strong $z>7$ Ly$\alpha$ emission.

We search the literature for $z>7$ galaxies with known high-confidence Ly$\alpha$ detections in the GOODS fields (i.e. S/N$>$10 in Ly$\alpha$ alone or S/N$>$10 of a systemic line plus a $>$7$\sigma$ Ly$\alpha$ detection), resulting in three objects: JADES-GS-z7-LA at $\zLya{} = 7.281$ \citep{Saxena2023}, z8\_GND\_5296 at $\zLya{} = 7.508$ \citep{Finkelstein2013,Tilvi2016,Jung2019,Jung2020}, and z7\_GND\_16863 at $\zLya{} = 7.599$ \citep{Jung2019,Jung2020}.
Two of these systems fall within the JADES/NIRCam footprint (z8\_GND\_5296 and JADES-GS-z7-LA) and we investigate whether either show a significant photometric overdensity.
The galaxy z8\_GND\_5296 is a UV-bright (\Muv{} = $-21.5$) system in the GOODS-N field with a most recent \Lya{} EW measurement of 33$\pm$4 \AA{} from \citet{Jung2020}.
JADES-GS-z7-LA is, on the other hand, a very UV-faint (\Muv{} = $-16.7$) galaxy in the GOODS-S field with a near-unity \Lya{} escape fraction yielding an extremely large EW of 400$\pm$90 \AA{} implying that it lies within a very large ionized bubble \citep{Saxena2023}.
Both of these strong LAEs exhibit very high \OIIIHb{} EWs of $\approx$1500 \AA{} placing them in the top $\approx$17\% and $\approx$7\% of the \OIIIHb{} EW distribution, respectively, given their UV luminosities (see \S\ref{sec:OIIIHb_EW_results}).

Here we focus on quantifying the photometric overdensities (following e.g. \citealt{Castellano2016,Endsley2021_LyA,Leonova2022,Whitler2023_overdensity}) of the two $z>7$ LAEs in JADES noted above given that deep spectroscopic follow-up of $z\sim7-8$ candidates surrounding these rare sources remains incomplete.
We thus aim to supplement our base $z\sim7-9$ selection criteria with additional cuts to better identify neighboring galaxies lying relatively close in redshift space to the two strong $z>7$ LAEs in JADES.
Fortunately, both of these LAEs lie at $z=7.0-7.6$ where relatively high-precision photometric redshifts can be obtained by exploiting the medium-band photometry, thereby improving our photometric overdensity estimates.
At $z\approx7.0-7.6$, the [OIII]$\lambda\lambda$4959,5007 doublet falls in the F410M filter yet outside of the F356W bandpass and thus galaxies in this redshift range with large \OIIIHb{} EWs will show remarkably red F356W$-$F410M colors.
From the inferred \OIIIHb{} EW distributions in the JADES dataset (see Table \ref{tab:inferredEWparameters} and Fig. \ref{fig:EWdistns}), we expect the majority of $\Muv{} < -18$ galaxies at $z\approx7.3$ will exhibit \OIIIHb{} EWs of $>$400 \AA{} which, assuming a flat rest-optical continuum (in $F_\nu$), will result in F356W$-$F410M colors of $\gtrsim$0.6 mag at $z\approx7.0-7.6$.
Such prominent colors should be relatively easy to identify for all but the faintest objects in our deep JADES imaging.
Moreover, at $z\approx7.0-7.6$ the F335M band is entirely redward of the Balmer break at $3648$ \AA{} and thus we would generally expect flat F335M$-$F356W colors given the typically little dust attenuation inferred for the sample ($A_V \lesssim 0.01$ mag).

\begin{figure}
\includegraphics{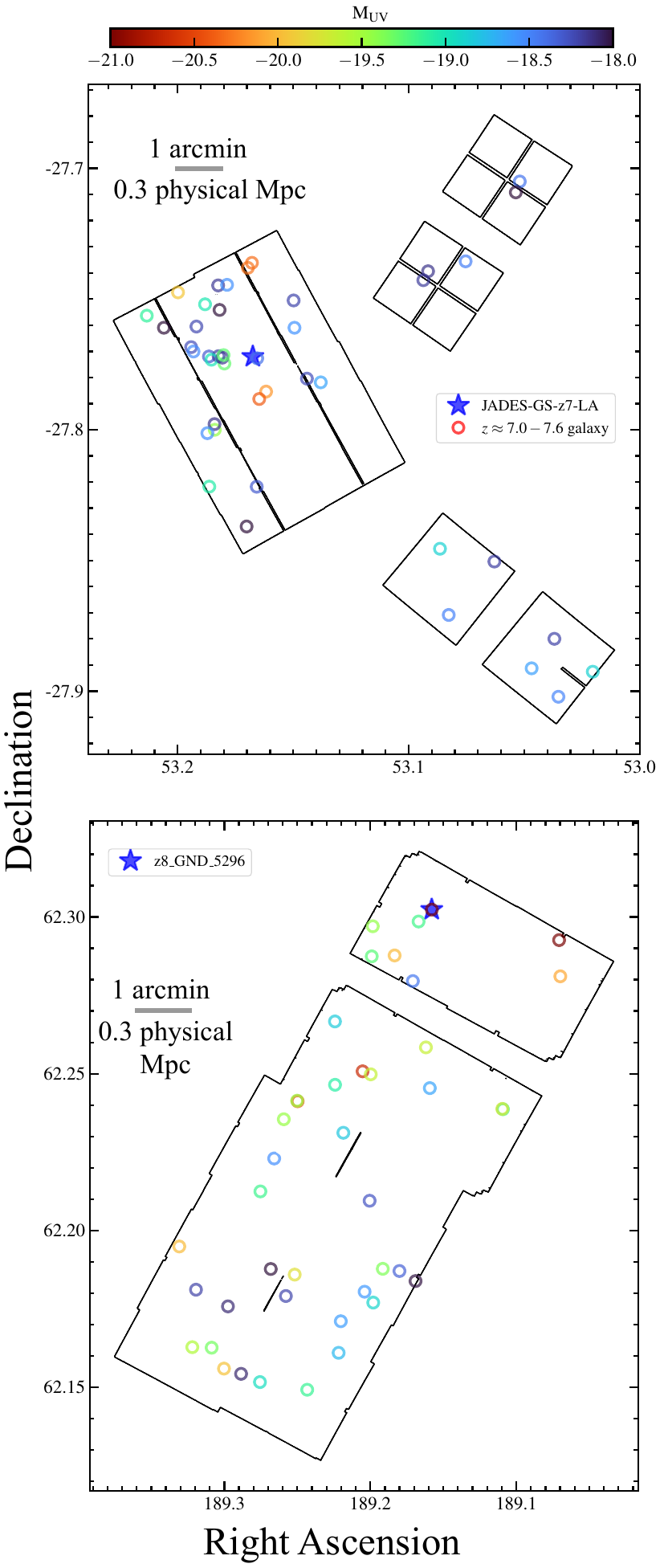}
\caption{On-sky distributions of candidate $z\approx7.0-7.6$ galaxies in the two GOODS fields (colored circles) with the known high-confidence $z>7$ Ly$\alpha$ emitter in each field shown as a blue star. JADES-GS-z7-LA is a very UV-faint ($\Muv{} = -16.7$) galaxy at $z=7.28$ with near-unity Ly$\alpha$ escape (EW=400$\pm$90 \AA{}; \citealt{Saxena2023}) while z8\_GND\_5296 is a UV-bright ($\Muv{} = -21.5$) galaxy at $z=7.51$ with a less extreme Ly$\alpha$ EW of 33$\pm$4 \AA{} \citep{Finkelstein2013,Jung2020}. We identify a dense group of six faint ($-19.3 \leq \Muv{} \leq -18.2$) neighboring $z\approx7.0-7.6$ galaxies approximately 1 arcmin to the East of JADES-GS-z7-LA, as well as two bright ($\Muv{} \approx -20.3$) such galaxies $\approx$1 arcmin to the South.}
\label{fig:LAE_nearbyMaps}
\end{figure}

Guided by the above expectations for the colors of $z\approx7.0-7.6$ galaxies, we identify potential neighbors of the very UV-faint (\Muv{} = $-16.7$) $z=7.28$ Ly$\alpha$ emitter JADES-GS-z7-LA and the UV-bright ($\Muv{} = -21.5$) $z=7.51$ Ly$\alpha$ emitter z8\_GND\_5296 by supplementing our F090W dropout selection criteria (\S\ref{sec:selection}) with the below additional cuts:
\begin{enumerate}
    \item F356W $-$ F410M $>$ 0.6
    \item F335M $-$ F356W $<$ 0.3
    \item F150W $<$ 29.0
\end{enumerate}
Here, we have added the condition of F150W $<$ 29 (corresponding to $\Muv{} \lesssim -18.0$) to ensure that our selection of candidate neighbors is fairly complete at $z\approx7.0-7.6$ given the different depth tiers of the JADES imaging \citep{Eisenstein2023}.
The on-sky positions of the candidate $z\approx7.0-7.6$ galaxies in the JADES GOODS-S and GOODS-N footprints are shown in Fig. \ref{fig:LAE_nearbyMaps}, along with the positions of the two strong $z>7$ LAEs covered by JADES.
In the GOODS-S field containing the very UV-faint LAE, we identify 43 potential $z\approx7.0-7.6$ galaxies with F150W $<$ 29 across the 42.2 arcmin$^2$ JADES area with F335M coverage.
Across the 47.2 arcmin$^2$ JADES/GOODS-N area with F335M coverage, we find 42 $z\approx7.0-7.6$ galaxies which includes the UV-bright ($\Muv{} = -21.5$) LAE that is the source of interest in this field (z8\_GND\_5296).
The very similar numbers of $z\approx7.0-7.6$ candidates in each field despite significantly shallower NIRCam exposure times over much of GOODS-N \citep{Eisenstein2023} is consistent with past findings that the (completeness-corrected) number density of $z\sim7$ Lyman-break galaxies is considerably higher over GOODS-N than GOODS-S \citep[e.g.][]{Bouwens2015_LF,Finkelstein2015_LF}.
Below, we first discuss the surface density of potential $z\approx7.0-7.6$ neighbors surrounding the LAEs selected from JADES, then proceed to calculate the surrounding photometric overdensities folding in the expected completeness of our selection.

It is immediately clear from Fig. \ref{fig:LAE_nearbyMaps} that the surface density of $z\approx7.0-7.6$ galaxies is unusually high near the GOODS-S LAE JADES-GS-z7-LA.
In particular, about 1 arcmin East of this very UV-faint LAE lies a complex of six distinct $z\approx7.0-7.6$ galaxies with a maximum separation of 23 arcsec from one another.
The surface density of $z\approx7.0-7.6$ galaxies in this very small (0.073 arcmin$^2$) region of GOODS-S is 87.3 galaxies/arcmin$^{-2}$, a factor of $\approx$70$\times$ higher than that on average over the deep imaging region covering JADES-GS-z7-LA.
The six objects comprising this dense concentration of $z\approx7.0-7.6$ galaxies have absolute UV magnitudes in the range $-19.3 \leq \Muv{} \leq -18.2$ and \OIIIHb{} EWs ranging between $\approx$400--2000 \AA{}.
Moreover, there are two UV-bright ($-20.4 \leq \Muv{} \leq -20.2$) galaxies located approximately 1 arcmin to the South of the UV-faint LAE JADES-GS-z7-LA.

We also identify five candidate $z\approx7.0-7.6$ galaxies lying near ($<$ 1.5 arcmin separation) the UV-bright LAE in GOODS-N (z8\_GND\_5296), all located roughly to the southeast (see Fig. \ref{fig:LAE_nearbyMaps}).
Considering the smallest rectangular area that encloses all six of these objects, we obtain a surface density of 3.6 galaxies/arcmin$^{-2}$ which is approximately 4$\times$ higher than the average over the full JADES/GOODS-N footprint with F335M.
These five relatively nearby objects have inferred \OIIIHb{} EWs of $\approx$500--1500 \AA{} and UV magnitudes of $-20.0 \leq \Muv{} \leq -18.4$, indicating that they are all substantially ($\geq4\times$) fainter than z8\_GND\_5296.

We now quantify the photometric overdensities around the two strong $z=7.3-7.5$ LAEs in JADES by comparing the nearby $z\approx7.0-7.6$ galaxy counts to that expected from literature luminosity functions at this epoch.
Our procedure largely follows that of \citet{Whitler2023_overdensity} which we outline below.
As a first step towards comparing with the expected cosmic mean density, we must first correct our measured $z\approx7.0-7.6$ galaxy counts for selection completeness.
To this end, we analytically estimate our selection completeness as a function of true redshift and absolute UV magnitude (as in \citealt{Endsley2021_OIII,Whitler2023_overdensity}) by generating mock SEDs each with a flat continuum (in $F_\nu$) and IGM attenuation incorporated using the analytic model from \citet{Inoue2014}.
Given that our selection for this overdensity analysis depends on \OIIIHb{} emission via the F356W$-$F410M and F335M$-$F356W color cuts above, we also add \OIIIHb{} emission into the mock SEDs following our results on the $z\sim7-9$ EW distribution in \S\ref{sec:OIIIHb_EW_results}, considering additional constraints on the $z\sim7$ EW distribution among the very bright ($-22.5 \lesssim \Muv{} \lesssim -21$) population from \citet{Endsley2023_CEERS}.
Specifically, for a given \Muv{}, we assume a log-normal \OIIIHb{} EW distribution defined by median EW log(\muEW{}/\AA{}) = max([2.87 , $-0.16\, (\Muv{}+20) + 2.87$]) and standard deviation \sigmaEW{} = min([0.25 , $0.08\,(\Muv{}+20) + 0.3$]) dex and adopt a fixed [OIII]$\lambda$5007/H$\beta$ ratio of 6 \citep[e.g.][]{Tang2019}.
In doing so, we account for our strong incompleteness to objects with relatively weak \OIIIHb{} emission (EW$\lesssim$400 \AA{}) for this overdensity analysis, though we note that such objects are expected to be in the minority at $F150W < 29$ ($\Muv{} \lesssim -18$) given the derived EW distributions. 

\begin{figure}
\includegraphics{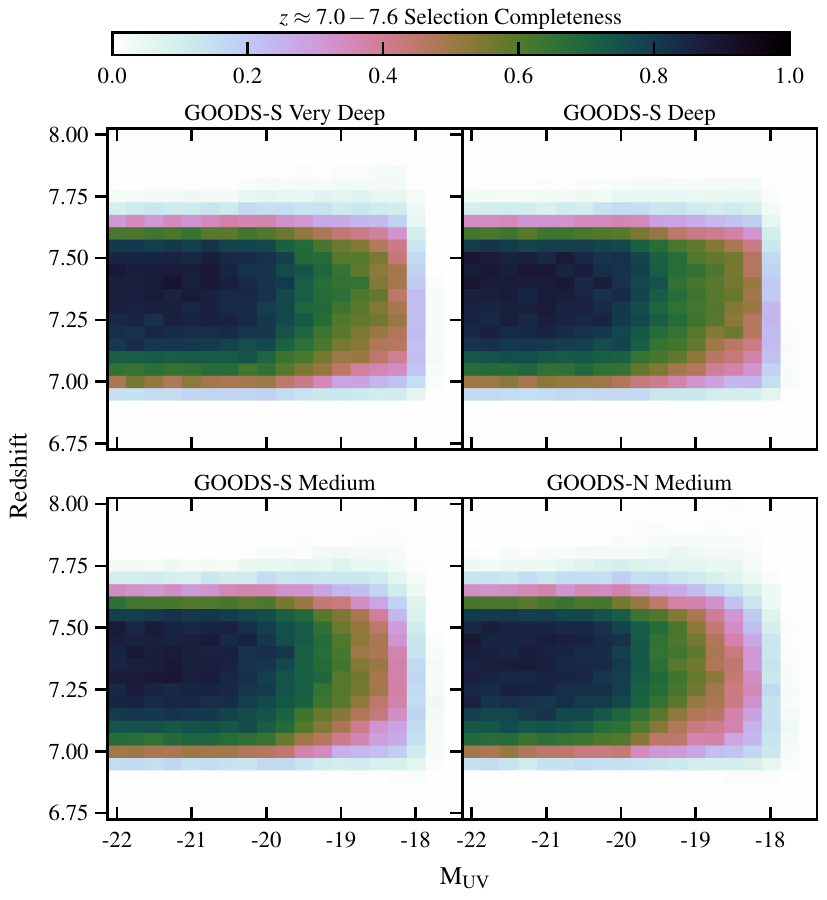}
\caption{Simulated completeness of the $z\approx7.0-7.6$ selection criteria where we supplement the base $z\sim7-9$ selection criteria (\S\ref{sec:selection}) with additional color cuts that utilize the long-wavelength medium band filters to identify \OIIIHb{} emitters within this redshift range (see \S\ref{sec:overdensity}). We show the simulated completeness as a function of \Muv{} and redshift in the four different JADES/NIRCam imaging regions where the cutoffs at $\Muv{} \approx -18$ are due to the requirement of F150W $<$ 29 for this selection.}
\label{fig:overdensity_selectionCompleteness}
\end{figure}

In the selection completeness simulations, for each grid point in redshift ($z=6.5-8.5$, 0.05 unit spacing) and absolute UV magnitude ($-22 \leq \Muv{} \leq -17$, 0.25 mag spacing), we generate 1000 mock SEDs each with a different \OIIIHb{} EW value pulled randomly from the parametrized distribution above.
To account for photometric scatter in our selection, we add Gaussian noise to the `true' photometry derived from each mock SED.
The photometric noise is computed as a function of \Muv{} separately for each of the four different imaging regions (medium, deep, and very deep in GOODS-S as well as medium in GOODS-N) by fitting a linear relationship between \Muv{} and the logarithm of the photometric error in each band using data from all $z\sim7-9$ galaxies selected in each imaging region.
We find that such a linear relationship adequately captures the general trend between photometric uncertainty and \Muv{} imposed by typically larger galaxy sizes (and hence Kron apertures) at brighter luminosities.

The resulting simulated completeness of our $z\approx7.0-7.6$ selection is shown in Fig. \ref{fig:overdensity_selectionCompleteness} for each of the four different JADES/NIRCam imaging regions.
Consistent with our targeted redshift interval, the estimated completeness is $\gtrsim$50\% at $z\approx7.05-7.60$ at the brightest luminosities and very low at redshifts $z<6.9$ and $z>7.7$ for all four regions.
We also find that the completeness is $\gtrsim$50\% at absolute UV magnitudes of $\Muv{} \lesssim -18.5$ in the very deep and deep imaging regions in GOODS-S, while such relatively high completeness is achieved at a slightly brighter magnitude threshold of $\Muv{} \lesssim -19$ in the medium-depth regions of GOODS-S and GOODS-N.
These selection completeness grids are convolved with the UV luminosity function of \citet{Bouwens2022_lensedLF} to estimate the expected surface density counts of $z\approx7.0-7.6$ galaxies in each imaging region under the assumption of cosmic mean density.
Following the methods of \citet{Whitler2023_overdensity}, we compute the amplitude of the photometric overdensities around JADES-GS-z7-LA and z8\_GND\_5296 as a function of angular separation by dividing the GOODS-S and GOODS-N footprints into rings of varying concentric circular radii centered on each $z>7$ Ly$\alpha$ emitter.
In each ring, the amplitude of the photometric overdensity at that angular separation is computed as the surface density of objects identified by our $z\approx7.0-7.6$ selection divided by the cosmic mean.

\begin{figure}
\includegraphics{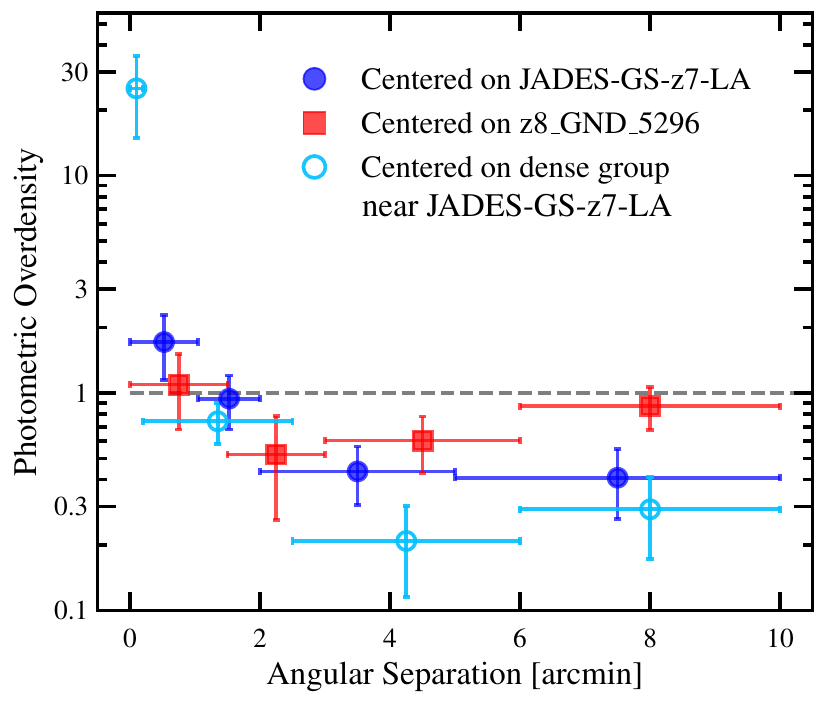}
\caption{Measurements of photometric overdensity as a function of concentric angular radii around the two strong $z>7$ Ly$\alpha$ emitters covered by the JADES/NIRCam imaging. Neither object shows a remarkably strong photometric overdensity across the $z\approx7.0-7.6$ redshift range probed by our photometric selection when centering the concentric rings on the LAEs themselves. However, there is a dense group of $z\approx7.0-7.6$ galaxies near JADES-GS-z7-LA (see Fig. \ref{fig:LAE_nearbyMaps}) where the photometric overdensity is very high ($\approx$25) on small scales ($<$0.2 arcmin). The underdensity in GOODS-S on large scales is consistent with previous findings \citep{Bouwens2015_LF,Finkelstein2015_LF}.}
\label{fig:photometricOverdensities}
\end{figure}

Given its near-unity Ly$\alpha$ escape fraction and relatively small \Lya{} velocity offset ($\approx$120 km/s), JADES-GS-z7-LA is expected to reside in a very large ($R \gtrsim 3$ physical Mpc) ionized bubble \citep{Saxena2023}.
As discussed in \citet{Saxena2023}, this object cannot have created such a large bubble on its own given its very low luminosity ($\Muv{} = -16.7$), implying that the local ionizing photon budget must be dominated by neighboring galaxies.
With the JADES/GOODS-S imaging, we have identified a very dense group of $z\approx7.0-7.6$ galaxies located $\approx$50 arcsec ($\approx$250 kpc in projection) East of JADES-GS-z7-LA (Fig. \ref{fig:LAE_nearbyMaps}) where the surrounding photometric overdensity is very large ($\approx$25) on small scales (see Fig. \ref{fig:photometricOverdensities}).
It is conceivable that these objects all lie at very similar redshifts as JADES-GS-z7-LA and are assisting in powering a large ionized bubble in the vicinity \citep[e.g.][]{Lu2023}. 
The two UV-bright ($\Muv{} \approx -20.3$) systems located at a very similar projected distance to the South of JADES-GS-z7-LA (Fig. \ref{fig:LAE_nearbyMaps}) may be significantly contributing as well.
While the surface density of $z\approx7.0-7.6$ galaxies measured in the $\lesssim$2 arcmin radius ($\approx$600 kpc projected) around JADES-GS-z7-LA itself is consistent with the cosmic mean (see Fig. \ref{fig:photometricOverdensities}), this can easily rise by a factor of $\sim$2--3 if the nearby galaxies are largely located near JADES-GS-z7-LA along the line of sight (i.e. $\Delta z \lesssim 0.1$).

In contrast to the UV-faint Ly$\alpha$ emitter described above, we do not identify any significant photometric overdensity near the UV-bright ($\Muv{} = -21.5$) $z=7.51$ Ly$\alpha$ emitter z8\_GND\_5296 (Fig. \ref{fig:photometricOverdensities}). 
The surface density of $z\approx7.0-7.6$ objects on both small ($<$1.5 arcmin; $<$450 kpc projected) and large (6--10 arcmin; $\sim2-3$ Mpc projected) separations from z8\_GND\_5296 are consistent with the cosmic mean.
This may indicate that the strong Ly$\alpha$ emission from z8\_GND\_5296 is due more to its physical properties.
With an estimated \OIIIHb{} EW of $1540^{+220}_{-160}$ \AA{}, z8\_GND\_5296 lies in the high-end tail of the EW distribution for UV-bright $z\sim7-9$ galaxies (see Fig. \ref{fig:EWdistns}).
Accordingly, this source has a large inferred ionizing photon production efficiency of \logxiionObs{} = 25.73$^{+0.06}_{-0.05}$ (from the TcSFH \textsc{beagle} SED fits) indicating that it has a relatively large intrinsic \Lya{} EW \citep[e.g.][]{Tang2019} thus requiring stronger attenuation (from either the host galaxy or IGM) to make it appear as a weak \Lya{} emitter.

The detection of the systemic [CIII]$\lambda$1908 line from z8\_GND\_5296 may also indicate that its \Lya{} photons are redshifted far into the damping wing (velocity offset $\approx$400 km/s; \citealt{Hutchison2019}) as commonly found among similarly bright $z\sim7$ galaxies (see \citealt{Endsley2022_REBELS} and references therein), resulting in relatively low \Lya{} opacity through the IGM \citep[e.g.][]{Stark2017,Mason2018_IGMneutralFrac}. 
However, this is speculative as it is unclear which component of the [CIII] doublet is currently detected given skyline contamination in the ground-based spectrum \citep{Hutchison2019}.
Detailed spectroscopic measurements (e.g., \citealt{Meyer2024,Witstok2024}) are required to better assess the relative role of internal properties to the strong \Lya{} emission seen from this UV-bright \Lya{} emitter, as well as measure the redshifts of the neighboring $z\approx7.0-7.6$ candidates to better constrain its surrounding overdensity (see also \citealt{Jung2020}).

\section{Summary} \label{sec:summary}

We have used deep nine-band NIRCam imaging taken as part of the JADES program to characterize the star-forming and ionizing properties of a very large (N=756) sample of Lyman-break $z\sim6-9$ galaxies.
This sample includes hundreds of reionization-era galaxies in the very UV-faint regime ($-18 \leq \Muv{} \lesssim -16.5$; see Fig. \ref{fig:MuvRedshiftDistns}) where it was effectively impossible to measure the rest-optical SEDs (at least among a statistical sample) with \Spitzer{}/IRAC.
Our main conclusions are summarized below.

\begin{enumerate}
    
    \item The faintest galaxies in our sample ($\Muv{} \sim -17$) tend to have inferred stellar masses of (1--3)$\times$10$^7$ \Msol{} while those at the brightest end of our sample ($-22 \lesssim \Muv{} \lesssim -21$) have $\sim$100$\times$ higher stellar masses (Fig. \ref{fig:MstarSFH}). Masses inferred from models with priors weighted towards extended SFHs (i.e. `continuity' priors) are systematically $\approx$0.5 dex higher than those assuming CSFH, though can rise to factors of 10--100$\times$ higher in the most extreme cases for galaxies with the youngest light-weighted ages ($\ageCSFH{} \sim 3$ Myr; see Fig. \ref{fig:MstarSFH}g).

    \item There are no galaxies in our sample where the photometric data clearly imply very large stellar masses ($>$3$\times$10$^{10}$ \Msol{}). There are only 13 galaxies in our sample with inferred stellar masses $>$3$\times$10$^{9}$ \Msol{} when adopting the continuity prior that favors extended SFHs (Fig. \ref{fig:mostMassive}). Some of these objects are relatively faint ($\Muv{} \sim -19$) and show strong Balmer breaks consistent with old stellar populations. Others are simply so luminous ($\Muv{} \sim -22$) such that a massive evolved stellar population can be hidden under the light of more recently-formed stars. The most massive candidate in our sample has an inferred mass of 1$\times$10$^{10}$ \Msol{} when including rest-frame near-infrared SED constraints from parallel MIRI imaging (Fig. \ref{fig:GS-28012}).

    \item The typical galaxy in our sample shows a NIRCam SED consistent with a young light-weighted age of $\ageCSFH{} \sim 50$ Myr (Fig. \ref{fig:typicalAge}). However, we confidently identify several individual galaxies showing strong Balmer breaks implying much older ages ($\ageCSFH{} \sim 250-1000$ Myr; Fig. \ref{fig:balmerBreaks}). Some of the objects with confident strong Balmer breaks show signs of recent star formation from significant emission line signatures while others show no indications of emission lines in the photometry. Follow-up spectroscopy is necessary to determine if any galaxies in this latter sub-population may be (temporarily) quenched. 

    \item We find a strong, highly-significant decline in the typical \OIIIHb{} EWs of $z\sim6-9$ galaxies towards lower UV luminosities (median EW$\approx$800 \AA{} at $\Muv{} = -20$ yet $\approx$350 \AA{} at $\Muv{} = -17.5$; see Fig. \ref{fig:EWdistns} and Table \ref{tab:inferredEWparameters}). We verify that this \OIIIHb{} EW trend with \Muv{} is reflected in the typical NIRCam colors of our sample (Fig. \ref{fig:monteCarloColors}). The \OIIIHb{} EW distribution is also found to broaden considerably at lower UV luminosities (standard deviation $\approx$0.3 dex at $\Muv{} = -20$ yet $\approx$0.5 dex at $\Muv{} = -17.5$). We infer a slight ($\approx$0.05--0.1 dex) decline in the typical \OIIIHb{} EW between the $z\sim6$ sample and the $z\sim7-9$ sample at fixed \Muv{}.

    \item In remarkable contrast to the strong decline in \OIIIHb{} EW with \Muv{}, we find that the H$\alpha$ EW distribution changes very weakly (if at all) across a factor of $\approx$10 in UV luminosity among our $z\sim6$ sample (where H$\alpha$ falls in F444W; Fig. \ref{fig:EWdistns}). If we assume that the $z\sim6$ H$\alpha$ EW distribution does not change with \Muv{} over the range probed by our $z\sim6$ sample, we infer a log-normal distribution with median EW=630$^{+10}_{-40}$ \AA{} and a standard deviation of 0.26$\pm$0.01 dex. 

    \item We demonstrate that the \OIIIHb{} and H$\alpha$ EW trends with \Muv{} can be explained by a combination of lower metallicity and systematically more recently-declining SFHs at lower UV luminosities (top panels of Fig. \ref{fig:metSFRratFescDistns}). In this interpretation, the brightest $z\sim6$ galaxies in our sample ($\langle \Muv{} \rangle = -20.0$) have a median inferred metallicity $\approx$3$\times$ higher than that of the faintest $z\sim6$ galaxies ($\langle \Muv{} \rangle = -20.0$; $Z \approx 0.06 Z_\odot$). Moreover, the median inferred ratio of SFR averaged over the past 3 Myr to that averaged over the past 50 Myr (SFR$_{3\ \mathrm{Myr}}$ / SFR$_{50\ \mathrm{Myr}}$) is found to be $\approx$3.5$\times$ higher among the brightest $z\sim6$ objects relative to the faintest galaxies. The brightest galaxies are frequently inferred to be experiencing a recent strong upturn in SFR (i.e. (SFR$_{3\ \mathrm{Myr}}$ / SFR$_{50\ \mathrm{Myr}} > 1$) while the faintest galaxies are inferred to be a more even mixture of objects experiencing a recent strong rise in SFR as a recent strong downturn in SFR. 

    \item There are two particular subsets of our sample which provide evidence in favor of bursty SFHs. A substantial fraction of galaxies in our sample ($\approx$10--20\% depending on \Muv{}) are inferred to exhibit extremely high \OIIIHb{} EWs ($>$1500 \AA{}; Fig. \ref{fig:EELGs}) implying that they have recently experienced a dramatic (factor $\gtrsim5$) increase in SFR over the past 3 Myr (upper panels of Fig. \ref{fig:exampleSFHs}). Another subset of galaxies in our JADES sample show relatively weak Balmer breaks yet also weak nebular line signatures (Fig. \ref{fig:youngWeakLines}) implying that they may be experiencing a lull in star formation activity that followed a burst of star formation that occurred $\sim$5--50 Myr ago (lower panels of Fig. \ref{fig:exampleSFHs}). These two sub-populations may therefore be caught during the peaks and trough of bursty star formation histories among reionization-era galaxies. We note that other recent studies are building empirical evidence for bursty SFHs among early galaxies \citep[e.g.][]{Dome2023,Dressler2023_jades,Looser2023,Looser2023b,Strait2023}.

    \item We also discuss how the \OIIIHb{} and H$\alpha$ EW trends with \Muv{} may be influenced by a strong correlation between UV luminosity and LyC escape fraction (bottom panels of Fig. \ref{fig:metSFRratFescDistns}). When allowing for substantial LyC escape yet enforce constant SFHs, we infer that the faintest $z\sim6$ galaxies in our sample ($\langle \Muv{} \rangle = -17.5$) are typically very efficient at leaking LyC photons into the IGM (median \fesc{}$\approx$0.5) while the brightest ($\langle \Muv{} \rangle = -20.0$) objects leak only a moderate fraction of their LyC photons (median \fesc{}$\approx$0.08). We discuss how this scenario has very different implications for the contribution of galaxies along the luminosity function to cosmic reionization compared to the interpretation of bursty SFHs. This highlights the need for deep spectroscopic follow-up to better determine the physical origin of the \OIIIHb{} and H$\alpha$ EW trends with UV luminosity found in this analysis.

    \item Finally, we quantify the photometric overdensities around two strong Ly$\alpha$ emitters at $z>7$ in the JADES footprint. The very UV-faint ($M_\mathrm{UV}=-16.7$) $z=7.28$ Ly$\alpha$ emitter lies close to a very dense concentration of $z\approx7.0-7.6$ galaxies (Fig. \ref{fig:LAE_nearbyMaps}) that may have helped generate a large ionized bubble leading to efficient Ly$\alpha$ transmission through the largely neutral IGM. However, the UV-bright ($M_\mathrm{UV}=-21.5$) Ly$\alpha$ emitter shows no significant nearby overdensity (Fig. \ref{fig:photometricOverdensities}) perhaps suggesting that efficient ionizing photon production and a large Ly$\alpha$ velocity offset strongly contributed to its Ly$\alpha$ detection, and that not all strong Ly$\alpha$ emitters in the reionization era necessarily occupy large ionized bubbles.
   
\end{enumerate}

The JADES results presented here provide a valuable observational baseline against which to compare predictions of reionization-era galaxy properties from models \citep[e.g.][]{Mutch2016,   Rosdahl2018,Wilkins2020,Wilkins2023,Wu2020_simba,Ceverino2021,Hutter2021,Kannan2021,Hirschmann2022,Lewis2022,Seeyave2023}. 
Such future work will help build insight into the nature and assembly of the faintest (and brightest) $z\sim6-9$ galaxies, and their role in cosmic reionization.

\section*{Acknowledgements}

The authors thank the anonymous referees for constructive comments which improved the quality of this paper.
RE thanks John Chisholm, Charlotte Mason, and Adele Plat for enlightening discussions that improved this work.
EE, DJE, BDJ, BR, GR, MR, FS, DPS, LW \& CNAW acknowledge support from JWST/NIRCam contract to the University of Arizona NAS5-02015.
DPS acknowledges additional support from the National Science Foundation through the grant AST-2109066.
LW acknowledges additional support from the National Science Foundation Graduate Research Fellowship under Grant No. DGE-2137419.
WB acknowledges support by the Science and Technology Facilities Council (STFC), ERC Advanced Grant 695671 "QUENCH".
KB acknowledges support by the Australian Research Council Centre of Excellence for All Sky Astrophysics in 3 Dimensions (ASTRO 3D), through project number CE170100013.
AJB, AJC, JC, IEBW, AS \& GCJ acknowledge funding from the "FirstGalaxies" Advanced Grant from the European Research Council (ERC) under the European Union’s Horizon 2020 research and innovation programme (Grant agreement No. 789056)
S.C acknowledges support by European Union’s HE ERC Starting Grant No. 101040227 - WINGS.
ECL acknowledges support of an STFC Webb Fellowship (ST/W001438/1)
A.L.D. thanks the University of Cambridge Harding Distinguished Postgraduate Scholars Programme and Technology Facilities Council (STFC) Center for Doctoral Training (CDT) in Data intensive science at the University of Cambridge (STFC grant number 2742605) for a PhD studentship.
DJE is also supported as a Simons Investigator.
TJL and RM acknowledge support by the Science and Technology Facilities Council (STFC) and by the ERC through Advanced Grant 695671 "QUENCH". 
RM also acknowledges funding from a research professorship from the Royal Society.
DP acknowledges support by the Huo Family Foundation through a P.C. Ho PhD Studentship.
RS acknowledges support from a STFC Ernest Rutherford Fellowship (ST/S004831/1).
The research of CCW is supported by NOIRLab, which is managed by the Association of Universities for Research in Astronomy (AURA) under a cooperative agreement with the National Science Foundation.
JW acknowledges support from the ERC Advanced Grant 695671, ``QUENCH'', and the Fondation MERAC.

This material is based in part upon High Performance Computing (HPC) resources supported by the University of Arizona TRIF, UITS, and Research, Innovation, and Impact (RII) and maintained by the UArizona Research Technologies department.
The authors acknowledge use of the lux supercomputer at UC Santa Cruz, funded by NSF MRI grant AST-1828315.

This work made use of the following software: \textsc{numpy} \citep{harris2020_numpy}; \textsc{matplotlib} \citep{Hunter2007_matplotlib}; \textsc{scipy} \citep{Virtanen2020_SciPy}; \textsc{astropy}\footnote{\url{https://www.astropy.org/}}, a community-developed core Python package for Astronomy \citep{astropy:2013, astropy:2018}; \textsc{Source Extractor} \citep{Bertin1996} via \textsc{sep} \citep{Barbary2016_sep}; \textsc{photutils} \citep{Bradley2022_photutils}; \textsc{beagle} \citep{Chevallard2016}; \textsc{multinest} \citep{Feroz2008,Feroz2009};  \textsc{prospector} \cite{Johnson2021}; \textsc{dynesty} \citep{Speagle2020}; \textsc{sedpy} \citep{johnson_sedpy}; and \textsc{fsps} \citep{Conroy2009_FSPS,Conroy2010_FSPS} via \textsc{python-fsps} \citep{johnson_python_fsps}.

\section*{Data Availability}

The \HST{} ACS data used in this work are available from the \textit{Hubble} Legacy Field archive (\url{https://archive.stsci.edu/prepds/hlf/}). 
All raw \JWST{} JADES data utilized in this work is now publicly available via the Mikulski Archive for Space Telescopes (\url{https://mast.stsci.edu/portal/Mashup/Clients/Mast/Portal.html}), with a subset released as part of high-level science products (\url{https://archive.stsci.edu/hlsp/jades}). 




\bibliographystyle{mnras}
\bibliography{main} 



\appendix

\section{Considering a Potential Population of Low-Redshift Contaminants} \label{app:photoz}

\begin{figure}
\includegraphics{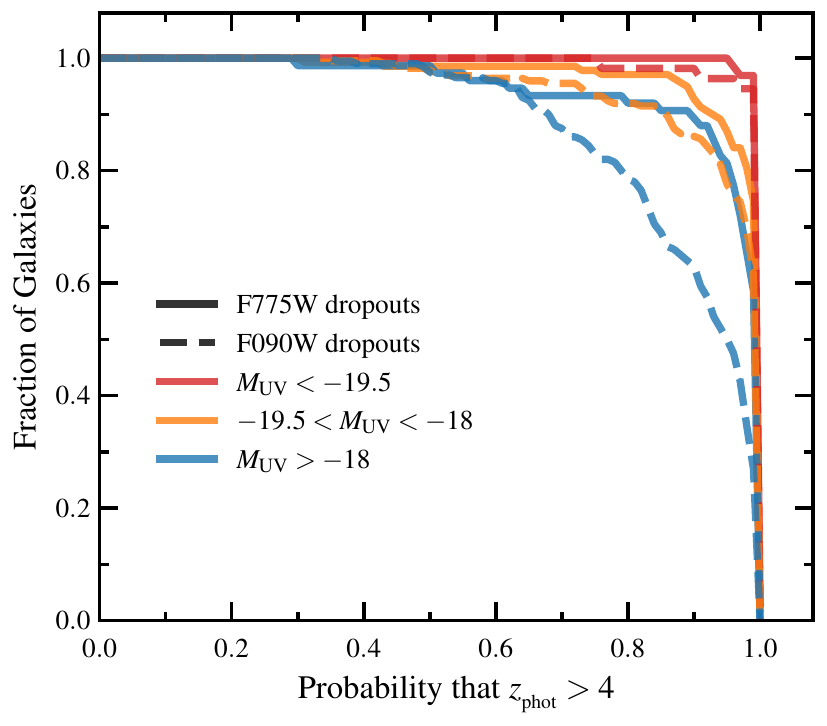}
\caption{Cumulative distribution of the fraction of galaxies in our final Lyman-break sample (\S\ref{sec:selection}) as a function of the probability that their photometric redshift is $z>4$. Here, we exclude photometric bands covering the rest-optical regime of each dropout sample (see Appendix \ref{app:photoz}). We plot each of the six galaxy sub-samples used in \S\ref{sec:distributions} separately.}
\label{fig:photozCdf}
\end{figure}

As discussed in \S\ref{sec:selection}, we choose to select galaxies based on Lyman-break color criteria rather than photometric redshifts. 
Utilizing photometric redshifts can bias our sample preferentially towards objects with strong rest-optical lines since such lines imprint unique long-wavelength NIRCam color patterns not easily reproduced with lower redshift solutions.
Nonetheless, we aim to quantify the potential low-redshift interloper fraction of our sample and assess whether the presence of such interlopers might impact our main conclusions.
To this end, we calculate the photometric redshifts of all galaxies in our final Lyman-break cut sample but excluding bands which fall in the rest-optical regime.
That is, we only fit using bands blueward of (and including) F150W and F200W for the F775W and F090W dropout samples, respectively.
We use the \textsc{beagle} two-component SFH setup described in \S\ref{sec:models} and allow redshifts in the range $z=0-10$ with a uniform prior.

For nearly every sample subset considered in \S\ref{sec:distributions}, $\gtrsim$90\% of galaxies have a high probability ($>$90\%) of lying at $z>4$ (see Fig. \ref{fig:photozCdf}).
The only exception is the very faint ($\Muv{} > -18$) subset of the F090W dropout sample, in which $\approx$80\% of the galaxies have a large probability ($>$80\%) of lying at high redshifts (Fig. \ref{fig:photozCdf}).

In \S\ref{sec:OIIIHb_EW_results}, we found that the median \OIIIHb{} EW of reionization-era galaxies is substantially lower at fainter UV luminosities. 
Here, we verify that this result is not likely due to a potential moderate population of low-redshift contaminants in the very faint subsets.
We account for the high-redshift probabilities of each object when inferring the \OIIIHb{} EW distributions by replacing Eq. \ref{eq:bayes} with the following:
\begin{equation} \label{eq:bayes_photoz}
    P \left(\theta\right) \propto \prod_i \left[ \sum_j P_{i,j}\left(\mathrm{EW}\right) P_j\left(\mathrm{EW}|\theta\right) \right]^{P_i (z>4)}
\end{equation}
By raising the term in brackets to the power of $P_i (z>4)$ (i.e. the probability that galaxy $i$ lies at $z>4$), we weight the contribution of each source to the EW distribution constraint based on the likelihood that it lies at high redshift.

With Equation \ref{eq:bayes_photoz}, we infer median \OIIIHb{} EWs of $400^{+70}_{-70}$ \AA{} and $320^{+40}_{-50}$ \AA{} among the very faint F775W and F090W dropout subsets, respectively, using the TcSFH model outputs.
These median EWs are entirely consistent with our fiducial values reported in Table \ref{tab:inferredEWparameters}.
This indicates that a potential moderate ($\lesssim$10\%) population of low-redshift interlopers in our faintest bins cannot explain the strong UV luminosity dependence on the median \OIIIHb{} found in \S\ref{sec:OIIIHb_EW_results}. 
We have also verified that adopting Eq. \ref{eq:bayes_photoz} does not change our conclusions on the lack of a strong UV luminosity dependence on the H$\alpha$ EW distribution at $z\sim6$ (\S\ref{sec:Halpha_EW_results}).

\section{Assessment of the Goodness of SED Fits} \label{app:chisq}

\begin{figure}
\includegraphics{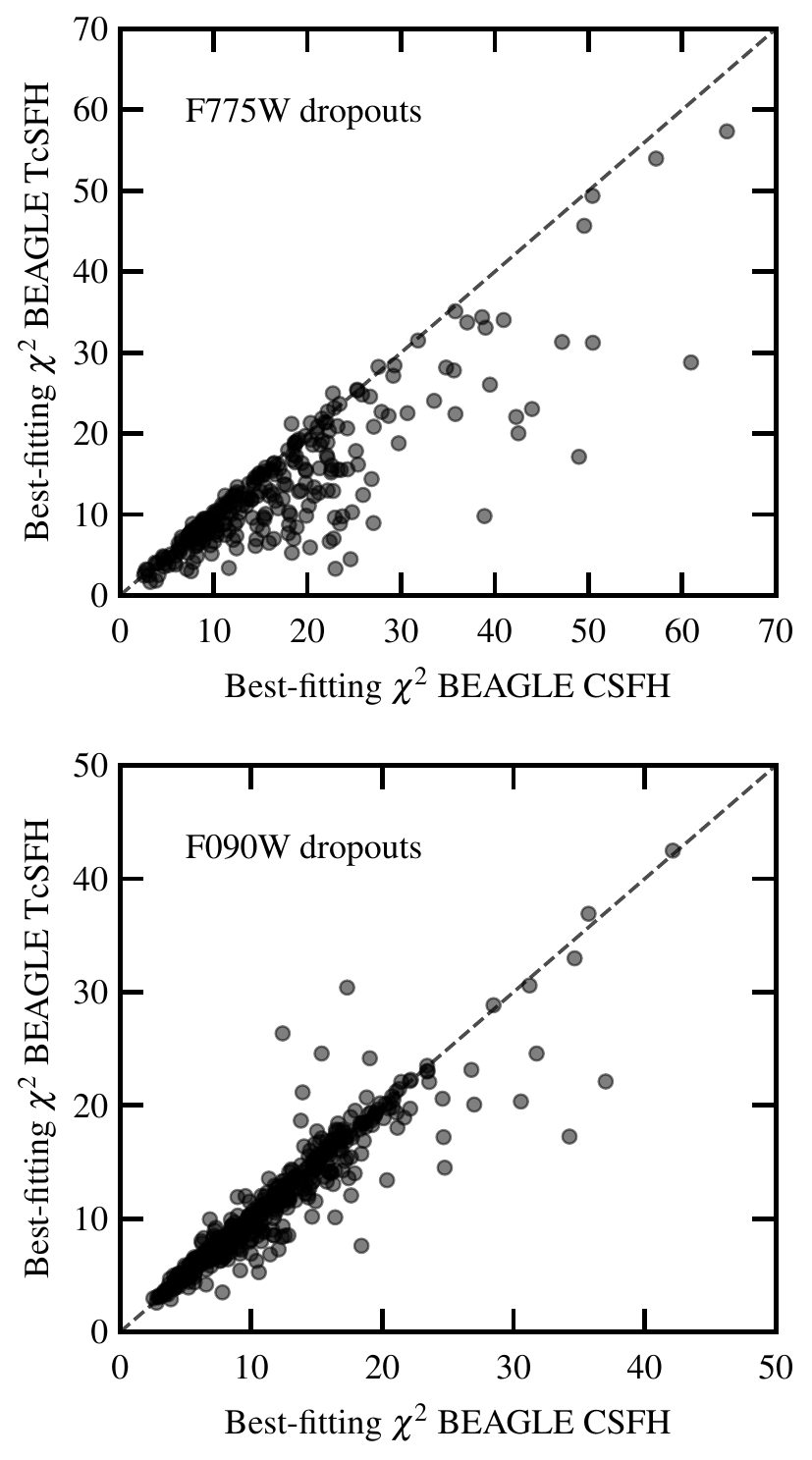}
\caption{A comparison of the best-fitting $\chi^2$ values (using all HST and NIRCam bands) between the \textsc{beagle} TcSFH model fits (y-axis) and the \textsc{beagle} CSFH fits (x-axis). We show the comparison for the F775W dropout ($z\sim6$) sample and the F090W dropout ($z\sim7-9$) sample in the top and bottom panels, respectively.}
\label{fig:chisq}
\end{figure}

\begin{figure}
\includegraphics{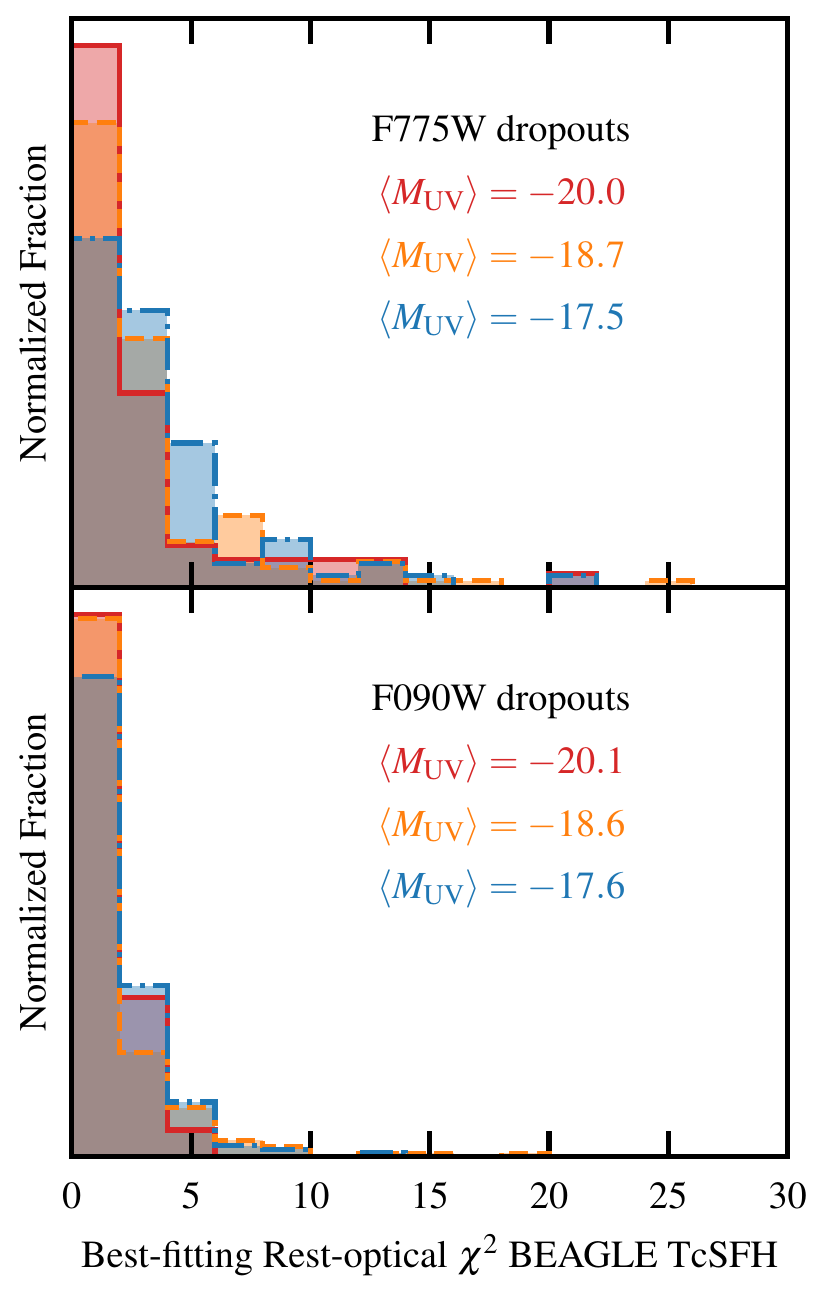}
\caption{The distribution of best-fitting $\chi^2$ values from the \textsc{beagle} TcSFH fits for each sample subset used in our \OIIIHb{} and H$\alpha$ EW distribution analyses (\S\ref{sec:distributions}). Here we are only considering the bands that probe the rest-optical, so F200W (F277W) and redder for the F775W (F090W) dropout samples, resulting in six (five) fitted bands. The rest-optical photometry is well fit by these models for the large majority of objects in each redshift and UV luminosity bin, with no strong differences in fit quality in the faintest bins.}
\label{fig:chisqRestOpt}
\end{figure}

Here we explore the goodness of our \textsc{beagle} SED fits which we utilize in our EW distribution analysis (\S\ref{sec:distributions}).
We show the best-fitting $\chi^2$ values of the F775W dropout ($z\sim6$) and F090W dropout ($z\sim7-9$) samples in the top and bottom panels, respectively, of Fig. \ref{fig:chisq}.
In the F775W dropout sample, it is clear that a significant fraction of objects are considerably better fit with the \textsc{beagle} TcSFH models than the CSFH models, motivating us to adopt the EW distribution results from the TcSFH fits as fiducial.
The reason why the TcSFH models provide substantially better fits to a subset of $z\sim6$ galaxies is discussed in detail in \S\ref{sec:bursty}.
Briefly, these models allow for solutions where the continuum is consistent with a young stellar population (age $\lesssim$30 Myr) while still yielding relatively weak \OIIIHb{} and H$\alpha$ emission from a recent downturn in SFR.
Such a contrast in fit quality is not seen in the F090W dropout sample given that the NIRCam data is not sensitive to H$\alpha$ emission at $z\gtrsim7$, which is necessary to determine if weak [OIII] emission can be explained by extremely low metallicity (see \S\ref{sec:bursty}).

We also check that the \textsc{beagle} TcSFH models generally deliver acceptable fits to the rest-optical photometry alone. 
In Fig. \ref{fig:chisqRestOpt}, we show the distributions of best-fitting $\chi^2$ values limited to the rest-optical bands, considering each redshift and UV luminosity subset adopted in \S\ref{sec:distributions} separately.
We find that the best-fitting rest-optical $\chi^2$ values are less than the number of fitted rest-optical bands (six for F775W dropouts, 5=five for F090W dropouts) for the large majority of objects ($\geq$85\%) in every subset.
Moreover, we find no strong differences in rest-optical fit quality among the faintest bins relative to the brighter bins.
 

\bsp	
\label{lastpage}
\end{document}